\documentclass[aps,prx,reprint,nofootinbib,twocolumn,superscriptaddress,showpacs,showkeys,longbibliography,floatfix,bibnotes]{revtex4-1}
\usepackage{eurosym}
\usepackage{amsmath,amssymb,amstext}
\usepackage{ulem}
\usepackage[usenames,dvipsnames]{color}
\usepackage{bm}
\usepackage{graphicx}
\usepackage{braket}
\usepackage{natbib}
\usepackage{comment}
\usepackage{dcolumn}
\usepackage[english]{babel}
\usepackage{wasysym}
\usepackage[colorlinks,bookmarks=false,citecolor=blue,linkcolor=red,urlcolor=blue]{hyperref}

\begin{document}
	
	\title{Fractional Quantum Zeno Effect Emerging from Non-Hermitian Physics}
	
	\author{Yue Sun}
	\affiliation{State Key Laboratory of Quantum Optics and Quantum Optics
		Devices, Institute of Laser Spectroscopy, Shanxi University, Taiyuan 030006,
		China} 
	\affiliation{Collaborative Innovation Center of Extreme Optics,
		Shanxi University, Taiyuan 030006, China}
	
	\author{Tao Shi}
	\thanks{tshi@itp.ac.cn}
	\affiliation{CAS Key Laboratory of Theoretical Physics, Institute of Theoretical Physics, Chinese Academy of Sciences, Beijing 100190, China}
	\affiliation{CAS Center for Excellence in Topological Quantum Computation, University of Chinese Academy of Sciences, Beijing 100049, China}

	\author{Zhiyong Liu}
	\affiliation{Shenyang National Laboratory for Materials Science, Institute
		of Metal Research, Chinese Academy of Sciences, Shenyang 110016, China} %
	\affiliation{School of Materials Science and Engineering, University of
		Science and Technology of China, Shenyang 110016, China}
	
	\author{Zhidong Zhang}
	\affiliation{Shenyang National Laboratory for Materials Science, Institute
		of Metal Research, Chinese Academy of Sciences, Shenyang 110016, China} %
	\affiliation{School of Materials Science and Engineering, University of
		Science and Technology of China, Shenyang 110016, China}
	
	\author{Liantuan Xiao}
	\affiliation{State Key Laboratory of Quantum Optics and Quantum Optics
		Devices, Institute of Laser Spectroscopy, Shanxi University, Taiyuan 030006,
		China} 
	\affiliation{Collaborative Innovation Center of Extreme Optics,
		Shanxi University, Taiyuan 030006, China}
	
	\author{Suotang Jia}
	\affiliation{State Key Laboratory of Quantum Optics and Quantum Optics
		Devices, Institute of Laser Spectroscopy, Shanxi University, Taiyuan 030006,
		China} 
	\affiliation{Collaborative Innovation Center of Extreme Optics,
		Shanxi University, Taiyuan 030006, China}
	
	\author{Ying Hu }
	\thanks{huying@sxu.edu.cn}
	\affiliation{State Key Laboratory of Quantum Optics and Quantum Optics
		Devices, Institute of Laser Spectroscopy, Shanxi University, Taiyuan 030006,
		China} 
	\affiliation{Collaborative Innovation Center of Extreme Optics,
		Shanxi University, Taiyuan 030006, China}
	
\begin{abstract}
Exploring non-Hermitian phenomenology is an exciting frontier of modern physics. However, the demonstration of a non-Hermitian phenomenon that is quantum in nature has remained elusive. Here, we predict quantum non-Hermitian phenomena: the fractional quantum Zeno (FQZ) effect and FQZ-induced photon antibunching. We consider a quantum optics platform with reservoir engineering, where nonlinear emitters are coupled to a bath of decaying bosonic modes whose own decay rates form band structures. By engineering the dissipation band, the spontaneous emission of emitters can be suppressed by strong dissipation through an algebraic scaling with fractional exponents—the FQZ effect. This fractional scaling originates uniquely from the divergent dissipative density of states near the dissipation band edge, different from the traditional closed-bath context. We find FQZ-induced strong photon antibunching in the steady state of a driven emitter even for weak nonlinearities. Remarkably, we identify that the sub-Poissonian quantum statistics of photons, which has no classical analogs, stems here from the key role of non-Hermiticity. Our setup is experimentally feasible with the techniques used to design lattice models with dissipative couplings.
\end{abstract}
	\maketitle

	\begin{figure}[tb]
		\centering
		\includegraphics[width=1\columnwidth]{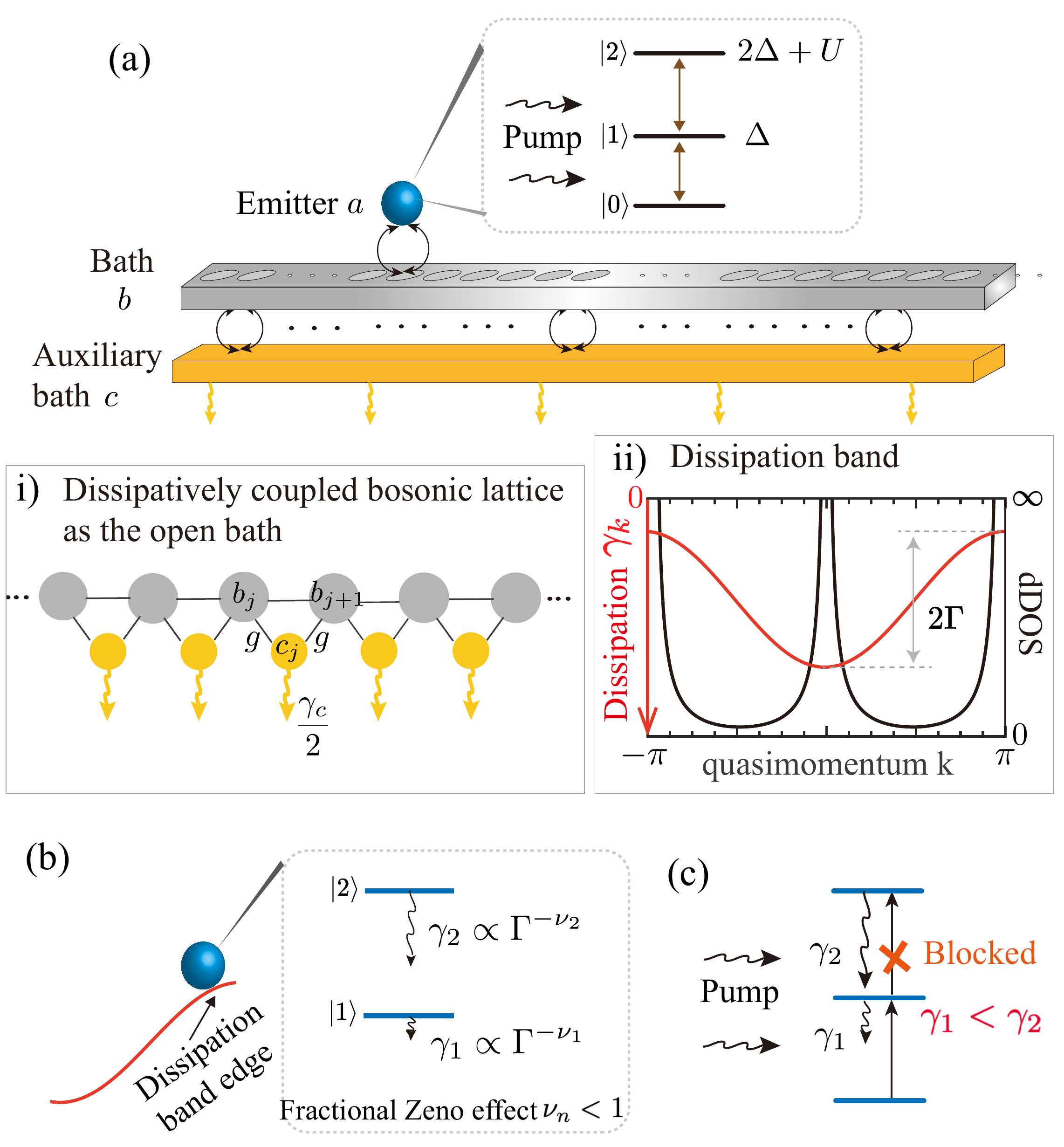}
		\caption{Illustration of a nonlinear emitter coupled to a 1D open bath and the fractional quantum Zeno (FQZ) effect. (a) Setup. An emitter $a$ (blue ball) is coupled to a bosonic bath $b$ (gray). The bath $b$ is coupled to the auxiliary bath $c$ (yellow). Top: energy levels of a driven emitter with $n$ excitations ($n=0,1,2$), where $\Delta$ is the detuning from the central frequency of the bath $b$ and $U$ is the strength of nonlinear interaction. (i) Engineered open bath for the emitter. Bosonic mode $b_j$ in a lattice of $N_b$ sites, in general, has the tunneling rate $J$ and phase $\protect\theta$. The auxiliary modes $c_j$ have the common, large loss rate $\gamma_c/2$. Both $b_j$ and $b_{j+1}$ are coupled to the lossy mode $c_j$ with the coupling rate $g$ ($j=1,...,N_b-1)$. This engineers a dissipative coupling between $b_j$ and $b_{j+1}$ with the rate $\Gamma/2$, where $\Gamma=4g^2/\gamma_c$ [cf.~Eq.~(\ref{eq:rho})]. (ii) The open bath exhibits the dissipation band $\gamma_k$ (red curve) with the bandwidth $2\Gamma$. The corresponding dissipative density of states (dDOS) is shown by the black curve. (b) FQZ effect. Strong dissipation $\Gamma\gg 2J$ confines the emitter to the dissipation band edge with divergent dDOS. This leads to the FQZ effect and a multiexcitation quasibound state: Their decay rate $\gamma_n\propto \Gamma^{-\nu_n}$ ($\nu_n<1$ for $n=1,2)$ is suppressed by large $\Gamma$ via the scaling with fractional exponent $\protect\nu_n$. (c) FQZ-induced photon antibunching. The $n$-dependent FQZ effects allow one to engineer $\gamma_1<\gamma_2$ in such a way to enable photon blockade in the emitter even for weak nonlinearity $U$. }
		\label{Fig:setup}
	\end{figure}
	
	\section{INTRODUCTION}
While enchanting non-Hermitian properties unattainable in Hermitian systems are widely revealed~\cite{JanReview2021,Wang2018,Kunst2018,Gong2018,Zhou2018,Kawabata2019,Xue2020,Nunnenkamp2020,Wang2021,Bender1998}, genuine quantum non-Hermitian phenomenology is still largely  uncharted territory. For a classical system, coupling to an environment induces dissipation and can be well described by a non-Hermitian Hamiltonian. On the quantum level, however, dissipation occurs with quantum fluctuations, something fundamentally absent in the classical regime. Exploring quantum non-Hermitian phenomena, thus, involves identifying observable consequences driven by the role of non-Hermiticity amid quantum fluctuations. In this direction, the possibility to  ingeniously design dissipation with quantum materials enables unique opportunities. Experimentally, non-Hermitian physics such as parity-time symmetry~\cite{Bender1998} were simulated with photons~\cite{Xiao2017,Fang2017,Ozturk2021}, atoms~\cite{Yanbo2022,Dongdong2022,Li2019,Takasu2020,antiPT2016}, electronic spins~\cite{Wu2019}, and superconducting qubits~\cite{Naghiloo2019}. Recently, realizations of dissipative couplings with an array of photonic resonators~\cite{FanSH2021}, atomic spin waves~\cite{Dongdong2022}, or polaritons~\cite{Pickup2020,Pernet2022} coupled to a reservoir have led to engineered dissipative lattice models with non-Hermitian bands. However, the observed phenomena to date were classical in nature. Theoretically, non-Hermitian effects under the quantum mechanical framework are being actively pursued~\cite{Nakagawa2018,Yamamoto2019,Gopalakrishnan2021,Roccati2022,Gong202201,Gong202202,Longhi2016}, but the difficulty to tackle the dynamical long-time behaviors of open many-body quantum systems has severely limited these studies to single-particle physics and short timescales. Though highly sought after, unambiguous quantum non-Hermitian phenomenon have remained elusive so far.  

Here, we predict quantum non-Hermitian phenomena—the fractional quantum Zeno (FQZ) effect and FQZ-induced photon antibunching, based on a quantum optics setup harnessing reservoir engineering~\cite{Poyatos1996,Diehl2008,Verstraete2009,Diehl2011,Muller2012,Tomadin2012,Douglas2015,Ramos2014,Cian2019,Chang2018,Harrington2022}. The system consists of nonlinear quantum emitters, such as atoms and artificial atoms, which host multiple excitations with nonlinear interactions. We exploit an engineered ``open bath'' represented by a continuum of decaying bosonic modes in a lattice with dissipative couplings, whose own dissipation rates form band structures. Different from previous reservoir engineering, which designs the energy dispersion of a closed medium, we manipulate the dissipation band of the open bath to dynamically tailor the system-bath interaction. As the central result, we show the FQZ effect emerging in the long-time emitter dynamics, where the spontaneous emission of emitters is dissipatively suppressed according to an algebraic scaling with fractional exponents. The scaling behaviors of $n=1,2,...$ excitations, moreover, are distinct, which can be controlled via the detuning of emitters. By analyzing the steady-state second-order correlation function of the weakly driven emitter, we show strong, FQZ-induced photon antibunching even for a weak nonlinearity. Different from conventional quantum light generation~\cite{Paul1982,Bamba2011,Kong2022}, the present sub-Poissonian quantum statistics of photons is driven by structured dissipation captured by a non-Hermitian Hamiltonian, which opens the door to exploring non-Hermitian quantum optics. 

The FQZ effect predicted here is conceptually different from the familiar nonanalytic phenomena in quantum optics and condensed matter physics. It generically results from the combination of strong dissipation and a divergent \textit{dissipative} density of states (dDOS) — the number of bath modes with a particular dissipation rate — near the dissipation band edge. In the closed-bath context, including the extensively studied spin-boson problems in quantum optics and various impurity models in condensed matter physics~\cite{Balatsky2006,Hur2012,Goldstein2013,Shitao2016}, the system-bath interaction is constrained by energy conservation. Instead, our setup has the distinctive feature that such energetic constraint is removed owing to the open nature of the bath. Thus, while gapless modes are crucial for conventional nonanalytic phenomena~\cite{Goldstein2013}, the present fractional scaling is possible regardless of whether or not the bath has gapless energy or dissipation spectra. It further has the unique property where the scalings of $n$-excitation states are different for various $n$, which cannot occur in platforms with the closed bath~\cite{Chang2018,Tudela2015,Sollner2015,Liu2016,fQHE2012,Shi2017,Bello2019,Nakazato1996,Tudela2018,John1990,Shitao2016,Shi2018,Luengo2019,Luengo2020,Luengo2021,Ramos2016}, including nanophotonic systems and solid-state band-gap materials, where the resonance condition locks the emitter with the bath energetically. As an important consequence, we show the sub-Poissonian quantum statistics of photons resulting from the key role of non-Hermiticity. 

We also emphasize a novel approach based on the Keldysh formalism for efficient solutions of the \textit{full} dynamics of nonlinear emitters immersed in the open bath. In particular, we are able to identify the role played by non-Hermitian Hamiltonians in the steady-state quantum correlation functions of the weakly driven emitters, capitalizing on a deep relation~\cite{Shitao2015,Yue2016} between the master equation and the scattering theory. Mathematically, compared with the established Green function approach associated with the closed bath, the open bath enriches the analytic structure of the emitter Green functions: The branch cut corresponding to the continuum generally becomes a ``branch circle" instead of a line, which challenges the studies of quantum few-body and many-body dynamics. We circumvent this difficulty by introducing the effective fictitious bath. 

The phenomena and principles described here exist for a wide class of dissipation band structures and for the open bath in arbitrary dimensions. 
Our results are also of relevance for current experiments with quantum materials, where realizing dissipative lattice models and engineering non-Hermitian band structures are state of the art, while demonstrating quantum non-Hermitian phenomena has remained an open challenge. 

We begin in Sec.~\ref{sec:main} by outlining the main results and their implications. In Sec.~\ref{sec:model}, we describe the 
quantum mechanical model of emitters coupled to a one-dimensional (1D) open bath. In Sec.~\ref{sec:formalism}, we develop
the general formalism for solutions of the dynamics and quantum correlations of weakly driven emitters. In Sec.~\ref{sec:FQZ}, we study the spontaneous emission of an emitter with single and two excitations, respectively, and show the FQZ effect. We also study the quantum correlation between two emitters. In Sec.~\ref{sec:weakdriving}, we analyze the second-order correlation function of a weakly driven emitter and show FQZ-induced sub-Poissonian
light generation. In Sec.~\ref{sec:arbitrary}, we present a scaling analysis for the arbitrary bath. In Sec.~\ref{sec:exp}, we discuss the experimental implementations, and we conclude our paper in Sec.~\ref{sec:con}.

	\section{Overview of main results}\label{sec:main}
	
	\label{sec:main}
	
	Our setup is illustrated in Fig.~\ref{Fig:setup}(a), where one or several emitters are coupled to an engineered bosonic open bath in one dimension. The emitter is nonlinear with multiple bosonic excitations and is weakly driven by an external field. This may be cavities with Kerr interactions or two-level systems in the limit of infinite repulsive interactions. The emitters and the open bath \textit{as a whole} constitute an open quantum system whose dynamics is governed by a master equation (see Sec.~\ref{sec:model}): It contains a time evolution governed by an effective non-Hermitian emitter-bath Hamiltonian and quantum jumps~\cite{Dalibard1992,Daley2014} associated with the emission of quanta from the open bath into the auxiliary reservoir [Fig.~\ref{Fig:setup}(a)]. The effective Hamiltonian associated with the open bath describes a dissipative lattice with an energy band $\epsilon_{k}=2J\sin (k+\theta )$ and a dissipation band $\gamma _{k}=\Gamma (1+\cos k)$. In the limit of $\Gamma /(2J)\rightarrow 0$, the closed bath is recovered, where the rich physics of various impurity models has been extensively studied ranging from quantum optics~\cite{Hur2012,Shitao2016,Goldstein2013} to condensed matter physics~\cite{Balatsky2006}. 
	
	Instead, we are interested in the unexplored physics in the strong dissipation limit $
	\Gamma /(2J)\gg 1$, where the interaction between the emitters and the open bath is no longer restricted by energy conservation, and the 
	dissipation band $\gamma _{k}$ is expected to play the central role in determining quantum emissions. Analogous to the density of states associated with energy dispersion, we characterize the dissipation band with the dDOS, i.e., the number of modes $D_s(\gamma)$ at the dissipation rate $\gamma$. In 1D, we have
	\begin{equation}
		D_s(\gamma)\propto \Big|\frac{1}{\partial_{k}\gamma_{k}}\Big|. \label{eq:dos}
	\end{equation}
We explore how $\gamma(k)$ and dDOS influence the spontaneous emissions of multiple excitations of emitters and their statistics as quantified by the second-order correlation function.
		
	It is challenging to solve the master equation of the entire open system consisting of the emitters and the open bath, when the emitters are highly nonlinear and externally driven. In particular, it is difficult to pinpoint the role played by the non-Hermitian Hamiltonian in the dynamical long-time behaviors of the emitters. We address this challenge by developing a framework based on the Keldysh formalism~\cite{Rammer2007,Sieberer2016} and the scattering theory. 
	
	Our road map consists of two steps (see Sec.~\ref{sec:formalism}). First, we develop an efficient approach to solve the spontaneous emission of multiple excitations in nonlinear emitters without driving, based on the Keldysh formalism (Fig.~\ref{Fig:method}). With respect to the Green function approach associated with the closed bath, the open bath significantly enriches the analytic structure of the emitter Green function. Here, the branch cut corresponding to the continuum generally becomes a ``branch circle" instead of a line, which separates the first Riemann surface to two disconnected regions, and challenges the studies of multiple excitations. This issue is solved by the analytic continuation,  which naturally introduces the concept of an effective fictitious bath. Second, we connect the steady-state quantum correlation functions of the weakly driven emitter with Green functions of the undriven case, based on a relation between the multiparticle scattering amplitudes and the steady state of the master equation as initially proposed in Refs.~\cite{Shitao2015,Yue2016}. 
	
	In particular, to demonstrate quantum non-Hermitian phenomena, we focus on the quantum nature of light as indicated by the second-order correlation function $g^{(2)}(0)<1$ of the driven emitter. We explicitly identify the role of non-Hermiticity by establishing the relation
	\begin{equation}
		g^{(2)}(0)=\left\vert \frac{1}{1-U\Pi _{f}(2\omega _{\mathrm{d}})}
		\right\vert ^{2},  \label{g2}
	\end{equation}
where $\omega_d$ is the driving frequency, $U$ is the strength of the nonlinear interaction, and the function $\Pi _{f}$ is directly related to Green functions determined by non-Hermitian Hamiltonians. This allows us to show how the sub-Poissonian photon statistics despite weak nonlinearity results from the non-Hermitian part of the Hamiltonian. 
	
	Applying the above formalism, we predict the FQZ effect, where the spontaneous emission
	rate $\gamma _{n}$ of $n$ excitations ($n=1,2,...$) in emitters scales with the bath dissipation rate $\Gamma$ as
	\begin{equation}
		\gamma _{n}\propto \Gamma^{-\nu_n},   \label{eq:fractionzeno}
	\end{equation}
where the exponent $\nu_n<1$. Equation (\ref{eq:fractionzeno}) indicates that the decay rate is suppressed by strong dissipation, similarly as the well-known QZ effect (see, e.g.,~Refs.~\cite{Scully1997,Misra1977,Itano1990,Wang2008,Maniscalco2008,Han2009,Syassen2008,Signoles2014,Froml2020,Hu2015}), but the algebraic scaling here has the fractional exponent, differently from the characteristic featureless $\Gamma^{-1}$ scaling of the QZ effect. Moreover, the scaling behavior of $\gamma_n$ varies with the number $n$ of excitations, which can be controlled via the detuning of the emitter. 
		
	 The physical picture for the FQZ effect is the following. When the open bath is strongly dissipative by itself, the emitters are dynamically imposed to mainly couple with the long-lived bath modes hosted near the dissipation
	band edge [Fig.~\ref{Fig:setup}(b)]. The dDOS of this region then determines the scaling behaviors of the emitters: Whenever the dDOS there diverges, fractional scaling arises as in Eq.~(\ref{eq:fractionzeno}), regardless of whether $\gamma_k$ is gapless or not. The scaling analysis for the open bath with arbitrary dissipation band structure and dimension is summarized in Table~\ref{table1} (see Sec.~\ref{sec:arbitrary}). 
	
	Specifically, for a single excitation, we identify the long-lived quasibound state as the superposition of the emitter excitation and
	the bosonic modes of the open bath, whose decay rates as given by Eq.~(\ref
	{eq:fractionzeno}) determine the emitter dynamics at long times (see Fig.~\ref{Fig:oneatom} and Sec.~\ref{subsec:one}). The bath modes in the quasibound state are confined to the dissipation band edge, forming a giant cloud around the emitters. The spatial size of the cloud grows with the bath dissipation rate $\Gamma$ through a nonanalytic
	scaling. Consequently, when two emitters
	are present, the open bath mediates simultaneous sizable and remote quantum correlations of
	emitters, with a correlation length increasing with $\Gamma$. These results are shown in Fig.~\ref{Fig:twoatom} and Sec.~\ref{subsec:two}. 
		
	For two excitations with nonlinear interactions, we find the emergence of quasibound states with the FQZ scaling $\nu_2<\nu_1<1
	$. For instance, $\nu_1=1/2$
	and $\nu_2=1/3$ for single and two excitations indicates a much faster spontaneous decay of two
	excitations compared to a single excitation. Moreover, the scalings can be independently controlled via
	the emitter detuning and the nonlinear interaction. These results are summarized in Fig.~\ref{Fig:U} and Sec.~\ref{sec:two}.
	
	As a remarkable manifestation of FQZ effect, we show it opens a new route toward the generation of strong antibunching, even in the limit of weak interactions [Figs.~\ref{Fig:setup}(c) and~\ref{Fig:wd}]. By using Eq.~(\ref{g2}) to analyze $g^{(2)}(0)$ of a single emitter under a weak driving field, we find significant antibunching behavior of excitations (see Sec.~\ref{sec:weakdriving}). We show how it results from the appropriately engineered $\gamma_1<\gamma_2$, enabled by the independently controllable, different FQZ effects of one and two excitations. In particular, we can achieve strong antibunching in the weak-interaction regime $U/2<2\gamma_1<\gamma_2$, even when interactions are so small as $U/2<\gamma_1$. 
	
	The FQZ-induced antibunching originates from the structured dissipation of the open bath and, therefore, is conceptually different from the conventional mechanisms including strong nonlinearities~\cite{Paul1982} and interferences~\cite{Bamba2011,Kong2022}. Notably, it represents a first genuine quantum phenomenon emerging from non-Hermitian bands.

	Our results have important implications for advancing the field of non-Hermitian physics into fully quantum regimes. Despite significant interest and ongoing efforts, state of the art experiments (see, e.g., Refs.~\cite{antiPT2016,Fang2017,Pickup2020,Pernet2022,FanSH2021,Yanbo2022,Dongdong2022,Fang2017,Naghiloo2019,Li2019,Ozturk2021}) on non-Hermitian phenomena in quantum systems have been limited to classical or single-particle physics. Theoretically (see, e.g., Refs.~\cite{Nakagawa2018,Yamamoto2019,Gopalakrishnan2021,Roccati2022,Gong202201}), it remains an open challenge to explicitly show quantum many-body phenomena purely governed by a non-Hermitian Hamiltonian in the full quantum dynamics including quantum jumps, without conditioning on the measurements such as postselections. Our work sheds light on how to surmount these challenges. 
			
	\section{Emitters coupled to an open bath}\label{sec:model}
	
	In this section, we describe in detail the theoretical model for the emitters coupled to a 1D open bath
	and outline the key quantities we are interested in. 
		
	Our setup in Fig.~\ref{Fig:setup}(a) consists of three ingredients: emitters 
	$a$ (blue ball), a bath of bosonic modes $b$ (gray), and an auxiliary bath of lossy 
	modes $c$ (yellow). (i) We consider the paradigm of one and two
	emitters. In the rotating frame with respect to the central frequency
	of $b$ modes, the Hamiltonian of emitters is written as 
	\begin{equation}
		H_{\text{emit}}=\sum_{l=1,2}\bigg[\bigg(\Delta a_{l}^{\dagger }a_{l}+\frac{U}{2}%
		a_{l}^{\dagger 2}a_{l}^{2}\bigg)+\varepsilon (a_{l}^{\dagger }e^{-i\omega _{%
				d}t}+\mathrm{H.c.})\bigg], \label{eq:Hemit}
	\end{equation}%
	where $\Delta $ is the detuning of emitters. The second term is the on-site Kerr interaction with strength $%
	U$, which, in the limits $U\rightarrow 0$ and $U\rightarrow \infty $
	describes a free boson mode and a two-level system, respectively. The third
	term describes the driving field with amplitude $\varepsilon $ and driving frequency 
	$\omega _{d}$. (ii) The free propagation of the bosonic mode $b$ in a
	lattice of $N_{b}$ sites is described by the tight-binding Hamiltonian $H_{%
		b}=\sum_{j=1}^{N_{b}}(Je^{i\theta }b_{j}^{\dagger }b_{j+1}+\mathrm{%
		H.c.})$ with the hopping strength $J$ and the nontrivial phase $\theta \neq
	0$, where $b_{j}$ ($b_{j}^{\dag }$) is the annihilation (creation) operator
	at site $j$. (iii) Finally, the lossy modes $c_{j}$ ($j=1,...,N_b-1$) have
	the common decay rate $\gamma _{c}/2$.   
		
	We assume that the coupling of emitters to local modes $b_{j=0,d}$ is
	described by the Jaynes-Cummings (JC) Hamiltonian $H_{\text{sb}}=\Omega
	(a_{1}b_{0}^{\dag }+a_{2}b_{d}^{\dag }+\text{H.c.})$ with a Rabi frequency $%
	\Omega $. As shown in (i) in Fig.~\ref{Fig:setup}(a), neighboring modes $b_{j}$ and $%
	b_{j+1}$ are both coupled to the mode $c_{j}$ with a coupling strength $g$, described by $H_{\text{bc}}=g\sum_{j}(b_{j}^{\dag }+b_{j+1}^{\dag
	})c_{j}+\text{H.c.}$
		
	When the decay rate $\gamma_c/2$ of the bath $c$ is much larger than all the other energy scales, it can be adiabatically eliminated~\cite{Kessler2012,Reiter2012} to yield
	a master equation for the reduced density matrix $\rho$ associated with the hybrid system that consists of the emitters 
	and the bath $b$, i.e.,
	\begin{equation}
		\partial _{t}\rho =-i[H_{\text{emit}}+H_{\text{sb}}+H_{b},\rho ]+%
		\mathcal{D}_{b}[\rho ]. \label{eq:rho}
	\end{equation}
	Here, the dissipator takes the form
	\begin{equation}
		\!\!\!\!\mathcal{D}_{b}[\rho ]=-\{H_{D},\rho \}+\Gamma
		\sum_{j}\left(b_{j+1}\rho b_{j}^{\dagger }+b_{j}\rho b_{j+1}^{\dagger
		}+2b_{j}\rho b_{j}^{\dagger }\right), \label{eq:D}
	\end{equation}
	where $H_{D}=\Gamma \sum_{j}
	[ (b_{j}^{\dagger }b_{j+1}+\mathrm{H.c.})/2+b_{j}^{\dagger }b_{j}]
	$ with the effective decay rate $\Gamma =4g^{2}/\gamma_c$. The dissipator $\mathcal{D}_b$ leads to a dissipative coupling rate $\Gamma/2$ between neighboring bath modes $b_j$ and $b_{j+1}$ in the lattice on top of the coherent coupling rate $J$, along with an on-site decay rate $\Gamma$ of $b_j$.  We note that dissipatively coupled lattices have been recently engineered with atoms~\cite{Yanbo2022,Dongdong2022} and photonic resonators~\cite{FanSH2021}. Subsequently, we denote the effective non-Hermitian Hamiltonian of Eq.~(\ref{eq:rho}) by $H_\textrm{eff}=H_{\text{emit}}+H_{\text{sb}}+H_{b}-iH_{D}$. 
		
		Thus, the master equation~(\ref{eq:rho}) describes a scenario where quantum emitters are coupled to an ``open bath'' $b$, which undergoes dissipation by itself as governed by the dissipator $\mathcal{D}_{b}$, apart from its own coherent evolution as governed by $H_{b}$.
	Competition of this two processes is characterized by the ratio $\Gamma /(2J)
	$. In the limit of $\Gamma /(2J)\rightarrow 0$, the traditional closed bath is recovered~\cite{Balatsky2006,Hur2012,Shitao2016,Goldstein2013}, whereas, in the opposite limit of $\Gamma /(2J)\gg 1$, the bath is dominated by its open nature. 
					
	Our goal is to study the spontaneous emissions of multiple excitations in emitters and the dynamics of quantum correlation functions. We specifically consider two paradigmatic cases.
	
(i) We first consider the case without the driving field ($\varepsilon =0$), and the emitter $a_{1}$ is initially populated with one
	and two excitations, while the bath $b$ is initially in the vacuum state. We study the spontaneous
	emissions of excitations at times $t>0$ characterized by 
	\begin{eqnarray}
		G(t) &=&-i\langle 0 |a_{l}(t)a_{1}^{\dagger }(0)|0\rangle ,  \notag \\
		D(t) &=&-i\frac{1}{2}\langle 0 |a_{1}^{2}(t)a_{1}^{\dagger
			2}(0) |0\rangle ,
	\end{eqnarray}%
	with $l=1,2$, where the average is taken on the vacuum state $|0\rangle$.
	
(ii) We then consider a single emitter $a_1$ driven by a weak 
	field ($\varepsilon \neq 0$). We analyze the statistics of the emitter
	excitations in the steady state, as quantified by the second-order correlation
	function 
	\begin{equation}
		g^{(2)}(\tau )=\frac{1}{n^{2}}\text{Tr}\left[a_{1}^{\dagger }a_{1}^{\dagger
		}(\tau )a_{1}(\tau )a_{1}\rho _{\mathrm{ss}}\right] \label{eq:g2}
	\end{equation}
	with $n=\text{Tr}
	(a_{1}^{\dagger }a_{1}\rho _{\mathrm{ss}})$ being the first-order correlation function, in the steady state $\rho_{\mathrm{ss}}$ of the master equation~(\ref
	{eq:rho}). Whenever $g^{(2)}(0)<1$, the statistics is sub-Poissonian~\cite{Paul1982}, which is genuine quantum statistics with no classical
	analogs. 
	
	\section{Formalism}
	
	\begin{figure}[tb]
		\centering
		\includegraphics[width=0.9\columnwidth]{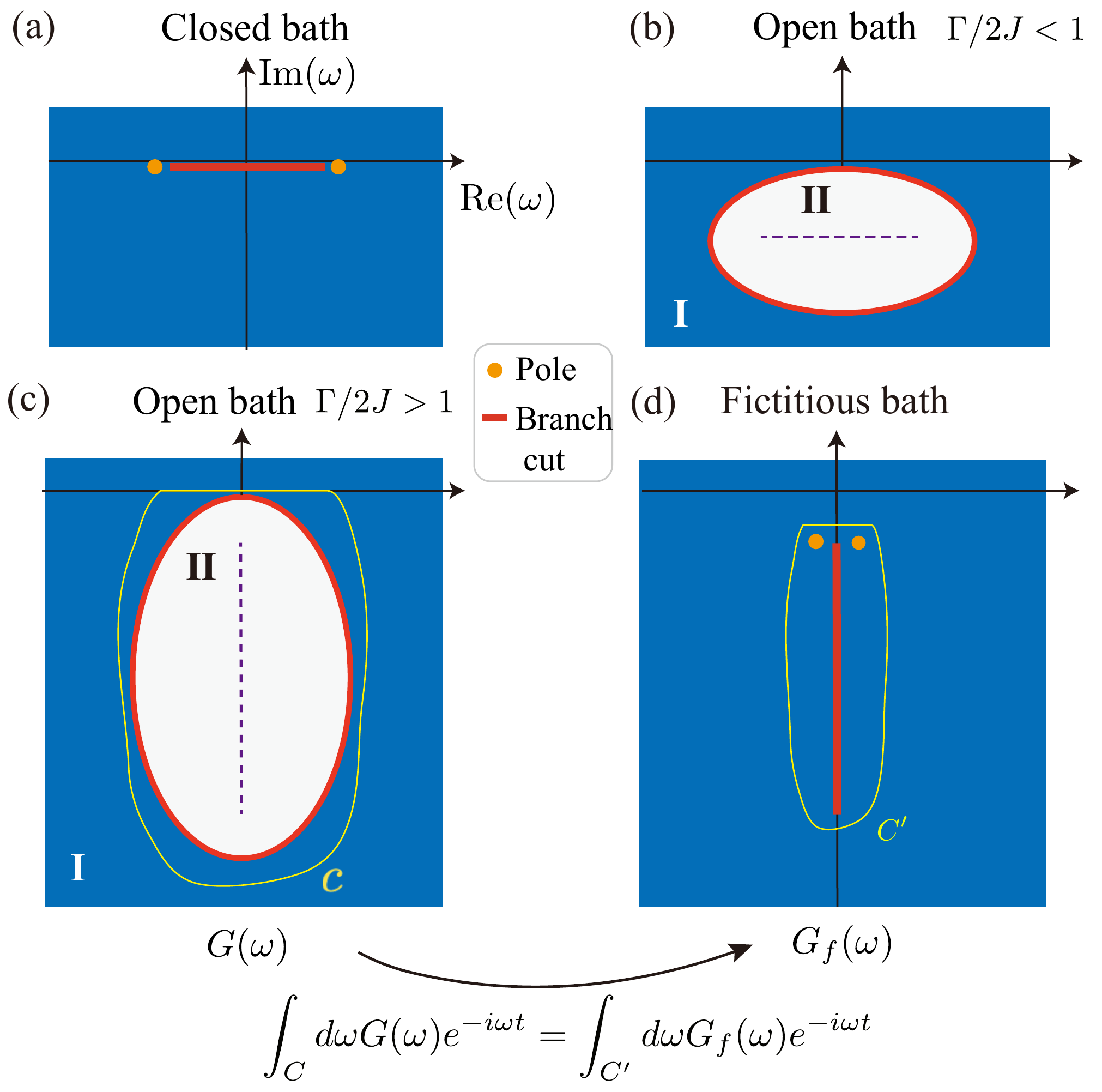}
		\caption{Illustration of the concept of the formalism for emitters in the open bath. The illustrated
			bath has the complex energy spectrum $\protect\epsilon_k-i\protect\gamma_k$, with $%
			\protect\epsilon_k=2J\sin k$ and $\protect\gamma_k=\Gamma (1+\cos k)$, for $k\in
			(-\protect\pi,\protect\pi]$. (a)—(c) Distinct analytical structures of the 
			Green function $G(\protect\omega)$ in the complex frequency plane $\protect%
			\omega=\text{Re}(\protect\omega)+i\text{Im}(\protect\omega)$, when the bath
			is (a) closed, (b) open with $\Gamma/(2J)<1$, and (c) open with $\Gamma/(2J)>1$. In (a), the orange dots denote the poles of 
			$G(\protect\omega)$, representing the bound states, and the red line denotes
			the branch cut $\in [-2J,2J]$, representing the scattering continuum. In (b)
			and (c), the red ellipses denote the branch cut parametrized by $(\protect%
			\epsilon_k, \protect\gamma_k)$, whose major axis is denoted by the dashed
			line. The self-energy~(\protect\ref{eq:Sigma0}) obtained from the residue
			theorem~\protect\cite{SM} is $\Sigma(\protect\omega)=0$ in region II (white)
			but is finite in region I (blue). Regions I and II are separated by the branch cut.
			The emitter dynamics is $G(t)=\int_C (d\omega/2\pi) G(\protect\omega%
			)e^{-i\protect\omega t}$ with an integration along a contour $C$ (yellow
			curve) in I. (d) Concept of the fictitious bath. We use the analytical
			continuation~\protect\cite{SM} to construct a fictitious bath with a simple
			spectrum, which generates the same emitter dynamics. The associated Green
			function $G_f(\protect\omega)$ has a line branch cut (red line) and two
			poles (yellow dots), so $G(t)=\protect\int_{C^{\prime }} (d\omega/2\pi)
			G_f(\protect\omega)e^{-i\protect\omega t}$ can be efficiently calculated.
			This concept is applicable for multiple excitations and wide 
			parameter regimes from $\Gamma/(2J)\ll 1$ to $\Gamma/(2J)\gg 1$. }
		\label{Fig:method}
	\end{figure}
	
	\label{sec:formalism}
		
	In this section, we systematically develop an efficient approach based on the Keldysh formalism and the scattering theory to solve for the dynamics and quantum correlation functions of the weakly driven and nonlinear emitters. In particular, it allows us to identify purely non-Hermitian scenarios. In principle, for small systems, the master equation (\ref{eq:rho}) can be solved numerically. However, to access the dynamical long-time behaviors of the highly nonlinear, driven emitters in the thermodynamic limit $N_{b}\rightarrow \infty$ requires intensive numerical calculations based on the finite-size scaling. Moreover, while the emitter and the open bath as a whole undergoes Markovian time evolution described by the master equation, we remark that the dynamics of the emitter, by itself, is non-Markovian because of the structured dispersion and dissipation of the open bath, and a successive elimination of the bath $b$ from Eq.~(\ref{eq:rho}) to obtain a master equation for the emitters is not valid.
	
Our approach to obtain the full dynamics of emitters hinges on two important elements. (i) In the master equation~(\ref{eq:rho}) without driving ($\epsilon=0$), the action of the jump operator depletes the excitations from the open bath, but the effective Hamiltonian $H_\textrm{eff}$ commutes with the number operator $N=\sum_{l} a_l^\dag a_l+\sum_{j} b_j^\dag b_j$, and. therefore, the steady state is the vacuum state. Consequently, the spontaneous emission dynamics of $n$ ($n=1,2,...$) excitations is fully captured by the $n$-particle retarded Green function in the vacuum state. In Secs.~\ref{subsec:self}—\ref{subsec:effective}, we develop the formalism for undriven emitters coupled to the open bath.  (ii) When emitters are weakly driven ($\epsilon\neq 0$), although the steady state of the master equation is out of equilibrium, which violates the dissipation-fluctuation theorem, we are able to connect the correlation functions of the weakly driven emitter with Green functions of the undriven case following the spirit of Refs.~\cite{Shitao2015,Yue2016}, as described in Sec.~\ref{subsec:wd}. Detailed derivations in our formalism can be found in Appendixes~\ref{sec:G} and \ref{sec:steady}.

	\subsection{Non-Hermitian emitter-bath Hamiltonian}\label{subsec:self}
	
Without the driving field ($\epsilon =0$), at zero temperature the steady state of the master equation (\ref{eq:rho}) is
the vacuum state $|0\rangle$. In the vacuum state, the time-ordered single-particle Green functions $G^{t}(t)=-i\langle 0\vert \mathcal{T}a_{l}(t)a_{1}^{\dagger
}(0)\vert 0\rangle $ and the retarded Green functions $G^{R}(t)=-i\langle 0\vert 
\mathcal{[}a_{l}(t),a_{1}^{\dagger }(0)]\vert 0\rangle
$ coincide with each
other, i.e., $G^{t}(t)=G^{R}(t)\equiv G(t)$. As a result, in the
frequency domain, all the emitter Green functions in the single-excitation subspace are determined by $
G(\omega )=\langle 0 |a[1/(\omega-H_\text{eff})]a^{\dag }|0\rangle$. In general, in the vacuum state, all the emitter Green functions in the $n$-excitation subspace are determined by the retarded $n$-particle ($n=1,2,...$) Green function. Hereupon, we drop ``retarded'' for convenience. 
	
	Thus, the dynamics of undriven emitters is purely 
	governed by the effective emitter-bath Hamiltonian $H_\textrm{eff}$ (see also Appendix~\ref{sec:steady}), which in the momentum space is written as
	\begin{eqnarray}
		H_\text{eff} &=&H_{a}+\sum_{k}(\epsilon_{k}-i\gamma
		_{k})b_{k}^{\dag }b_{k}  \notag \\
		&+&\frac{\Omega }{\sqrt{N_{b}}}\sum_{k}(a_{1}^{\dag }+a_{2}^{\dag
		}e^{ikd})b_{k}+\mathrm{H.c.} \label{eq:H}
	\end{eqnarray}%
with $b_{k}=(1/\sqrt{N_{b}})\sum_{j}b_{j}e^{-ikj}$ ($k\in (-\pi,\pi]$). In Eq.~(\ref{eq:H}), the second term is the non-Hermitian Hamiltonian describing the open bath, which exhibits structured dispersion relation $%
	\epsilon_{k}=2J\cos (k+\theta )$ and dissipation rate $\gamma _{k}=\Gamma (1+\cos k)$. It is easy to check that $[H_\textrm{eff},N]=0$ with $N=\sum_l a_l^\dag a_l+\sum_k b_k^\dag b_k$; i.e., the number of excitations is a good quantum number. In subsequent
	discussions, we assume $\theta=-\pi/2$ without loss of generality. 
	
	\subsection{Property of the Green function associated with the open bath}\label{subsec:green}
	
	To derive the dynamics of undriven emitters, we obtain the single-particle Green function $
	G(\omega )$ analytically by integrating out bath modes $b$. For instance, for a single emitter, we obtain 
	\begin{equation}
		G(\omega
		)=\frac{1}{\omega -\Delta -\Sigma (\omega )}
	\end{equation}
	which is determined by the self-energy
	\begin{equation}
		\Sigma (\omega)=\Omega^{2}\int \frac{dk}{2\pi }\frac{1}{\omega -\epsilon
			_{k}+i\gamma _{k}}.  \label{eq:Sigma0}
	\end{equation}%
	One can then further obtain the two-particle Green function 
	\begin{equation}
	D(\omega )=\frac{1}{\Pi^{-1}(\omega )-U}\label{Dw2}
	\end{equation}
	 with $\Pi (\omega
	)=i\int d\omega^{\prime }G(\omega ^{\prime })G(\omega -\omega
	^{\prime })/(2\pi )$. The general
	expression of Green functions for many emitters is derived in Appendix~\ref{sec:G}. 
		
	As a reference point, let us first recall how to obtain the emitter dynamics in a closed bath where $\gamma _{k}=0$. There, it follows from the Lehmann spectral representation $%
	G(\omega )=\int_{-\infty }^{\infty }\mathcal{A}(x)/(\omega -x+i0^{+})dx$
	that the emitter dynamics is determined by the analytic structure of the
	Green function $G(\omega )$ in the complex $\omega $ plane shown in Fig.~%
	\ref{Fig:method}(a). There, two isolated poles $\epsilon_{s}$ [i.e., $%
	G^{-1}(\epsilon_{s})=0$] correspond to the energies of bound states, and the
	branch cut $x\in \lbrack -2J,2J]$ represents the continuum of the bath,
	i.e., the scattering states. Thus, $G(\omega )$ in the entire complex
	plane is determined only by the nonzero spectral weight $\mathcal{A}(x)$ in
	the vicinity of poles and branch cut that can be obtained straightforwardly~\cite{SM}. Based on the Lehmann representation, the Fourier transform of $%
	G(\omega )$ can be obtained efficiently; the contribution from poles
	leads to the long-term oscillation of the remnant excitation in the emitter,
	while depending on the energy dispersion of the bath, the branch cut gives
	rise to (non-)Markovian decay [e.g., the power-law (exponential) decay $%
	\sim 1/t^{\delta }$ ($e^{-\gamma t}$)].
	
        For the open bath, however, the momentum-dependent decay rate $\gamma _{k}\neq 0$ generally leads to nontrivial and rich analytic structure of $G(\omega )$. As illustrated in Figs.~\ref{Fig:method}(b) and~\ref{Fig:method}(c), according to Eq.~(\ref{eq:Sigma0}), the branch cut deforms from the structureless straight line to an ellipse $%
	(\epsilon _{k},\gamma _{k})$ (red curve) parametrized by $k$. It 
	separates the first Riemann surface (RS) to two disconnected regions: The
	self-energy is $\Sigma (\omega )=0$ in the white region ($\omega \in \text{II%
	}$) but finite in the blue region ($\omega \in \text{ I}$).
	Depending on the ratio $\Gamma /(2J)$, the elliptical branch cut undergoes
	an interesting deformation [Figs.~\ref{Fig:method}(b) and~\ref{Fig:method}(c)]: Its major
	axis (dashed line) shrinks from a line on the real axis for $\Gamma /(2J)<1$
	to a point for $\Gamma /(2J)=1$ and then expands in the orthogonal
	direction in the negative imaginary axis for $\Gamma /(2J)>1$. 
	
	When $G(\omega)$ exhibits a branch circle, the 
two-particle Green function (\ref{Dw2}) contains an even more complicated analytic structure consisting of a
``branch area.'' As such, the computation of emitter dynamics is nontrivial, where it is generally hard to perform the Fourier
	transform (yellow curve) and the convolution directly. 
		
	\subsection{Efficient solution via a fictitious bath}\label{subsec:effective}
	
	In order to efficiently solve the emitter dynamics, we introduce the ``fictitious bath'': As we prove, an appropriate \textit{fictitious bath}
	with a \textit{simple} spectrum 
	\begin{equation}
		\bar{\omega}_{k}=\left\{ 
		\begin{array}{c}
			-i\Gamma -2J_{\mathrm{eff}}\sin k,\text{ for }\Gamma/2J<1 \\ 
			-i\Gamma -2iJ_{\mathrm{eff}}\cos k,\text{ for }\Gamma/2J>1%
		\end{array}%
		\right.  \label{obk}
	\end{equation}
	with $J_{\mathrm{eff}}=
	\sqrt{\left\vert J^{2}-\Gamma ^{2}/4\right\vert }$ generates exactly the \textit{same} dynamics of the emitters [Fig.~\ref{Fig:method}(d)].
	
	The dispersion~(\ref{obk}) of the fictitious bath is obtained using the
	analytic continuation, as illustrated in Figs.~\ref{Fig:method}(c) and~\ref{Fig:method}(d), which allows one to collapse the complex elliptical branch cut to a line coinciding with its major axis. Mathematically, the Fourier
	transforms $F(t)=\int_{C}(d\omega/2\pi)F(\omega )e^{-i\omega t}$ ($F=G,D$) are integrals
	along the yellow contour $C$ in region I, which is not contractible due
	to the elliptical branch cut in the first RS. However, it turns out that 
	 the self-energy $\Sigma (\omega )$ can be properly defined in the
	second RS by analytic continuation (see Appendix~\ref{sec:G}). As an example, for one emitter, in region II of the first RS and region I of the second RS, $\Sigma
	(\omega )\equiv \Sigma_{f}(\omega )$ is unified as%
	\begin{equation}
		\Sigma _{f}(\omega )=\Omega^{2}\int \frac{dk}{2\pi }\frac{1}{\omega -\bar{%
				\omega}_{k}}. \label{sigmaf}
	\end{equation}
	The advantage of the analytic continuation is that one can further deform
	the integral contour $C$ to $C^{\prime }$ (yellow curve) in the second RS.
	
	Thus, the emitter dynamics is completely
	determined by the fictitious bath
	\begin{equation}
		F(t)=\int_{C^{\prime }}\frac{d\omega}{2\pi} {F}%
		_{f}(\omega )e^{-i\omega t}
	\end{equation}
	for ($F=G,D$) through the simple analytic structure of ${G}_{f}(\omega )=1/(\omega
	-\Delta -\Sigma _{f})$ and $D_{f}(\omega )=[\Pi _{f}^{-1}(\omega )-U]^{-1}$
	with $\Pi _{f}(\omega )=i\int d\omega ^{\prime }G_{f}(\omega ^{\prime
	})G_{f}(\omega -\omega ^{\prime })/(2\pi )$. Figure~\ref{Fig:method}%
	(d) showcases the significantly simplified analytical structure of the Green function ${G}_{f}(\omega )$ associated with the fictitious bath. There, two poles $\bar{\epsilon}_{s}$ (orange dots) and a branch cut $\bar{%
		\omega}_{k}$ (red line) remarkably connect two foci of the original
	elliptical branch cut of ${G}(\omega)$ in the original model. The spectral
	weights $\bar{Z}_{s}^{-1}=1-\partial _{\omega }\Sigma_{f}|_{\omega =\bar{%
			\epsilon}_{s}}$ and $\mathcal{A}(x)$ of poles and the branch cut can be
	obtained analytically, giving rise to single- and two-excitation dynamics
	described by ${G}(t)$ and ${D}(t)$, respectively~\cite{SM}.
	
        The fictitious-bath approach can be applied to efficiently compute the dynamics
	of multiexcitations or multiple emitters in a generic 1D open bath in the
	thermodynamic limit. We emphasize, however, that the bath dynamics, e.g.,
	the multiexcitation scattering off the emitter and the propagation in the
	bath, \textit{cannot} be studied via the fictitious bath.

	Interestingly, the spectrum $\bar{\omega}_k$ of the fictitious bath in Eq.~(\ref{obk}) coincides with that of the original bath under open boundary conditions (OBCs)  in the thermodynamic limit. It is worth noting that the original open bath under OBCs exhibits a skin effect~\cite{Wang2018}. In contrast, the effective Hamiltonian of the fictitious bath exhibits completely different eigenstates, without a skin effect.  	
		
	\begin{figure*}[tb]
		\centering
		\includegraphics[width= 1.0\textwidth]{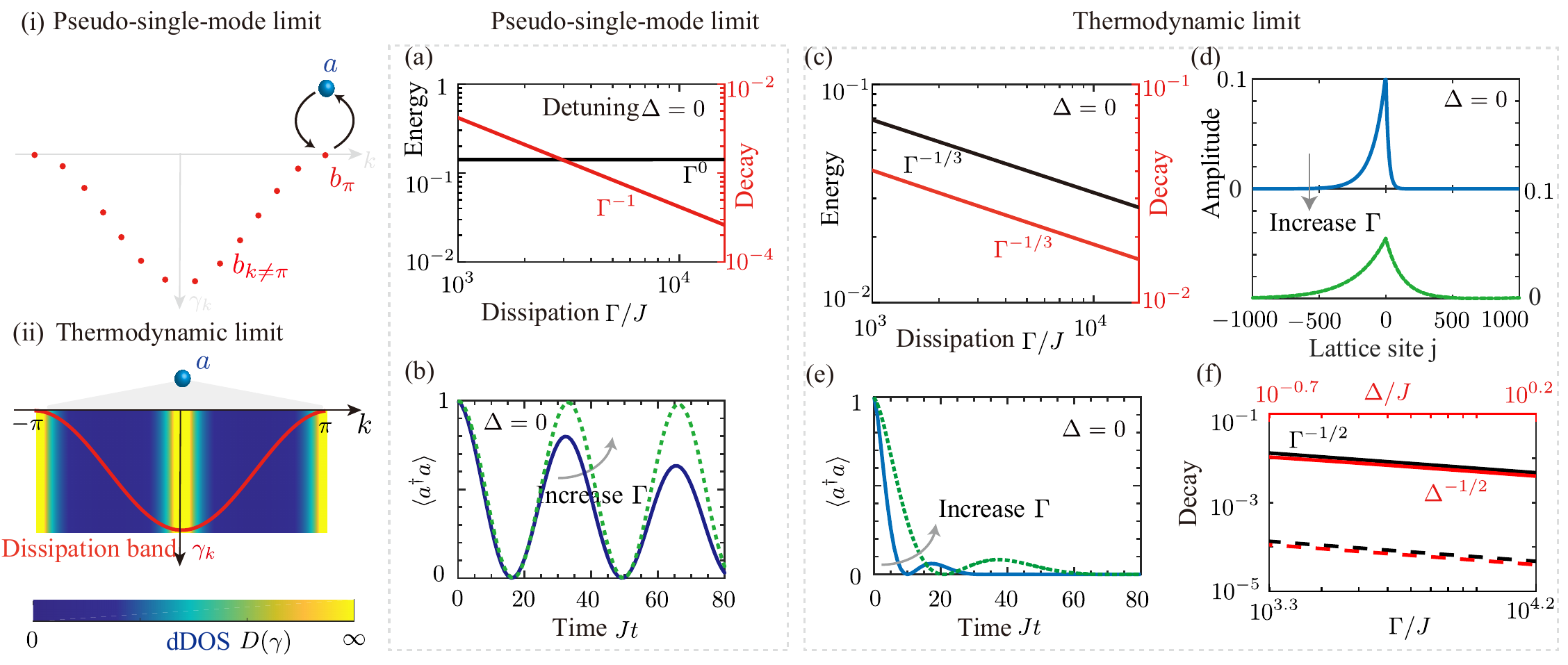}
		\caption{Spontaneous emission of an emitter with single excitation in a strongly dissipative open bath. Left [(i),(ii)]: physical picture. (i)
			Pseudo-single-mode regime: In a finite-size bath, the gap in the dissipation
			spectrum (red dots) isolates the long-lived bath mode $b_\protect\pi$ from
			$b_{k\neq\protect\pi}$. Eliminating $b_{k\neq \protect\pi}$ 
			yields an effective model where the emitter $a$ (blue ball) couples only to $b_\protect\pi$. (ii) Thermodynamic limit: Because of divergent
			dDOS (\protect\ref{eq:dos}) (colored square) at $k=\protect\pi$
			of the dissipation band $\protect\gamma_k$ (red curve), the emitter couples
			strongly to the dissipation band edge. Center [(a),(b)]: QZ
			effect in the pseudo-single-mode regime. (a) Scalings with the bath
			dissipation $\Gamma/J$. By diagonalizing the Hamiltonian~(%
			\protect\ref{eq:H}) with the Rabi frequency $\Omega/J=1$, size $N_b=50$, and
			detuning $\Delta=0$, we find two eigenstates agreeing with that of the
			effective model in (i). The (positive) energy and decay rate are shown as a
			function of $\Gamma/J$. (b) Population dynamics $\langle
			a^\dag(t)a(t)\rangle $. Numerical results are obtained using Eq.~(\protect
			\ref{eq:H}) with $\Omega/J=0.3$ and $N_b=10$, for $\Gamma/J=20, 500$ (solid and dashed line, respectively). Right [(c)—(f)]: FQZ effect in the thermodynamic limit. (c)
			Scaling behavior of the quasibound states when $\Delta=0$. By numerically
			solving the poles of the function ${G}_{f}(\protect\omega)$ using Eq.~(%
			\protect\ref{eq:Sigmaf}) with $\Omega/J=1$, we show the (positive) energy
			and decay rate of the quasibound state as a function of $\Gamma/J$. (d) Giant quasibound state. Its wave function is found by diagonalizing Eq.~(%
			\protect\ref{eq:H}) with $N_b=2000$, $\Delta=0$, and $\Omega/J=1$. The
			amplitude of its bath component at lattice site $j\in [-1000,1000]$ is shown
			for $\Gamma/J=100,1000$ (solid and dashed line, respectively). (e) Population dynamics $\langle
			a^\dag(t)a(t)\rangle=|G(t)^2|$ when $\Delta=0$. Here, $G(t)$ is obtained from
			the numerical Fourier transform of ${G}_{f}(\protect\omega)$ with $%
			\Omega/J=1 $ and $\Gamma/J=100, 1000$ (solid and dashed line, respectively). (f) Scaling behavior
			of the quasibound state when $\Delta\neq 0$. By numerically solving the
			poles of ${G}_{f}(\protect\omega)$ with $\Omega/J=1, 0.1$ (solid and dashed line, respectively), we plot the decay rate of the quasibound state, whose
			energy is $\Delta$, as a function of $\Gamma$ (black line) for $\Delta/J=-0.7$
			and a function of $\Delta$ (red line) for $\Gamma/J=10000$, respectively. (a),
			(c), and (f) are double-log plots. In (b) and (e), the initial
			conditions are $\langle a^\dag(0)a(0)\rangle=1$ and the vacuum state of the
			bath.}
		\label{Fig:oneatom}
	\end{figure*}

         \subsection{Dissipative density of states}
         
In the context of closed baths with structured energy dispersions $\epsilon_k$, the energy density of states (DOS) of the bath has played an important role in the emitter-bath interaction. For an open bath with also structured dissipation described by $\gamma_k$, analogously, we introduce dDOS labeled by $D_s(\gamma)$, namely, the number of bath modes with a dissipation rate $\gamma$. The 1D dDOS is defined as
 \begin{equation}
D_s(\gamma):=\int\frac{dk}{2\pi}\delta (\gamma-\gamma_k). \label{Dr}
\end{equation}
Explicit calculation of Eq.~(\ref{Dr}) leads to Eq.~(\ref{eq:dos}). In Sec.~\ref{sec:arbitrary}, we introduce dDOS in arbitrary dimensions and analyze the behavior of the self-energy through dDOS for arbitrary open baths with strong dissipations.

	\subsection{Second-order correlation function of a weakly driven emitter}
	
	\label{subsec:wd}
	
We turn to calculate the steady-state correlation functions of an emitter $a$ driven by a weak field with frequency $\omega_d$. By weak, we mean the
	driving strength $\varepsilon$ is finite but much smaller than the spectral
	gap of the system without the driving field. In the presence of driving ($\epsilon\neq 0$), the steady state of Eq.~(\ref{eq:rho}) is nonequilibrium, for which the dissipation-fluctuation theorem no longer holds. However, we can connect the steady-state correlation functions with the Green function of the undriven case. 
	
	Such a connection is enabled by a relation between scattering and the master equation formalism as first proposed in Ref.~\cite{Shitao2015}, where the intuitive picture is the following: In the weakly driven
system, the driving field just pumps the system by injecting multiple photons; 
thus, the pumping process can be considered as the few-photon scattering off
an undriven system. This idea has been successfully applied to many quantum optical systems~\cite{Yue2016,Shi2011,Shi2013}. Here, we follow the similar spirit, as detailed in Appendix~\ref{sec:steady}.

In particular, we obtain the first- and second-order correlation functions of the weakly driven emitter in terms of Green functions of the undriven case as
\begin{eqnarray}
\frac{\left\langle a^{\dagger }a\right\rangle _{\mathrm{ss}}}{\varepsilon ^{2}} &=&\left\vert \int_{-\infty}^\infty dt e^{-i\omega _{d%
		}t}\left\langle 0\right\vert \mathcal{T}
a(0)a^{\dagger }(t)\left\vert 0\right\rangle \right\vert ^{2}\nonumber\\
\frac{\left\langle a^{\dagger }a^{\dagger }(\tau )a(\tau)a\right\rangle _{\mathrm{ss}}}{\varepsilon ^{4}}&=&\left|\int_{-\infty }^{+\infty }dt_{1}dt_{2}e^{-i\omega _{d%
		}(t_{1}+t_{2})}G(\tau ;t_{1},t_{2})\right|^2\nonumber
\end{eqnarray}
with $G(\tau; t_{1},t_{2})=-i\left\langle 0\right\vert \mathcal{T%
	}a_{1}(\tau )a_{1}(0)a_{1}^{\dagger }(t_{1})a_{1}^{\dagger
	}(t_{2})\left\vert 0\right\rangle /2$. Here, $a(t)=e^{iH_{\mathrm{eff}}^\dag t}ae^{-iH_{\mathrm{eff}}t}$ is
	governed by the non-Hermitian Hamiltonian~(\ref{eq:H}) without driving.  
	
As the key result, by applying the Dyson expansion to calculate Green functions of the undriven emitter (see Fig.~\ref{SubFig3}, Appendix~\ref{sec:G}), we obtain 
		\begin{equation}
		g^{(2)}(\tau )=\left\vert 1+\bar{\Pi}_{f}(\tau )T(2\omega _{d})\right\vert ^{2},
		\label{g2t}
	\end{equation}
	with the scattering matrix $T(2\omega _{d})=[U^{-1}-\Pi _{f}(2\omega _{%
		d})]^{-1}$ and the two-particle Green function
	\begin{equation}
		\bar{\Pi}_{f}(\tau )=i\int \frac{d\omega ^{\prime }}{2\pi }G_{f}(\omega _{
			d}+\omega ^{\prime })G_{f}(\omega _{d}-\omega ^{\prime
		})e^{-i\omega ^{\prime }\tau }. \nonumber 
	\end{equation}
	In the asymptotic limit $\tau \rightarrow \infty $, $g^{(2)}(\infty
	)\rightarrow 1$ as $\bar{\Pi}_{f}(\tau )$ tends to zero. In the limit $
	\tau \rightarrow 0$, we obtain $	g^{(2)}(0)$ in Eq.~(\ref{g2}) from the identity $\bar{\Pi}_{f}(0)=\Pi _{f}(2\omega _{d})$. 
	
	Since $g^{(2)}(0)<1$ indicates the quantum nature of light, Eqs.~(\ref{g2}) and (\ref{g2t}) provide us the central principle to explicitly identify the role played by the non-Hermiticity of the effective Hamiltonian in generating sub-Poissonian quantum light.
	
	We emphasize that the formalism developed in this section is general, applicable for arbitrary non-Hermitian band structures $\epsilon_{k}-i\gamma _{k}$ of the bath.

	\section{Fractional quantum Zeno effect}\label{sec:FQZ}
	
	  Based on the above formalism, in this section, we explore the emitter physics in the strong dissipation regime $\Gamma /(2J)\gg 1$, for $\epsilon _{k}=2J\sin k$ and $\gamma
	_{k}=\Gamma (1+\cos k)$, and reveal the FQZ
	effects for different numbers of excitations. Cases for the open bath with arbitrary forms of dissipation bands and dimensions are discussed in Sec.~\ref{sec:arbitrary}.
		
	\subsection{Single excitation}\label{sec:single}
	
	We begin with studying the dynamics in the single-excitation
	subspace, where the on-site interaction $U$ does not play any role and the
	non-Hermitian emitter-bath Hamiltonian~(\ref{eq:H}) becomes quadratic. We consider the cases
	with one and two emitters, respectively. 	
	
	\subsubsection{Single emitter}
	
	\label{subsec:one}
	
	When the open bath is dominated by its intrinsic dissipation for $\Gamma/(2J)\gg 1$, vital for
	the emitter dynamics at long times are bath modes with small
	dissipation rates. To gain 
	intuitions into the physics, let us first consider a bath with the finite
	size $N_{b}$ and, thus, a discrete dissipation spectrum [Fig.~\ref{Fig:oneatom}(i)], where there opens a gap $\delta\gamma =2\pi^{2}{\Gamma}/{N_{b}^{2}}$ between the mode $b_\pi$ with $\gamma_k=0$ and the rest modes $b_{k\neq\pi}$. Under the condition $\Omega/\delta \gamma \ll 1$, the fast-decaying modes $b_{k\neq\pi}$ can be adiabatically eliminated in the lowest-order perturbation treatment, yielding an
	effective Hamiltonian $H_{\text{eff}}^{\prime }=\Omega (a_{1}^{\dag }b_{\pi
	}+b_{\pi }^{\dag }a_{1})/\sqrt{N_{b}}+(\Delta-i\gamma _{1})a_{1}^{\dag
	}a_{1} $. It describes that the emitter, which has the effective decay rate  $\gamma
	_{1}\approx (\Omega ^{2}/N_{b}\Gamma )\sum_{k\neq \pi }1/(1+\cos
		k)\propto \Gamma^{-1}$ (i.e., standard QZ effect), is coupled to a single bath mode $%
	b_{\pi }$ with an effective coupling strength $\Omega/\sqrt{N_{b}}$. Thus, the emitter with $\Delta=0$ is expected to undergo a Rabi oscillation at frequency $2\Omega/\sqrt{N_b}$,
	which is protected by the QZ effect. As such, we refer to the regime of $\Omega/\delta \gamma \ll 1$ as the pseudo-single-mode regime. 
	
	The above analysis for the finite-size limit is verified by numerical simulations using the original
	non-Hermitian Hamiltonian $H_\text{eff}$ in Eq.~(\ref{eq:H}) with a finite
	size $N_b$ under the condition $\Omega/\delta \gamma \ll 1$. Specifically,
	we find two eigenstates of $H_\text{eff}$ coincide with that of the above
	effective model, whose eigenvalues show the expected $\Gamma^{-1}$ scaling [Fig.~\ref{Fig:oneatom}(a)]. The numerical result of time-dependent
	population $\langle a_1^\dag(t) a_1(t)\rangle$ in Fig.~\ref{Fig:oneatom}(b)
	clearly shows QZ dynamics, where the strong bath dissipation constrains the
	emitter to a coherent evolution in the subspace of the emitter and the mode $b_\pi$.
	
	However, in the thermodynamic limit [Fig.~\ref{Fig:oneatom}(ii)], the structured 
	dDOS associated with the dissipation band, which diverges near the edge, invalidates the above single-mode picture. To reveal the resulting emitter dynamics, we employ the formalism developed in Sec.~\ref%
	{sec:formalism} to obtain the self-energy~\cite{SM}
	\begin{equation}
		\Sigma_{f}(\omega )=\frac{-i\Omega^{2}}{\sqrt{-(\omega +i\Gamma )^{2}-4J_{ 
					\mathrm{eff}}^{2}}}  \label{eq:Sigmaf}
	\end{equation}
	and the Green function ${G}_{f}(\omega )=1/[\omega -\Delta -\Sigma
	_{f}(\omega )]$ associated with the fictitious bath. In the regime $%
	\Gamma /(2J)\gg 1$, where the bath undergoes strong dissipation by itself, the contribution to the emitter dynamics from the branch cut is found to decay as $\sim e^{-\Gamma t}$ (see Appendix~\ref{sec:G}). On the long times $Jt\gg 1/\Gamma$, as shown, the emitter dynamics is determined by
	the poles $\epsilon_{s}\equiv\epsilon _{b}-i\gamma _{b}$ of $G_f$, i.e., ${G}_{f}^{-1}(\epsilon_{s})=0$, where $%
	\epsilon_{b}$ and $\gamma _{b}$ represent the energy and the decay rate of
	the quasibound state, respectively. Interestingly,
	the quasibound states exhibit different behavior depending on the detuning $\Delta$.
	
	\begin{figure*}[tb]
		\centering
		\includegraphics[width=1\textwidth]{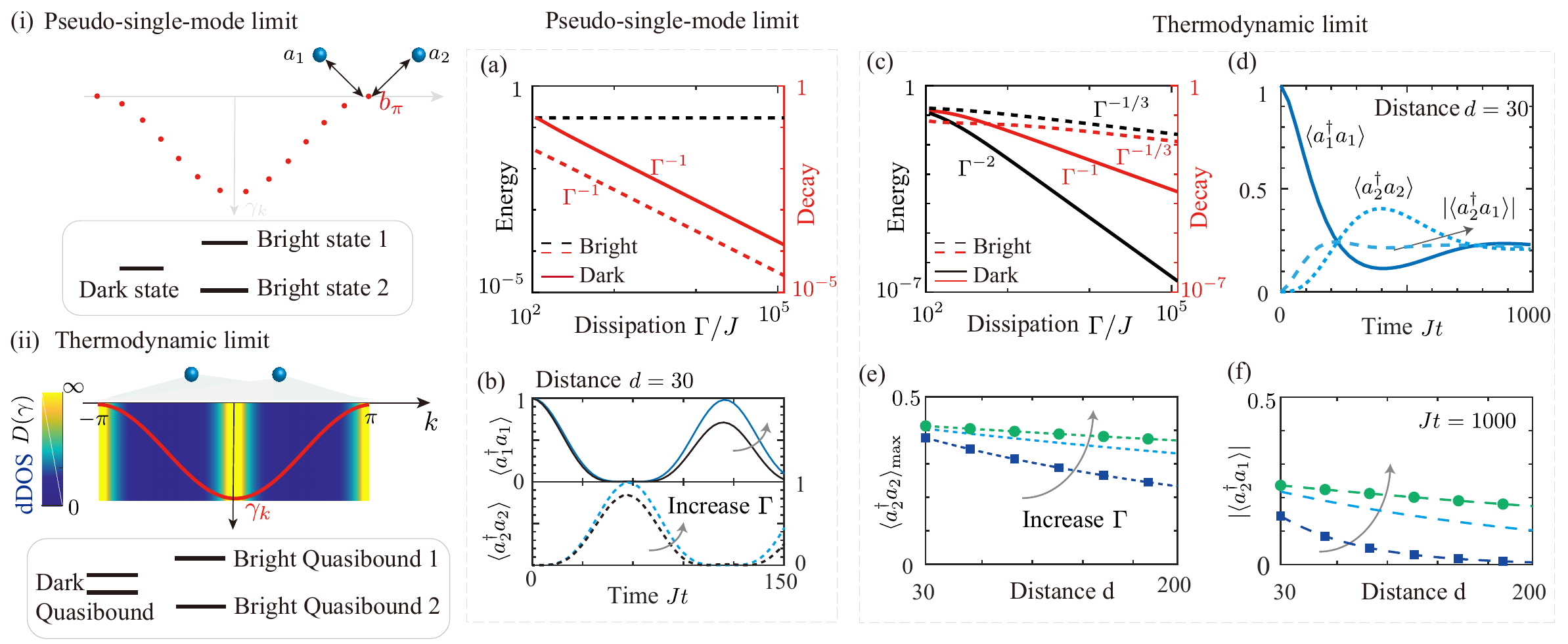}
		\caption{Remote state transfer and quantum correlation between two emitters mediated by a strongly dissipative open bath. Left [(i),(ii)]: physical picture. (i)
			Pseudo-single-mode regime: For a finite-size bath, elimination of
			bath modes $b_{k\neq \protect\pi}$ yields an effective model where
			emitters $a_1$ and $a_2$ interact via a single mode $b_\protect\pi$, forming a dark state and two bright states. (ii) Thermodynamic limit: Because of divergent dDOS (colored square) at $k=\pi$ of $\gamma_k$(red curve), two emitters interact
			strongly with the dissipation band edge, yielding two dark and two bright
			quasibound states. Center [(a),(b)]: QZ effect in the
			pseudo-single-mode regime. (a) Scaling with $\Gamma/J$.
			By diagonalizing Eq.~(\protect\ref{eq:H}) with
			Rabi frequency $\Omega/J=1$ and $N_b=70$ sites, we find a dark state (solid line)
			and two bright states (dashed line) described by the effective mode in (i).
			Their energy (black line) and decay rate (red line) are shown as a function of $%
			\Gamma/J$. The energy of the dark state is approximately $ 10^{-15}J$ and negligible.
			(b) Population dynamics $\langle a_l^\dag(t)a_l(t)\rangle$ ($l=1,2$).
			Results are numerically obtained using Eq.~(\protect\ref{eq:H}) with $%
			\Omega/J=0.3$ and $N_b=60$, for $\Gamma/J=500, 10000$ (black and blue line, respectively). Right [(c)—(f)]: FQZ effect in the thermodynamic limit. (c) Scaling behavior
			of the dark and bright quasibound states. By numerically solving the poles
			of $G_\pm(\protect\omega)$~\protect\cite{SM} using Eq.~(%
			\protect\ref{eq:Sigmaf2}) with $\Omega/J=1$ and distance $d=30$, we plot the
			energy and decay rate as a function of $\Gamma/J$. (d) Dynamics of
			populations and correlations. Results are obtained via numerical Fourier
			transform of $G_\pm (\protect\omega)$ with $\Omega/J=0.3$ and $%
			\Gamma/J=40000 $. (e) Maximum population $\langle a_2^\dag a_2\rangle_{\text{max}}$
			and (f) long-time quantum correlation $|\langle
			a_2^\dag (t) a_1(t)\rangle|$ at $Jt=1000$ as a function of $d$
			(even number), with $\Omega/J=0.3$ and $\Gamma/J=10000, 40000, 100000$
			(square-dashed, dashed, and dot-dashed line, respectively). In (a)—(f), the detuning is $\Delta=0$.
			In (b) and (d)—(f), the initial conditions are $\langle a_1^\dag(0)
			a_1(0)\rangle=1$, $\langle a_2^\dag(0) a_2(0)\rangle=0$, and the vacuum state
			of the bath. (a) and (c) are double-log plots. }
		\label{Fig:twoatom}
	\end{figure*}
	
	For zero detuning $\Delta=0$, two quasibound state solutions
	\begin{equation}
		\epsilon_{s}\equiv \epsilon _{b}-i\gamma _{b}=\frac{1}{2}\left( \pm \sqrt{3}%
		-i\right) \left( \frac{\Omega^{4}}{2}\right)^{1/3}\Gamma^{-1/3}
		\label{eq:Ebone}
	\end{equation}
	exhibit the same nonanalytic scalings with $\Gamma$, in both the energy and
	the decay rate. Thus, the quasibound state displays the FQZ effect.
	Note that $\epsilon _{b}/\gamma _{b}=\sqrt{3}$ is a constant. Equation (%
	\ref{eq:Ebone}) is confirmed by the numerical solutions of ${G}%
	_{f}^{-1}(\epsilon_{s})=0$ in Fig.~\ref{Fig:oneatom}(c). 
	
	The long-lived quasibound state is a superposition of the emitter and the bosonic modes of the open bath, i.e., $|B\rangle =c_{1}a_{1}^{\dagger }|0\rangle
+\sum_{k}f_{k}b_{k}^{\dag }|0\rangle $, where $c_1$ and $f_k$ are the coefficients.  As detailed in Appendix~\ref{sec:scaling}, we find $f_k=c_1(\Omega/\sqrt{N_b})\{1/[\epsilon_s-(\epsilon_k-i\gamma_k)]\}$. According to Eq.~(\ref{eq:Ebone}), the average momentum of the bath component is \textit{sharply} localized at $
	k_b\rightarrow \pi$, as indicated by the fact that $\epsilon_s\rightarrow 0$ for $\Gamma\rightarrow \infty$. In real space, the bath modes localized around the emitter form a giant cloud with a large localization length $l_b\propto \left( \Gamma/\Omega \right)
	^{2/3}$ [see Eqs.~(\ref{fj})—(\ref{lb}), Appendix~\ref{sec:scaling}], which increases with $\Gamma$ through a nonanalytic scaling. The giant size of the cloud as shown in Fig.~\ref{Fig:oneatom}(d) extends impressively over hundreds of lattice
	sites at large $\Gamma$. Note that the asymmetry is due to the asymmetry of the
	bath under $k\rightarrow -k$. 
 
	Thus, the emergence of the FQZ effect can be understood: The strong bath dissipation effectively
	tailors the coupling of emitter with the continuum to the
	dissipation-band edge near $k=\pi$; there, the divergent dDOS gives rise to a strong emitter-bath interaction, leading to the fractional scaling behavior of the emitter. Indeed, as we rigorously prove later in Sec.~\ref{sec:arbitrary}, the fractional scaling always emerges if the dDOS near the dissipation band edge diverges, irrespective of whether $\gamma_k$ is gapless or not. This is in strong contrast to what happens in the closed-bath case, where the bound state is created if the emitter is on resonance with the edge of the energy band $\epsilon_{k}$ (in our case, this corresponds to the limit of $\Gamma\rightarrow 0$ and when tuning $\Delta\approx \pm 2J$ near resonance with the edge of $\epsilon_{k}=2J\sin k$ at $k=\pm \pi/2$).
	
	The excitation population $\langle a_1^\dag(t) a_1(t)\rangle=|G(t)|^{2}$ on the emitter is determined by the Fourier transform of the Green function ${G}_{f}(\omega )$. In the long-term limit, the result~\cite{SM}	\begin{equation}
		G(t)\approx \frac{4}{3}e^{-\gamma _{b}t}\cos (\sqrt{3}\gamma _{b}t)
		\label{eq:G1t}
	\end{equation}
	from the double-pole approximation represents the long-lived oscillation
	between two quasibound states. It indicates that both the revival time and the
	lifetime of the emitter become longer when $\Gamma$ increases, while the maximum
	revival population is almost constant as a result of $\epsilon _{b}/\gamma _{b}=\sqrt{3}$. In Fig.~\ref{Fig:oneatom}%
	(e), we show the dynamics of $\langle a_1^\dag(t) a_1(t)\rangle$ obtained from the
	numerical Fourier transform of ${G}_{f}(\omega )$. It
	agrees with Eq.~(\ref{eq:G1t}) very well at times $t>1/\Gamma$.
	
	For a finite detuning $\Delta \neq 0$, however, the two quasibound states show different scalings, i.e.,
	\begin{eqnarray}
		\epsilon_{s}^{(1)}&=&\Delta -i\frac{\Omega ^{2}}{\sqrt{2\Delta }}e^{i(\pi/4)}\Gamma ^{-1/2},  \label{eq:Delta1} \\
		\epsilon_{s}^{(2)}&=&-i(\frac{\Omega ^{4}}{ 2\Delta^{2}}+2J^{2})\Gamma ^{-1}.
		\label{eq:Delta2}
	\end{eqnarray}%
	The decay rate of the first quasibound state in Eq.~(\ref{eq:Delta1}) exhibits the FQZ scaling $\propto \Gamma^{-1/2}$, in contrast to the other one in Eq.~(\ref{eq:Delta2}) with the QZ scaling $\propto \Gamma^{-1}$. Interestingly, the former also scales fractionally as $\propto \Delta^{-1/2}$ with $\Delta$. In Fig.~\ref%
	{Fig:oneatom}(f), we present the numerical solutions for the decay rate of
	the first quasibound state, which confirm the predicted nontrivial
	scalings. The long-term dynamics of the emitter is primarily determined by
	Eq.~(\ref{eq:Delta1}), which indicates an oscillation with frequency $\Delta$
	and a decay dynamics that can be fractionally suppressed via enhancing both $\Gamma$
	and $\Delta$.
	
	\subsubsection{Two emitters}
	
	\label{subsec:two}

		We now show that the FQZ effect leads to remarkable remote, long-term quantum correlation between two emitters. Tunable long-range correlation has been actively pursued in the closed-bath context such as using atoms coupled to a photonic crystal~\cite{Chang2018,Douglas2015,Shi2018}. Therein, however, a trade-off exists between the correlation length and the correlation strength, because the increase of photonic localization length in the atom-photon bound state is accompanied with reduced atomic population. Here, we show that the strongly dissipative open bath can mediate simultaneous substantial and long-range quantum correlation, due to the formation of a dark quasibound state composed of two emitters and the bath component with a very large spatial size.
	
	We focus on the interesting case $%
	\Delta=0$ and assume the distance $d$ of two emitters to be an even number
	without loss of generality. Again, we start from the pseudo-single-mode regime [Fig.~\ref{Fig:twoatom}(i)] to gain some intuition. In this case,
	adiabatic elimination of the bath modes $b_{k\neq \pi}$ yields the effective
	non-Hermitian Hamiltonian $H_{\text{eff}}^{\prime
	}=\sum_{l=1,2}[-i\gamma_{l}a_{l}^{\dag }a_{l}+\Omega/\sqrt{N_{b}}
	(a_{l}^{\dag }b_{\pi }+\text{H.c.})]$, where $\gamma_{1}$ is the same as before
	and $\gamma_{2}\approx (\Omega^{2}/N_{b}\Gamma)\sum_{k\neq \pi }
		e^{-ikd}/(1+\cos k)$. It describes a three-level system where two emitters
	are coupled to the bath mode $b_\pi$, protected by the standard QZ effect.
	In the limit $\Gamma/(2J) \rightarrow \infty $, the antisymmetric
	superposition of two emitters in the odd channel forms a dark state $|\psi
	_{-}\rangle =(a_{1}^{\dagger }-a_{2}^{\dagger })|0\rangle/\sqrt{2}$
	decoupled from all bath modes, whereas in the even channel the symmetric
	superposition of emitters hybridizes with the $b_\pi$ mode to form two bright
	states, $|\psi _{+}\rangle =c_{1}(a_{1}^{\dagger }+a_{2}^{\dagger
	})|0\rangle /\sqrt{2}+c_{2}b_\pi^{\dag }|0\rangle $, where $c_1$ and $c_2$ are coefficients. This physical picture
	indicates that, due to the dark state, excitation initially populating the first
	emitter is transferred to the second emitter at some time $t<\Gamma
	^{-1}$ even when they are remotely separated.

	In the thermodynamic limit, however, two emitters are strongly coupled to
	the dissipation band edge with divergent
	dDOS [Fig.~\ref{Fig:twoatom}(ii)]. For the separation $d$
	within the spatial size of the localized bath modes, analogy with the
	pseudo-single-mode case suggests potential creations of dark (bright)
	quasibound states in odd (even) channels.
	
	Mathematically, we determine the energy and the decay rate of the two-emitter quasibound
	states from the poles of Green functions $G_{\pm }(\omega
	)=1/(\omega -\Sigma _{f}^{\pm })$ in the even (odd) channels $+(-)$, where
	the self-energies $\Sigma _{f}^{\pm }$ are~\cite{SM}
	\begin{equation}
		\Sigma _{f}^{\pm }(\omega )=-i\frac{\Omega ^{2}[1\pm z_{+}^{d}(\omega )]}{
			\sqrt{-(\omega +i\Gamma )^{2}-4J_{\mathrm{eff}}^{2}}}  \label{eq:Sigmaf2}
	\end{equation}%
	with $z_{+}(\omega )=i(\omega +i\Gamma)/2J_{\mathrm{eff}}+\sqrt{-
		[(\omega +i\Gamma )^{2}/4J_{\mathrm{eff}}^{2}]-1}$. The approximate
	solution for the complex energy $\epsilon _{s}$ of the quasibound state can
	be obtained analytically via the Taylor expansion. For $\Omega /J>\sqrt{2/d}$%
	, we find two solutions for the ``dark" quasibound states 
	\begin{equation}
		\epsilon _{s}=\pm R-id\Omega ^{2}\Gamma ^{-1}  \label{eq:darkb}
	\end{equation}%
	in the odd channel with $R\sim \Gamma^{-2}$ and two solutions for the
	bright quasibound states in the even channel:%
	\begin{equation}
		\epsilon _{s}=\frac{1}{2}\left( \pm \sqrt{3}-i\right) (2\Omega
		^{4})^{1/3}\Gamma ^{-1/3}.  \label{eq:bright}
	\end{equation}%
	
	The above results indicate that the decay rate of the dark quasibound state
	is controlled by the distance $d$ and exhibits QZ scaling. Instead, the two
	bright quasibound states are blind to $d$ and feature the FQZ
	scaling $\Gamma^{-1/3}$ the same as the single-emitter case, except that the
	prefactor is enhanced by a factor of $4^{1/3}$. These analytical results are
	confirmed by numerical solutions of the poles of the Green functions $G_{\pm
	}(\omega )$, as shown in Fig.~\ref{Fig:twoatom}(c). Note that the dark quasibound state relies on strong dissipation and divergent dDOS of the dissipation band edge; hence, it is different in nature from the bound states with small decay rates in the context of a closed bath with multiple emitters~\cite{Tudela2018,Zhang2019}, where the energy resonance mechanism plays a fundamental role.

         According to Eqs.~(\ref{eq:Sigmaf2})—(\ref{eq:bright}), the bright quasibound state 	of the complex energy $\epsilon_s$ exhibits the localization length $l_b\sim
	1/\log[|z_+(\epsilon_s)|]$, which sets a characteristic length scale for the
	interaction of two emitters.  Remarkably, $|z_+|\rightarrow 1$ in the limit $%
	\Gamma /(2J)\rightarrow \infty$ indicates a remote correlation mediated by the
	bath mode near the dissipation-band edge, where the FQZ effect gives rise to
	a nonanalytic scaling of the correlation length $l_b\propto
	(\Gamma/\Omega)^{2/3}$ with $\Gamma$.

		 \begin{figure}[htbp]
		\centering
		\includegraphics[width=0.84\columnwidth]{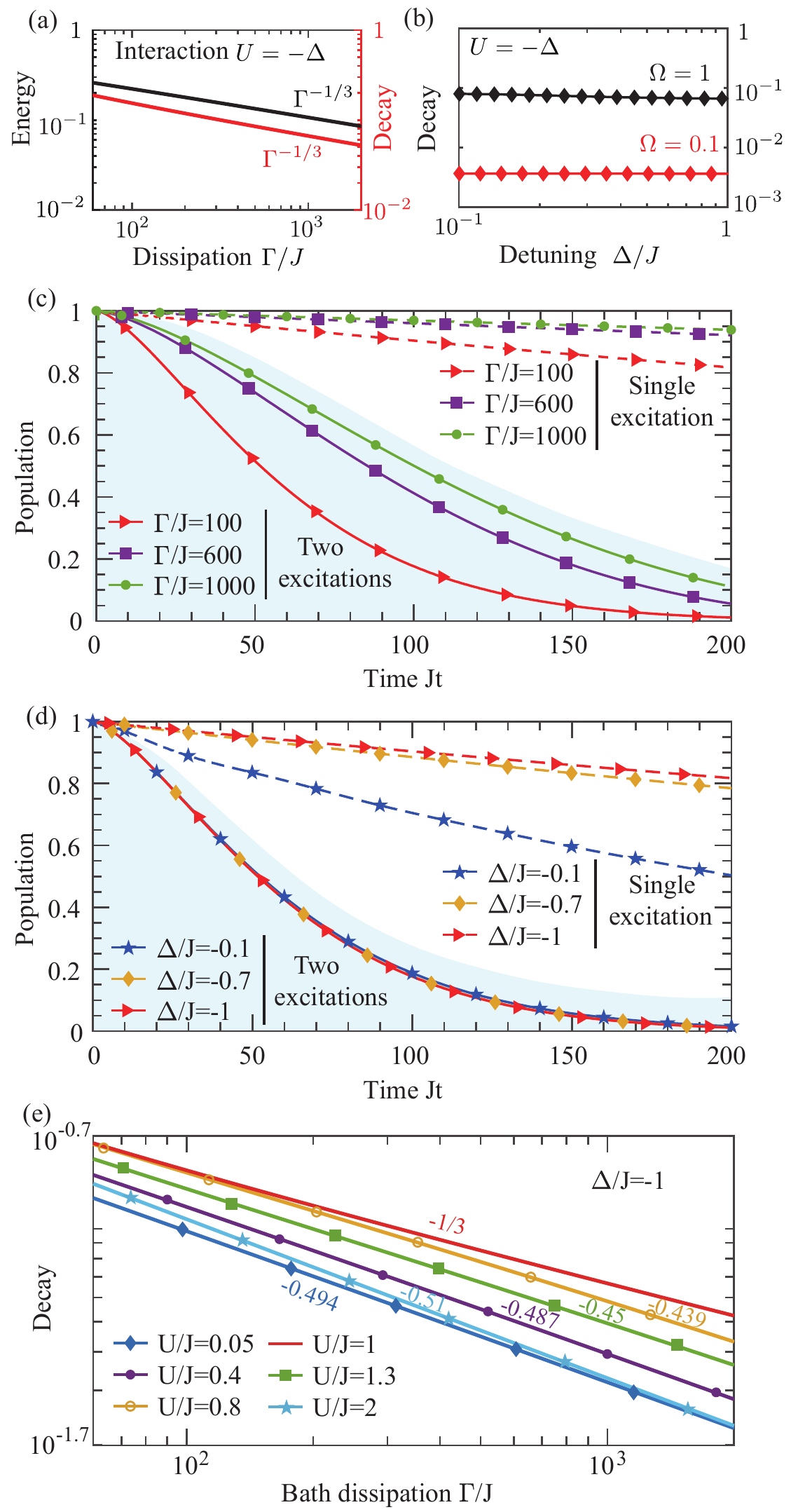}
		\caption{FQZ effect of two excitations in a nonlinear emitter
			and comparison with the single-excitation case. For two excitations, their interaction strength $U$ and the detuning $\Delta$ satisfy $U=-\Delta$ in (a)-(d). All numerical results are obtained via the approach in Sec.~\protect\ref{sec:formalism}. (a)
			Scalings of the energy (black line) and decay rate (red line) of the two-excitation
			quasibound states with the bath dissipation rate $\Gamma$. The Rabi
			frequency is $\Omega/J=1$ and $\Delta/J=-1$. (b) Decay rate of the quasibound
			state as a function of $\Delta$, when $\Gamma/J=1000$ and $\Omega/J=0.1, 1$
			(red and black line, respectively). (a) and (b) are double-log plots. (c),(d) Comparisons between
			spontaneous emissions of single and two excitations, when (c) $\Gamma$
			is changed and $\Delta/J=-1$ and (d) $\Delta$ is changed and $\Gamma/J=100$. In (c)
			and (d), $\Omega/J=0.1$. (e) FQZ scaling $\Gamma^{-\nu}$ ($\nu<1$) of two-excitation
			quasibound states for $U\neq -\Delta$, when $\Delta/J=-1$ and $\Omega/J=1$. Beside each curve, the extracted value of $-\nu$ is labeled.  }
		\label{Fig:U}
	\end{figure}
	
	The dynamics of two emitters directly follows from the Fourier
	transforms $G_{1n}(t)\sim \int (G_{+}\pm G_{-})e^{-i\omega t}d\omega /(4\pi
	) $ for $n=1,2 $ as~\cite{SM}
	\begin{equation}
		iG_{1n}(t)=\frac{2}{3}e^{-\bar{\gamma}t}\cos (\sqrt{3}\bar{\gamma}t)\pm 
		\frac{1}{2}e^{-(\Omega ^{2}d/\Gamma )t},  \label{eq:G12}
	\end{equation}
	where $\bar{\gamma}=\Omega^{4/3}/(4\Gamma)^{1/3}$. In Fig.~\ref{Fig:twoatom}%
	(d), we numerically perform the Fourier transforms to obtain the time-dependent population $\langle a_n^\dag (t)
	a_n(t)\rangle=|G_{1n}(t)|^2$ and the correlation $|\langle
	a_2^\dag (t) a_1(t)\rangle|$ of two remote emitters separated by $d=30$.
	The results at long times $t>1/\Gamma$ agree very well with Eq.~(\ref{eq:G12}).

	When $d/l_b\ll 1$, Eq.~(\ref{eq:G12}) predicts the maximal transferred
	population on the second emitter is $\langle a_2^\dag a_2\rangle_\text{max}%
	\rightarrow 0.36$ [Fig.~\ref{Fig:twoatom}(d)]  at the time $t_s\sim \pi /(\sqrt{3}\bar{\gamma})$. When $d$
	increases, $\langle a_2^\dag a_2\rangle_\text{max}$ diminishes [Fig.~\ref%
	{Fig:twoatom}(e)] due to the increased decay rate of the dark quasibound state [see Eq.~(\ref%
	{eq:darkb})]. However, as long as $d/l_b\ll 1$, the
	population decreases as $\propto -d/l_b$, leading to a state transfer that can occur across
	hundreds of lattice sites. 
	
	Interestingly, due to the presence of the dark quasibound state, 
	a remote correlation $|\langle a_2^\dag (t)
	a_1(t)\rangle|\sim 0.25$ between two emitters can exist for a remarkably long time, as expected from Eq.~(\ref{eq:G12}) under the condition $%
	t<\Gamma/(d\Omega^2)$; see Fig.~\ref{Fig:twoatom}(d). The remoteness of the long-term correlation is showcased in Fig.~\ref%
	{Fig:twoatom}(f). When $
	d<\Gamma/(1000\Omega^2 )$, we see that the correlation diminishes linearly and slowly with 
	$d$, remaining substantial over a distance $d\sim 200$ even at
	such a long time $%
	Jt=1000$. These results suggest the possibility to flexibly engineer simultaneous significant and remote correlations in practice where the finite-size bath is generally in between the thermodynamic and the quasi-single-mode limits.  

	\bigskip
	
	\subsection{Two excitations}\label{sec:two}
	
	In this section, we study the spontaneous emission of two excitations in an
	emitter with the on-site interaction $U$. We show that the decay rate of two
	excitations exhibits distinct FQZ scalings from the single-excitation
	counterpart, which can be tuned via $U$ and $\Delta$. 
	
       Since the effective
	emitter-bath Hamiltonian~(\ref{eq:H}) commutes with $N$, we can expand it in the two-excitation subspace spanned by the basis 
	$\{ a_{1}^{\dagger 2}\left\vert 0\right\rangle /\sqrt{2}\equiv
	\left\vert d\right\rangle ,a_{1}^{\dagger }b_{k}^{\dagger }\left\vert
	0\right\rangle \equiv \left\vert k\right\rangle _{e},b_{k}^{\dagger
	}b_{k^{\prime }}^{\dagger }\left\vert 0\right\rangle \equiv \left\vert
	kk^{\prime }\right\rangle \} $. We obtain%
	\begin{eqnarray}
		H_{2}^{\prime } &=&(U+2\Delta )|d\rangle \langle d|+\sum_{k}(\Delta
		+E_{k})|k\rangle _{e}\langle k|  \notag \\
		&+&\sum_{k}\frac{\sqrt{2}\Omega }{\sqrt{N_{b}}}\left( |k\rangle _{e}\langle
		d|+\text{H.c.}\right) +\sum_{kk^{\prime }}(E_{k}+E_{k^{\prime }})|kk^{\prime
		}\rangle \langle kk^{\prime }|  \notag \\
		&+&\sum_{kk^{\prime }}\frac{\sqrt{2}\Omega }{\sqrt{N_{b}}}\left( |k\rangle
		_{e}\langle kk^{\prime }|+\text{H.c.}\right)  \label{H2}
	\end{eqnarray}%
	where $E_{k}=\varepsilon _{k}-i\gamma _{k}$ is the complex energy of the
	bath mode $b_{k}$. 
	
	Because the bath mode $b_\pi$ has zero complex energy $E_\pi=0$, Eq.~(\ref{H2}) indicates two
	kinds of resonant processes. (i) For the interaction $U=-\Delta $, the
	doublon state $|d\rangle $ (i.e., two excitations in the emitter) is
	resonant with the state $|\pi \rangle _{e}$ (i.e., one excitation
	in the emitter and one excitation at $k=\pi $ in the bath). (ii) When $%
	U=-2\Delta $, the resonant coupling occurs between the doublon state $%
	|d\rangle $ and the state $|\pi \pi \rangle $ (i.e., two excitations at $%
	k=\pi $ in the bath).
	
	It turns out that, in the resonant case $U=-\Delta$, the decay rates of two
	excitations have different scaling behaviors from the single-excitation
	sector. Under the condition $\Omega^{2}/\Delta \Gamma \ll 1$, the states $%
	|kk^{\prime }\rangle $ (i.e., two excitations in the bath) can be
	adiabatically eliminated (see Appendix~\ref{sec:scaling}). To leading order, the dynamics is
	governed by the effective non-Hermitian Hamiltonian 
	\begin{equation}
		{H}_{2}^{\prime\prime}=\frac{\sqrt{2}\Omega }{\sqrt{N}}\sum_{k}\left(
		|k\rangle _{e}\langle d|+\text{H.c.}\right) +\sum_{k}E_{k}|k\rangle
		_{e}\langle k|  \label{eq:Heff}
	\end{equation}%
	in the rotating frame, which is exactly the Hamiltonian in the
	\textit{single-excitation} sector with $\Delta =0$ and $\Omega \rightarrow \sqrt{2}%
	\Omega $. 
	
	Equation (\ref{eq:Heff}) immediately allows us to use the earlier results of the single excitation to understand the physics of two excitations with the interaction $U=-\Delta $. Specifically, it indicates the existence of two giant
	two-excitation quasibound states whose
	complex energies are 
	\begin{equation}
		\epsilon _{s}\equiv \epsilon _{b2}-i\gamma _{b2}=\frac{1}{2}\left( \pm \sqrt{%
			3}-i\right) (2\Omega ^{4})^{1/3}\Gamma ^{-1/3}. \label{eq:Ebtwo}
	\end{equation}%
	We thus conclude that the spontaneous emission of two excitations is characterized by%
	\begin{equation}
		D(t)\approx \frac{4}{3}e^{-\gamma _{b2}t}\cos (\sqrt{3}\gamma _{b2}t)
		\label{eq:Dt}
	\end{equation}%
	at long times, which exhibits the FQZ effect \textit{without} explicit
	dependence on $\Delta $ and $U$.
	
	To validate above analysis, we apply the approach in Sec.~\ref{sec:formalism}
	and derive the two-particle Green function $D_{f}(\omega )=[\Pi
	_{f}^{-1}(\omega )-U]^{-1}$ associated with the fictitious bath. By numerically solving the poles of $D
	_{f}(\omega)$, we find the energy and decay rate of two-excitation
	quasibound states under the condition $U=-\Delta$. The numerical results
	shown in Figs.~\ref{Fig:U}(a) and~\ref{Fig:U}(b) confirm the $\Gamma^{-1/3}$ scaling and the insensitivity to $\Delta$. By
	numerically performing the transformation $D(t)=\int D_{f}(\omega
	)e^{-i\omega t}d\omega /2\pi $, we obtain the spontaneous emission of two
	excitations, as shown in Fig.~\ref{Fig:U}(c) for various $\Gamma$ and Fig.~%
	\ref{Fig:U}(d) for various $\Delta$. These results clearly corroborate Eq.~(%
	\ref{eq:Dt}); in particular, variations in the finite detuning barely
	influence the emitter dynamics.
	
	That the spontaneous emission rate of two excitations with interaction $U =-\Delta$ scales as $\Gamma^{-1/3}$ and is independent of $\Delta$ is in strong
	contrast to the single-excitation counterpart, where the decay rate scales as
$\Gamma^{-1/2}$ and can be controlled by $\Delta$ [cf. Eq.~(\ref{eq:Delta1}) and Fig.~\ref{Fig:oneatom}(f)]. As shown
	in Fig.~\ref{Fig:U}(c), the single excitation undergoes a significantly slower decay compared to two excitations. The large difference between the decay rates
	of one and two excitations is even more dramatic in Fig.~\ref{Fig:U}%
	(d). There, increasing $\Delta$ further suppresses the single-excitation
	decay as $\Delta^{-1/2}$, in contrast to the ``frozen" evolution trajectory of
	two excitations.
	
	For $U\neq -\Delta$, the fractional scalings appear generically; see Fig.~\ref{Fig:U}(e) for $\Delta/J=-1$. Since in the noninteracting limit two excitations exhibit similar scaling behaviors as the single excitation, we anticipate the FQZ scaling to cross over from $\Gamma^{-1/2}$ to $\Gamma^{-1/3}$ when the interaction is tuned from $U=0$ to $U=-\Delta$, as observed in Fig.~\ref{Fig:U}(e).
	
		\begin{figure}[tb]
		\centering
		\includegraphics[width=1\columnwidth]{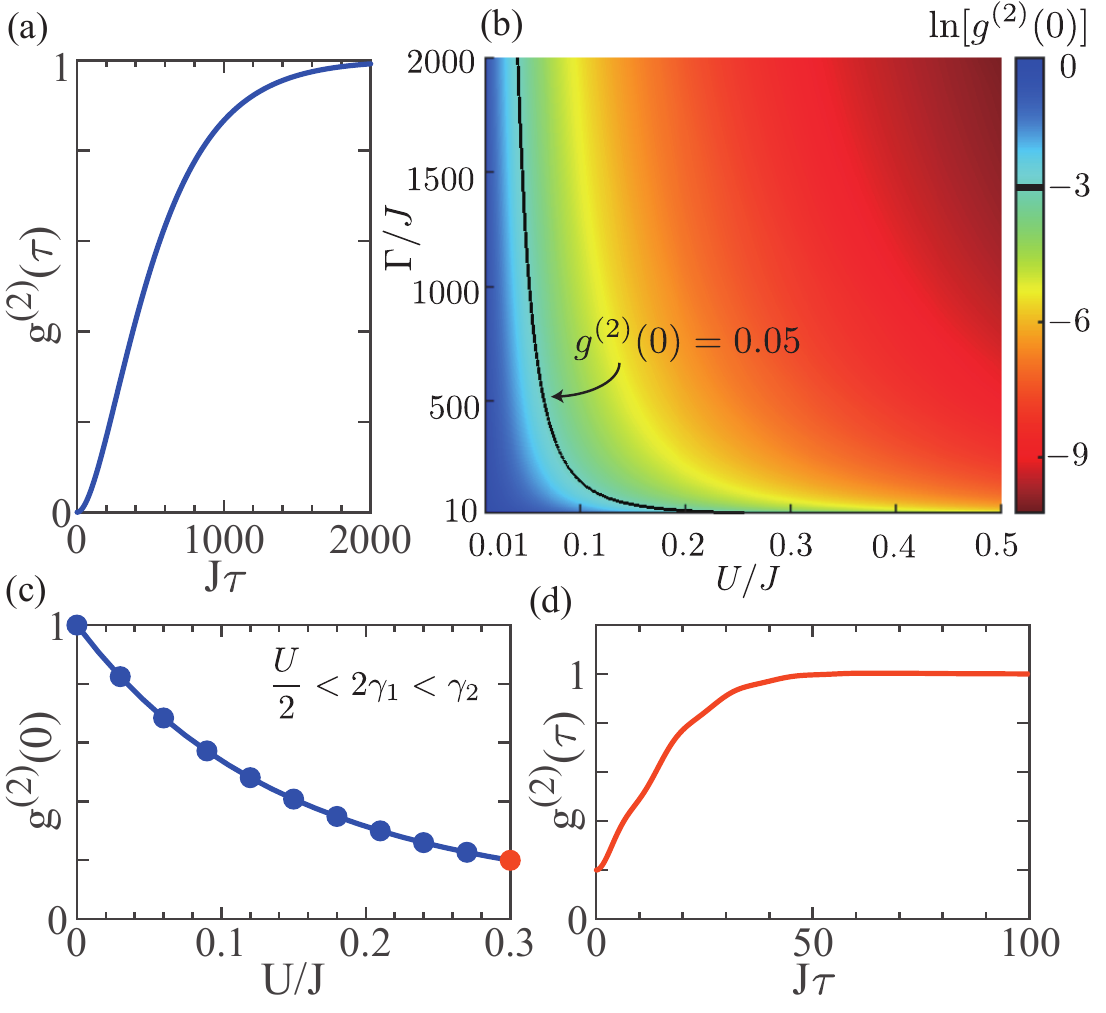}
		\caption{The FQZ-induced antibunching in a driven emitter. (a) Second-order correlation function $%
			g^{(2)}(\protect\tau)$ as a function of time $\protect\tau$. We calculate $%
			g^{(2)}(\protect\tau)$ in Eq.~(\ref{eq:g2}) based on Eq.~(\ref{g2t}), with the interaction
			$U/J=-\Delta/J=0.3$ and the bath dissipation rate $\Gamma/J=1000$. The driving
			frequency is $\protect\omega_d/J=-0.303$. (b) Minimal $g^{(2)}(0)$ in the
			parameter spaces of interaction $U/J$ and dissipation $\Gamma/J$ when $U/J=-\Delta/J$. Calculations are based on Eq.~(\ref{g2}). The black curve denotes where $%
			g^{(2)}(0)=0.05$. In (a) and (b), the JC coupling
			strength is $\Omega/J=0.3$. (c),(d) Strong
			antibunching in the weak-interaction regime $U/2<2%
			\protect\gamma_1<\gamma_2$. (c) Minimal $g^{(2)}(0)$ as
			a function of interaction $U$. (d) $g^{(2)}(\protect\tau)$ for $U/J=0.3$ and $\protect%
			\omega_d/J=-0.42$. In (c) and (d), we use $\Gamma/J=38.8$, $\Omega/J=0.79$, and $%
			\Delta/J=-0.3$. }
		\label{Fig:wd}
	\end{figure}

Intriguingly, the \textit{independently tunable} FQZ scalings for different
	numbers of excitations points to the possibility to tailor the emitter
	dynamics into the desired excitation subspaces. For instance, we can
	engineer the detuning $\Delta =-U$ and bath dissipation $\Gamma $ in such a
	way that the two excitations decay much faster than the single excitation.
	This has the direct consequence of the hierarchical Zeno effect; namely, any
	weak pump field cannot populate the two-excitation subspace in the
	characteristic timescale $\sim \Gamma ^{1/2}$, leading to confined dynamics
	in the single-excitation subspace. 	
	
	\section{FQZ-induced antibunching}
	
	\label{sec:weakdriving}
	
	In this section, we study the statistics of emitter excitations in the
	presence of a weak driving field. As predicted in Sec.~\ref{sec:two}, due
	to the FQZ effect, the dynamics is expected to be frozen in the single-excitation subspace even with a weak nonlinearity. Conventionally, the
	strong single-photon nonlinearity relies on strong Kerr interactions~\cite{Paul1982} or
	interference~\cite{Bamba2011,Kong2022}. Here, we show that the FQZ effect
	presents a new mechanism in the limit of weak interactions.
	
	Consider the driving light is on resonance with the single excitation, i.e., $%
	\omega _{d}=\text{Re}[\epsilon _{s}^{(1)}]$. It is instructive to first obtain some estimation for the second-order correlation $g^{(2)}(\tau)$ in Eq.~(\ref{g2t}) in the regime $\Gamma /J\gg 1$. Assuming the approximation $G_{f}(\omega )\sim
	1/(\omega -\epsilon _{s}^{(1)})$~\cite{footnote}, which leads to $\bar{\Pi}_{f}(\tau )\sim
	e^{-i(\epsilon _{s}^{(1)}-\omega _{d})\tau }/[2(\omega _{d%
	}-\epsilon _{s}^{(1)})]$, the resulting analytical expression reads
	\begin{equation}
		g^{(2)}(\tau )\sim \left\vert 1+\frac{Ce^{-i(\epsilon _{s}^{(1)}-\omega _{%
					d})\tau }}{1-C}\right\vert ^{2}. \label{g2app}
	\end{equation}%
	The ratio $C=U/[2(\omega _{d}-\epsilon _{s}^{(1)})]$, as a figure
	of merit, determines the statistics of the emitter excitations. For $\Delta\neq 0$, we obtain from Eq.~(\ref{eq:Delta1}) that $
	C=iU\sqrt{\left\vert \Delta \right\vert \Gamma }/\Omega ^{2}$, and Eq.~(\ref{g2app}) suggests $g^{(2)}(0)\sim 1/(1+|C|^2)<1$.
	
	As an example, we analyze the photon statistics in the case $U=-\Delta $, where the
	decay rate $\gamma_2\sim \Gamma ^{-{1}/{3}}$ of two excitations is larger than that $\gamma_1
	\sim \Gamma^{-{1}/{2}}$ of the single excitation. 
In the limit $|C|\gg 1$, Eq.~(\ref{g2app}) reduces to
	\begin{equation}
		g^{(2)}(0)\sim\frac{\Omega ^{4}}{U^{3}\Gamma }\ll 1,  \label{g20a}
	\end{equation}
	which indicates the sub-Poissonian statistics. The result (\ref{g20a}) explicitly provides the scaling of $g^{(2)}(0)$ on $%
	\Omega $, $U$, and $\Gamma $. In addition, $g^{(2)}(\tau )$
	saturates to unity in the timescale $\tau =1/\gamma _{1}=1/ \textrm{Im}[\epsilon
	_{s}^{(1)}]$.
	The condition (\ref{g20a}) can be understood using an intuitive picture: As
	the decay rate of two excitations is $\sim \Omega^{4/3}\Gamma ^{-1/3}$ [see
	Eq.~(\ref{eq:Ebtwo})], the condition (\ref{g20a}) indicates the interaction $%
	U$ is stronger than the two-excitation decay rate suppressed by the FQZ
	effect.

	Figures~\ref{Fig:wd}(a) and~\ref{Fig:wd}(b) numerically validate the strong antibunching
	when $U=-\Delta$. There, the Rabi coupling $\Omega
	/J=0.3$ is fixed, and the driving light is on resonance, i.e., $\omega _{%
		d}=\text{Re}[\epsilon_{s}^{(1)}]$. In Fig.~\ref{Fig:wd}(a) for $%
	\Gamma /J=10^{3}$ and $U/J=-\Delta /J=0.3$, the numerical result of $%
	g^{(2)}(\tau )$ explicitly displays the antibunching behavior. In Fig.~\ref%
	{Fig:wd}(b), we plot the numerical values of $g^{(2)}(0)$ in the $\Gamma $-$U
	$ plane for $U=-\Delta $. As shown by the black curve, to realize a desired $%
	g^{(2)}(0)=0.05$, the required nonlinearity $U$ becomes weaker as $\Gamma $
	increases, and in the large $\Gamma $ limit $U\sim \Gamma ^{-1/3}$,
	confirming the analysis based on Eq.~(\ref{g20a}).
	
When $U\neq -\Delta $, remarkably, capitalizing on the rich controllability over the FQZ scalings
	and, thus, the decay rates in different excitation subspaces, strong
	antibunching arises even for sufficiently weak nonlinearity $U/2<2\gamma _{1}<\gamma_2
	$, as numerically shown in Fig.~\ref{Fig:wd}(c). This can be understood by noting that, without the nonlinearity $U$, the decay rates of the two
	excitations have the same scaling relation as the
	single excitation, i.e., $\gamma_2=2\gamma_1\sim (|\Delta|\Gamma)^{-1/2}$. When increasing $U$ to the resonant point $-\Delta$,
	however, the two-excitation decay rate is gradually enhanced to $\gamma_2\sim
	\Gamma^{-1/3}$ (see Fig.~\ref{Fig:U}). In the crossover regime $0<U<-\Delta $, therefore, one
	expects antibunching even in the weak
	nonlinearity regime $U/2<2\gamma _{1}<\gamma_2$. In Fig.~\ref{Fig:wd}(c), we show
	sub-Poissonian statistics, i.e., $g^{(2)}(0)<1$, for $\Gamma /J=38.8$, $%
	\Omega /J=0.79$, and $\Delta /J=-0.3$, where $%
	g^{(2)}(0)$ monotonically decays to $0.2$ in the weak nonlinearity regime $%
	U/2<2\gamma _{1}$. In Fig.~\ref{Fig:wd}(d), $g^{(2)}(0)<g^{(2)}(\tau )$
	unambiguously displays antibunching behavior. We remark that the ability to engineer $\gamma_2>2\gamma_1$ due to the FQZ hierarchy allows for $g^{(2)}(0)<0.5$ even when the interaction is so small as $U/2<\gamma_1$. This cannot be accessed through the featureless QZ effect, where $\gamma_2\approx 2\gamma_1$ results in $g^{(2)}(0)>0.5$ for $U/2<2\gamma _{1}$.

\begin{table*}[tb]
	\caption{\label{table1}  Scaling analysis of the complex energy $\epsilon_s$ of the longest-living quasibound state for an arbitrary open bath in $d$ dimension. $C$ is a general notation for prefactor. }
	\begin{ruledtabular}
		\begin{tabular}{cccc}
			\textrm{$ {d}/{\mu} $}&
			\textrm{dDOS }&
			{\textrm{Gapped open bath ($ \gamma_{\text{min}}\neq0 $)}}&
			\textrm{Gapless open bath ($ \gamma_{\text{min}}=0$)}\\
			\colrule
			 $\frac{d}{\mu} <1 $ & $\lim_{\gamma\rightarrow \gamma_\textrm{min}}D_s(\gamma)=+\infty$ & $ \Delta+C(\gamma_{\text{min}}-i\Delta)^{-1+(d/\mu)}\Gamma^{-d/\mu} $ & $\begin{cases}C\Gamma^{\frac{1}{1-2\mu/d}} &(\Delta=0)\\
			 		\Delta+C(-i\Delta)^{-1+(d/\mu)}\Gamma^{-d/\mu} &( \Delta\neq0 )   \end{cases}$ \\
		  
			$\frac{d}{\mu} =1$ & $ \lim_{\gamma\rightarrow \gamma_\textrm{min}}D_s(\gamma)=\textrm{Const}\neq 0$ & $\Delta-iC \Gamma^{-1}\ln(\frac{\Gamma}{\gamma_{\text{min}}}) $ &$\Delta-iC \Gamma^{-1}\ln(\frac{\Gamma}{\epsilon}) $ \\
			
			$\frac{d}{\mu}  >1 $ & $\lim_{\gamma\rightarrow \gamma_\textrm{min}}D_s(\gamma)=0$ & $\Delta-iC\Gamma^{-1}$ & $\Delta-iC\Gamma^{-1}$\\
		\end{tabular}
	\end{ruledtabular}
\end{table*}

\section{Scaling behaviors for the arbitrary open bath}\label{sec:arbitrary}
	
In previous sections, we illustrate the FQZ effect for the 1D open bath with $\gamma_k=\Gamma(1+\cos k)$, which is gapless. To further understand the physics of the FQZ effect, we now present a general scaling analysis for the open bath with arbitrary dissipative bands $\gamma({\bf k})$ in dimensions $d=1,2,3$. As we show, the FQZ effect generically occurs as the result of strong dissipation and divergent dDOS near dissipative band edges, regardless of whether the bath spectrum is gapless or not. This makes the present fractional scaling intrinsically different from the conventional nonanalytic phenomena for which gapless modes are crucial. 
 		
Without loss of generality, we concentrate on the purely dissipative open bath [i.e., $\epsilon({\bf k})=0$]. In general, the dissipative band $\gamma({\bf k})$ has three characteristics. (i) The minimum dissipation rate is $\gamma_\textrm{min}\equiv\textrm{min}[\gamma({\bf k})]$, which necessarily sits at the dissipation band edge with the quasimomentum ${\bf k}_0$. When $\gamma_\textrm{min}\neq 0$, it provides the dissipative gap of the open bath. (ii) For a spatially homogeneous bath, the dissipative dispersion near the dissipative gap (edge) at ${\bf k}_0$ can be approximated as
\begin{equation}
\gamma({\bf k})= \gamma_\textrm{min}+c\Gamma |{\bf k}-{\bf k}_0|^\mu. \label{gammak0}
\end{equation}
Here, the power $\mu$ depends on the specifics of $\gamma({\bf k})$, the coefficient $c$ is such that $c{\bf k}^\mu$ is dimensionless, and $\Gamma$ characterizes the dissipation bandwidth. (iii) The dDOS in $d$ dimensions is defined as 
\begin{equation}
D_s(\gamma):=\int\frac{d^d k}{(2\pi)^d}\delta [\gamma-\gamma({\bf k})].
\end{equation}
According to Eq.~(\ref{gammak0}), the dDOS near $\gamma_\textrm{min}$ is given by
\begin{equation}
D_s(\gamma)= \frac{A}{(2\pi)^d}\frac{(\gamma-\gamma_\textrm{min})^{-1+(d/\mu)}}{\mu(c\Gamma)^{d/\mu}},\label{eq:Dgamma0}
\end{equation}
where the coefficient $A$ depends on the dimensions. We immediately see that the dDOS near $\gamma_\textrm{min}$ diverges when $d/\mu<1$ but vanishes when $d/\mu>1$. 

	\begin{figure}[tb]
		\centering
		\includegraphics[width=1\columnwidth]{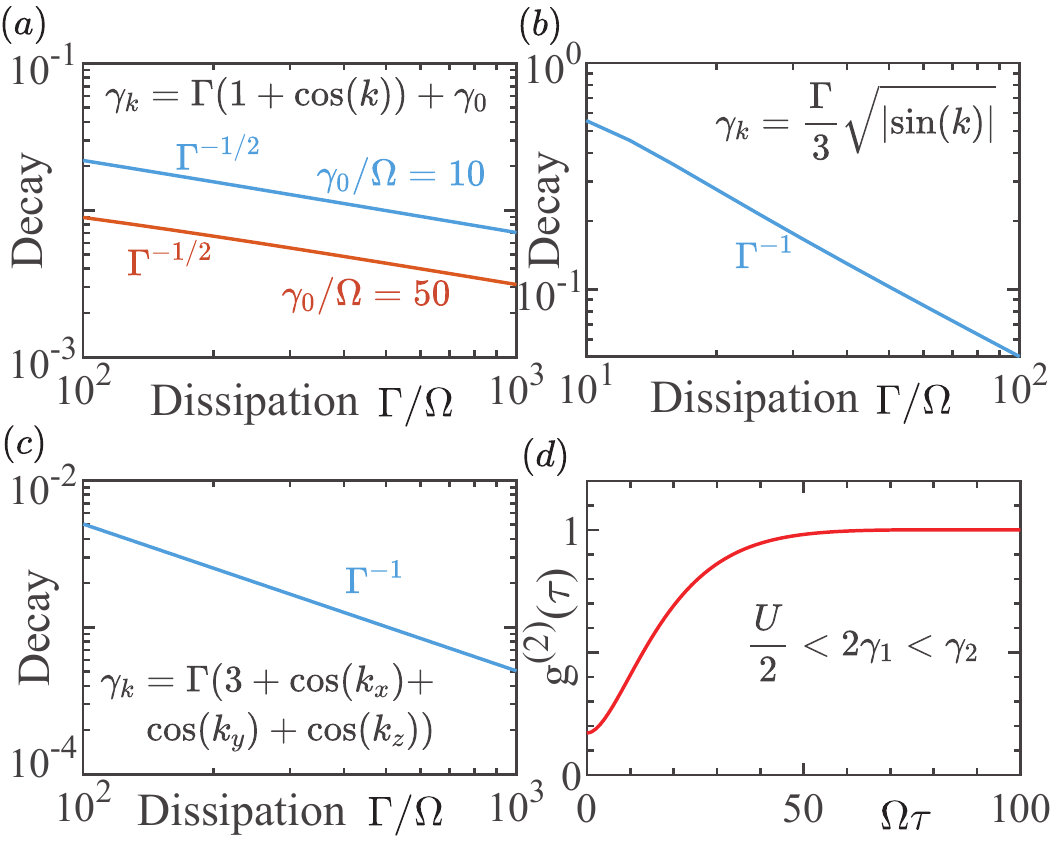}
		\caption{Scaling behavior of the emitter for various forms of $\gamma({\bf k})$ and dimension of a purely dissipative bath. (a) $\gamma(k)=\gamma_0+\Gamma(1+\cos k)$, for $\gamma_0/\Omega=10, 50$, respectively, (b) $\gamma(k)=(\Gamma/3) \sqrt{|\sin (k)|}$, and (c) $\gamma({\bf k})=\Gamma(3+\cos k_x+\cos k_y+\cos k_z)$. (d) FQZ-induced antibunching in a gapped bath with $\gamma(k) = \gamma_0+\Gamma(1+\cos k)$.
			In (a)—(c), we take $\Delta = 0$. In (d), we choose $\Delta/\Omega =-0.2$, $\Gamma/\Omega = 25$, interaction $U/\Omega = 0.2$, $\gamma_0/\Omega = 5$, and $\omega_d/\Omega = -0.23$, and $J=0$.}
		\label{Fig:scaling}
	\end{figure}

Now consider an emitter coupled with the open bath as before, and we are interested in the scaling behaviors of the complex energy $\epsilon_s$ of the quasibound states. We illustrate our analysis for $\Delta=0$ and single excitation. By extending the formalism in Sec.~\ref{sec:formalism} to the $d$-dimensional bath, we aim to solve $s+\Sigma(s)=0$, where $s=-i\omega$ and the self-energy reads as 
\begin{equation}
\Sigma(s)=\Omega^2\int \frac{d^d k}{(2\pi)^d} \frac{1}{s+\gamma({\bf k})}.\label{sigd}
\end{equation}
Since relevant for the long-time dynamics are the bath modes in the vicinity of $\gamma_\textrm{min}$, we use Eq.~(\ref{gammak0}) to obtain 
\begin{eqnarray}
\Sigma&=&\frac{A\Omega^2}{(2\pi)^d}\frac{1}{\mu(c\Gamma)^{d/\mu}}\int_{\gamma_\textrm{min}}^{\Lambda}\frac{1}{s+\gamma}\frac{1}{(\gamma-\gamma_\textrm{min})^{1-(d/\mu)}}d{\gamma} \nonumber\\
&=&C\left(\frac{\Lambda'}{\Gamma}\right)^{d/\mu} \frac{1}{s'}F\left(1,\frac{d}{\mu},\frac{d}{\mu}+1;-\frac{\Lambda'}{s'}\right). \label{sig0}
\end{eqnarray}
Here, we introduce a cutoff $\Lambda$ and redefine $s'=s+\gamma_\textrm{min}$ and $\Lambda'=\Lambda-\gamma_\textrm{min}\propto \Gamma$. Moreover, $F(\alpha,\beta,\zeta; z)$ is the hypergeometric function, and the coefficient $C$ depends on the dimensions. When the dissipation scale $\Gamma$ is largest compared to all the other relevant energy scales, we can expand the self-energy (\ref{sig0}) in terms of the small parameter $|s'/{\Lambda'}|$. We refer to Appendix~\ref{sec:all scaling} for detailed analysis. 

At the leading order, we find (i) fractional scaling, $\Sigma\propto \Gamma^{-d/\mu}$ for $d/\mu<1$, (ii) logarithmic behavior, $\Sigma\propto (\ln\Gamma)/\Gamma$ for $d/\mu=1$, and (iii) integer scaling, $\Sigma\propto \Gamma^{-1}$ for $d/\mu>1$. Finally, by solving $s+\Sigma(s)=0$, we analytically derive the scaling relations for the complex energy $\epsilon_s$ of quasibound states. The scalings for $\Delta\neq 0$ are obtained in a similar way. 

In Table~\ref{table1}, we collect the results of $\epsilon_s$ of the quasibound states with the smallest decay rate, for $\gamma_\textrm{min}\neq 0$ and $\gamma_\textrm{min}=0$, respectively. We see that fractional scalings always arise whenever $D_s(\gamma_\textrm{min})\rightarrow \infty$, whereas the standard QZ effect emerges if $D_s(\gamma_\textrm{min})\rightarrow 0$.  Note that, although singular dDOS may appear at other places of the Brillouin zone, in the strong dissipation regime, only the dDOS near $\gamma_\textrm{min}$ is important for the dynamical long-time behaviors of the emitters.

To validate above analysis, we numerically solve the poles of the single-particle Green function for three examples of $\gamma({\bf k})$. The results for $\Delta=0$ are presented in Fig.~\ref{Fig:scaling}. The first example is $\gamma(k)=\gamma_0+\Gamma(1+\cos k)$, corresponding to the gapped version of the 1D case considered in previous sections. If gapless modes are necessary for the fractional scalings, one would expect the FQZ effect to disappear. Instead, in Fig.~\ref{Fig:scaling}(a), we find the FQZ effect under various gap sizes of $\gamma_0$, where the $\Gamma^{-1/2}$ scaling agrees with what is predicted from the quadratic dissipative dispersion $\gamma(k)=\gamma_0+\Gamma(k-\pi)^2/2$ with the divergent dDOS at $\gamma_0$. As the second example, we consider a gapless 1D dissipation band $\gamma(k)=\Gamma\sqrt{|\sin (k)|}/3$. We see that standard QZ effect emerges [Fig.~\ref{Fig:scaling}(b)]. This can be understood, because $\gamma(k)\propto |k|^{1/2}$ near $\gamma_\textrm{min}=0$, so that the dDOS vanishes at $k=0$, leading to the linear scaling according to Table \ref{table1}. The third example is $\gamma({\bf k})=\Gamma(3+\cos k_x+\cos k_y+\cos k_z)$ in 3D. Similarly as its 1D counterpart analyzed previously, this is a gapless spectrum with $\gamma({\bf k})\propto|{\bf k}-\pi|^2$ near $\gamma_\textrm{min}=0$ at ${\bf k}=(\pi,\pi,\pi)$. However, Fig.~\ref{Fig:scaling}(c) reveals a completely different behavior from the 1D case, where the QZ effect, instead of the FQZ effect, emerges, due to the vanishing 3D dDOS at  $\gamma_\textrm{min}=0$. 

	In summary, we arrive at the physical picture that the FQZ effect occurs whenever the open bath itself undergoes strong dissipation and $D_s(\gamma_\textrm{min})$ diverges, irrespective of whether there are gapless modes or not. This conclusion is applicable also for the case with multiple excitations. As an application, in Fig.~\ref{Fig:scaling}(d), we show the FQZ-induced strong photon antibunching for a weak nonlinearity when the bath has the gapped dissipation spectrum $\gamma(k)=\gamma_0+\Gamma(1+\cos k)$. Note that, as shown by Table~\ref{table1}, manipulation of dDOS can tune the scaling behavior, e.g., from fractional scalings to logarithmic behavior or to integer scalings. 
	
	\section{Experimental implementation}\label{sec:exp}
	
\begin{figure}[tb]
		\centering
		\includegraphics[width=0.75\columnwidth]{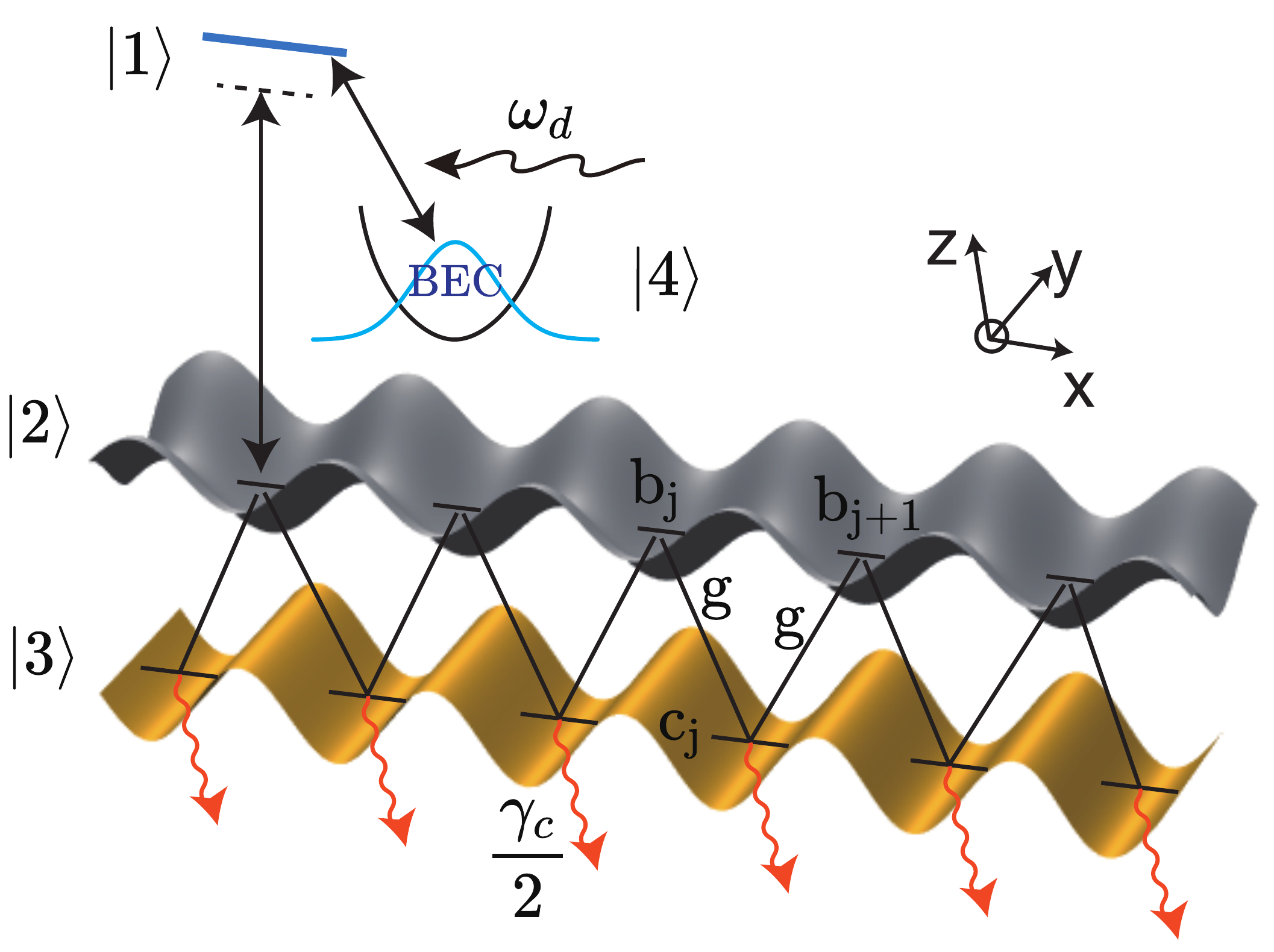}
		\caption{Implementation scheme with cold atoms for the master equation (\ref{eq:rho}). We use three hyperfine levels of the bosonic atom, labeled as $|i\rangle$ ($i=1,2,3$), respectively, to encode the emitter $a$, the bath $b$, and the auxiliary bath $c$. In $|1\rangle$, atoms can undergo Feshbach resonance. In $|2\rangle$, atoms are deeply trapped in the 3D lattice (gray). In $|3\rangle$, atoms are free in the $y-z$ directions but deeply trapped in a 1D optical lattice (orange) in the $x$ direction. We prepare the atomic BEC in another hyperfine state labeled by $|4\rangle$ and use the laser with frequency $\omega_d$ to induce the transition between $|4\rangle$ and $|1\rangle$. States $|1\rangle$ and $|2\rangle$ are coupled by lasers. Both $b_j$ and $b_{j+1}$ are near-resonant coupled to $c_j$ with the coupling rate $g$. When atoms are excited to $|3\rangle$, they are quickly lost from the system in the $y-z$ direction with a large loss rate $\gamma_c/2$.}
		\label{Fig:coldatom}
	\end{figure}

	Although the FQZ effect is predicted in the thermodynamic limit, it can be observed for the open bath with the finite size $N_b$, provided $N_b$ and the system parameters are such that the condition $\Omega /\delta \gamma \gg 1$ with $\delta\gamma=2\pi^2\Gamma/N_b^2$ is satisfied (see Appendix~\ref{sec:condition}). In this section, we present the microscopic setup for realizing the FQZ effect by using ultracold atoms in the state-dependent optical lattices, as schematically illustrated in Fig.~\ref{Fig:coldatom}. Recently, engineered lattice models with dissipative couplings have been demonstrated with the momentum-space lattice of cold atoms~\cite{Yanbo2022} as well as an ensemble of photonic resonators~\cite{FanSH2021} or atomic spin waves~\cite{Dongdong2022} coupled to the auxiliary reservoir. Our implementation of the master equation (\ref{eq:rho}) is in line with these experiments.
			
We encode the emitter $a$, the bath $b$, and the auxiliary bath $c$ in three ground-state hyperfine levels of the bosonic atom, labeled as $|1\rangle$, $|2\rangle$, and $|3\rangle$, respectively. 

(i) In state $|1\rangle$, atoms can undergo Feshbach resonance and realize the nonlinear term $(U/2)a^{\dag2} a^2$. To realize the driving field, we can prepare the atomic Bose-Einstein condensate (BEC) in another hyperfine state labeled by $|4\rangle$ and use the external laser field with frequency $\omega_d$ to induce the transition between $|4\rangle$ and $|1\rangle$. This implements the driving term $\varepsilon a^{\dagger }e^{-i\omega _{d}t}+\mathrm{H.c.}$, where $\varepsilon $ is related to the mean-field wave function of the BEC. 

(ii) In state $|2\rangle$, atoms are deeply trapped in the 3D optical lattice with $N_b$ sites. For realizing the FQZ effect, the term $\sum_j Je^{i\theta}b_j^\dag b_{j+1}+\textrm{H.c.}$ is not necessary, as shown previously. Note that, for general purpose, this term can be readily realized via a two-photon Raman transition between the adjacent lattice sites~\cite{Goldman2016}, where the phase $\theta$ is controlled via the relative phase of the coupling lasers. The term $\Omega a^\dag b_0+\textrm{H.c}.$ is implemented by using the laser to induce the transition between $|1\rangle$ and $|2\rangle$. 

(iii) In state $|3\rangle$, atoms are free in the $y-z$ directions but are deeply trapped in a 1D optical lattice in the $x$ direction, where the tunneling rate is ignorable. When atoms are excited to $|3\rangle$, they are quickly lost from the system in the $y-z$ directions with the loss rate $\gamma_c/2$. Both $b_j$ and $b_{j+1}$ are near-resonant coupled to $c_j$ via lasers ($j=1,...,N_b-1$), with the coupling rate $g$. For large loss rate $\gamma_c/2$, modes $c_j$ can be adiabatically eliminated to realize the nonlocal dissipator in Eq.~(\ref{eq:D}) with $\Gamma=4g^2/\gamma_c$. To observe the FQZ effect requires one to tune the parameters to satisfy $\Omega /\delta \gamma \gg 1$. 

	An alternative atomic platform may be
	provided by thermal atoms in a vapor cell~\cite{antiPT2016}. A unique
	feature of such setup is that atomic spin waves created by the
	electromagnetic-induced transparency in spatially separated optical channels is naturally dissipatively coupled via flying atoms. By further controlling the separation and laser beams, a dissipative atomic-spin-wave lattice has been realized~\cite{Dongdong2022}. The light interacting with the spin waves in an optical channel, thus, represents an emitter coupled to the open bath, whose properties can be detected via the transmission spectroscopy.\newline
	
	\section{Conclusion}\label{sec:con}

	In this work, we predict quantum non-Hermitian phenomena, the FQZ effect and the FQZ-induced sub-Poissonian photon statistics, based on a paradigm where nonlinear emitters interact with an engineered open bath. The FQZ effect generally arises from the combination of strong dissipation and divergent dDOS near the dissipation band edge and has no immediate counterpart in the closed-bath context. Capitalizing on its unique, excitation-number-dependent scaling behaviors, we are able to judiciously design a hierarchy of decay rates for the emitters. This opens a new route toward the generation of strong photon antibunching in the limit of weak nonlinearities. Remarkably, we identify that the present sub-Poissonian quantum statistics of photons is driven by the key role of non-Hermiticity. Our result presents a first step toward the exploration of non-Hermitian quantum optics. It is also of relevance in the context of recent experiments for non-Hermitian lattice models, where demonstrating quantum non-Hermitian phenomena remains an open challenge. 
			
	Our work offers a new way to design the system-bath interaction by engineering the intrinsic dissipation band structure of the open bath. When the bath undergoes strong dissipation by itself, the emitters are dynamically enforced to mainly couple with the weakly dissipating modes hosted near the dissipation band edge, whose dDOS plays a central role in the dynamical long-time behaviors of emitters. This route complements the conventional way to engineer the emitter-bath interaction in the closed-bath context which crucially relies on the energy resonance. It also opens a new path to realize interesting quantum non-Hermitian physics, as well as quantum simulations of many-body systems. Beyond engineering either an energy or a dissipation band of the bath, it is interesting to explore how their combinations may give rise to intriguing quantum effects. 
	
	In summary, our work provides a feasible route in the highly desired, yet challenging, quest for non-Hermitian quantum many-body effects. Beyond the general fundamental interests,
	ultimately, understanding the role played by non-Hermiticity in fully quantum regimes will enable us to leverage recent advances in
	non-Hermitian Hamiltonian engineering for actual quantum applications.

	\bigskip

	\section{Acknowledgements}
	
	We thank enlightening discussions with Carlos Navarette Benlloch, Hannes Pichler, Mikhail A. Baranov, Wei Yi, and Yanhong Xiao. This research is funded by National Key Research and Development Program of China (Grants No. 2022YFA1404201, No. 2022YFA1203903, and No. 2022YFA1404003), the National Natural Science Foundation of China (Grants No. 12034012 and No. 11874038), and NSFC-ISF (Grant No. 12161141018). T.S. is supported by National Key Research and Development Program of China (Grant No. 2017YFA0718304) and the National Natural Science Foundation of China (Grants No. 11974363, No. 12135018, and No. 12047503). Z.L. is supported by the National Natural Science Foundation of China (No. 52031014).

	Y.S. and T.S. contributed equally to this work.

	\appendix
	
	\section{Green functions and analytic continuations}
	
	\label{sec:G}
	 
	In this section, we derive the retarded Green functions $G^{R}(\omega )$ and $%
	D^{R}(\omega )$ for the undriven emitters. By the appropriate analytic continuation, we introduce the
	Green function $G_f(\omega )$ also available in the second RS, which naturally gives rise to the effective Hamiltonian $\bar{H%
	}_{\mathrm{eff}}$ describing emitters coupled to the bath with the simple
	dispersion relation and decay rates. In the first and second subsections, we
	study the situations for single and two emitters. In the third section, we
	derive $D^{R}(\omega )$ using the ladder diagram. The retarded Green
	function $D(t)$ in the time domain can be efficiently calculated using the
	analytic continuation $D_f(\omega )$ of $D^{R}(\omega )$. 
	
Before proceeding, we remark that, at zero temperature, the steady state of the master equation (\ref{eq:rho}) without the driving field ($\epsilon=0$) is the equilibrium state represented by the vacuum state $|0\rangle$ of excitations. There, the fluctuation-dissipation theorem applies. Specifically, time-ordered single-particle Green functions $G^{t}(t)=-i\langle 0\vert \mathcal{T}a_{l}(t)a_{1}^{\dagger
}(0)\vert 0\rangle$ and the retarded Green functions $G^{R}(t)=-i\langle 0\vert 
\mathcal{[}a_{l}(t),a_{1}^{\dagger }(0)]\vert 0\rangle \theta (t)
$ coincide with each other. Similarly, the time-ordered two-particle Green function $D^{t}(t)=-i\langle 0\vert 
\mathcal{T}a_{1}^{2}(t)a_{1}^{\dagger 2}(0)\vert 0\rangle /2$ coincides with the retarded two-particle Green function $D^{R}(t)=-i\langle 0\vert \mathcal{[}a_{1}^{2}(t),a_{1}^{%
\dagger 2}(0)]\vert 0\rangle \theta (t)/2$. That is, $G^{t}(t)=G^{R}(t)\equiv G(t)$, and $D^{t}(t)=D^{R}(t)\equiv
D(t)$, which have simple relations $G(t)=G^{K}(t)\theta (t)$ and $%
D(t)=D^{K}(t)\theta (t)$\ with Keldysh Green functions $G^{K}(t)=-i%
\langle 0\vert \mathcal{\{}a_{l}(t),a_{1}^{\dagger }(0)\}\vert
0\rangle $ and $D^{K}(t)=-i\langle 0\vert \mathcal{\{}%
a_{1}^{2}(t),a_{1}^{\dagger 2}(0)\}\vert 0\rangle /2$, respectively. We show later in Appendix~\ref{undriven} how one can efficiently study the spontaneous emission of $n$ excitations ($n=1,2,...$) through the $n$-particle retarded Green function in $|0\rangle$.

	\subsection{Single emitter}
		
\begin{figure}[tb]
		\centering
		\includegraphics[width=0.8\columnwidth]{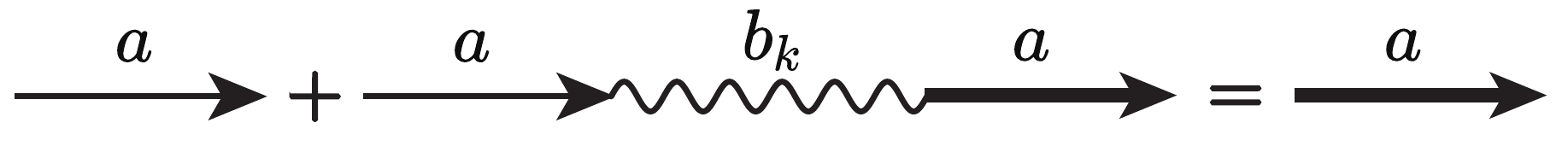}
		\caption{Feynman diagram for the single-excitation Green function. The thin
			(bold) arrow denotes the bare (exact) single-excitation Green function of
			the emitter, and the curvy line denotes the bare propagator of the bath mode.
		}
		\label{SFig1}
	\end{figure}

	For the open bath, the three Green functions, i.e., the retarded $G_{%
		b}^{R}$, the advanced $G_{b}^{A}$, and the Keldysh $G_{%
		b}^{K}$ Green functions, in the frequency domain are%
	\begin{align}
		G_{b}(\omega )=&\left(
		\begin{array}{cc}
			G_{b}^{K}(\omega ) & G_{b}^{R}(\omega ) \\
			G_{b}^{A}(\omega ) & 0%
		\end{array}%
		\right)  \notag \\
		=&\left(
		\begin{array}{cc}
			\frac{-2i\gamma _{k}}{(\omega -\varepsilon _{k})^{2}+\gamma _{k}^{2}} &
			\frac{1}{\omega -\varepsilon _{k}+i\gamma _{k}} \\
			\frac{1}{\omega -\varepsilon _{k}-i\gamma _{k}} & 0%
		\end{array}%
		\right) \notag
	\end{align}%
	in the Keldysh space, where the dispersion relation $\varepsilon _{k}=2J\cos
	(k+\theta )$ and the decay rate $\gamma _{k}=\Gamma (1+\cos k)$. As shown in
	Fig.~\ref{SFig1}, the Dyson expansion of the Rabi coupling term gives rise
	to the Green function%
	\begin{align}
		G(\omega )=&\left(
		\begin{array}{cc}
			G^{K}(\omega ) & G^{R}(\omega ) \\
			G^{A}(\omega ) & 0%
		\end{array}%
		\right)  \notag \\
		=&\frac{1}{(\omega -\Delta )\sigma ^{x}-\frac{\Omega ^{2}}{N}%
			\sum_{k}\sigma^{x}G_{b}\sigma ^{x}},  \label{Gs}
	\end{align}%
	where $\sigma ^{x}$ is the Pauli matrix. More explicitly, the analytic
	structure of the retarded Green function%
	\begin{equation}
		G^{R}(\omega )=\frac{1}{\omega -\Delta -\frac{\Omega ^{2}}{N}\sum_{k}\frac{1%
			}{\omega -\varepsilon _{k}+i\gamma _{k}}}\notag
	\end{equation}%
	fully determines the dynamics of the emitter. In the main text, we focus on
	the case $\theta =-\pi /2$.
	
	Since the bath has the mode-dependent dispersion relation $\varepsilon _{k}$
	and decay rate $\gamma _{k}$, the branch cut $\omega _{k}=\varepsilon
	_{k}-i\gamma _{k}$ forms an ellipse rather than collapsing into a line like
	that in the closed system ($\gamma _{k}=0$). The ellipse centered at $%
	(0,-\Gamma )$ has the major axis $2J$ and $\Gamma $ for $2J>\Gamma $ and $%
	2J<\Gamma $, respectively. In the special case $2J=\Gamma $, the branch cut
	becomes a circle. In the thermodynamic limit, the self-energy%
	\begin{equation}
		\Sigma(\omega )=\Omega ^{2}\int \frac{dk}{2\pi }\frac{1}{\omega -\omega
			_{k}}=\Omega ^{2}\int_{\left\vert z\right\vert =1}\frac{dz}{2\pi iz}\frac{1}{%
			\omega -\lambda (z)} \notag
	\end{equation}%
	becomes the contour integral in the $z$ plane ($z=e^{ik}$), which is
	completely determined by the poles $z_{0}$, i.e., $\omega -\lambda (z_{0})=0$, and the corresponding residues.
	
\begin{figure}[tb]
		\includegraphics[width=0.8\columnwidth]{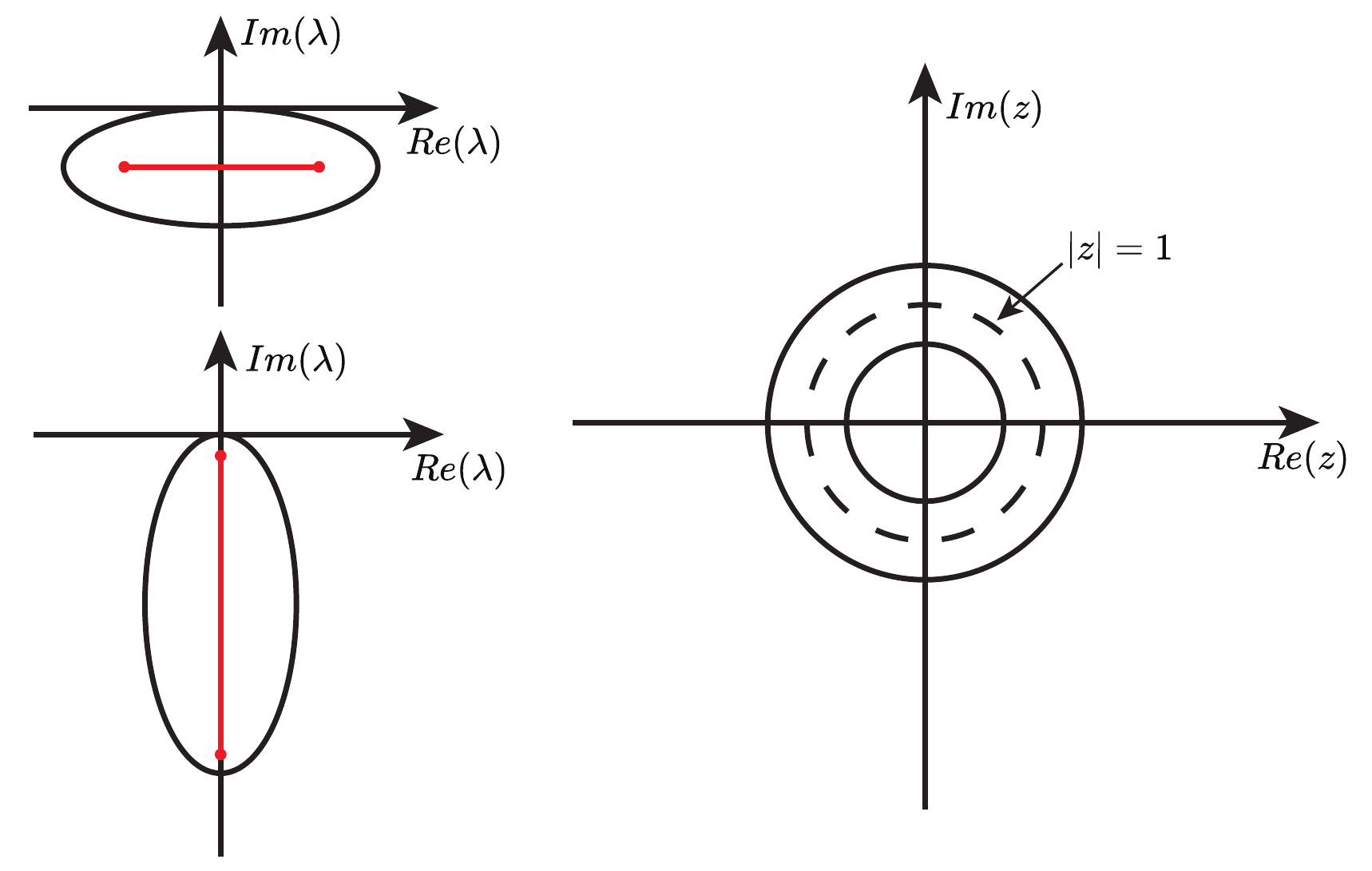}
		\caption{Conformal map from the $\protect\lambda$ plane to the $z$ plane.}
		\label{SubFig2}
	\end{figure}

	Before performing the lengthy calculation, we notice that the complex
	function $\lambda (z)$ conformally maps the annular region $1<\left\vert
	z\right\vert <z_{\text{max}}=\left\vert [J+(\Gamma/2)]/[J-(
		\Gamma/2)]\right\vert $ into the inner region of the first RS, i.e., the
	region inside the ellipse, where, in particular, the contour $\left\vert
	z\right\vert =\sqrt{z_{\text{max}}}$\ is mapped to a line connecting the
	foci of the ellipse (see Fig.~\ref{SubFig2}). As a result, for $\omega $ localized inside the
	ellipse, the poles $z_{0}$ are always in the region $1<\left\vert
	z\right\vert <z_{\text{max}}$, which results in the vanishing $\Sigma
	(\omega )$ in the first RS. The analytic continuation can be performed
	by the deformation of the integral contour from $\left\vert z\right\vert =1$
	to $\left\vert z\right\vert =\sqrt{z_{\text{max}}}$, which reproduces the
	same self-energy $\Sigma(\omega )$ in the first RS and extends it to the
	second RS. More specifically, for $2J>\Gamma $%
	\begin{eqnarray}
		&&\Sigma_f(\omega )  \notag \\
		&=&\Omega ^{2}\int_{\left\vert z\right\vert =\sqrt{z_{\text{max}}}}\frac{dz}{%
			2\pi iz}\frac{1}{\omega +i\Gamma -i(J-\frac{\Gamma }{2})z+i(J+\frac{\Gamma }{%
				2})z^{-1}}  \notag \\
		&=&\Omega ^{2}\int_{\left\vert z\right\vert =1}\frac{dz}{2\pi iz}\frac{1}{%
			\omega +i\Gamma -iJ_{\mathrm{eff}}z+iJ_{\mathrm{eff}}z^{-1}}  \notag \\
		&=&\Omega ^{2}\int \frac{dk}{2\pi }\frac{1}{\omega +i\Gamma +2J_{\mathrm{eff}%
			}\sin k},  \label{Sb}
	\end{eqnarray}%
	where $J_{\mathrm{eff}}=\sqrt{J^{2}-\Gamma ^{2}/4}$ and in the last step we
	use the relation $z=e^{ik}$. The comparison between $\Sigma (\omega )$
	and the last line in Eq. (\ref{Sb}) shows that the bath of the emitter can
	be effectively replaced by that with a much simpler spectrum $\bar{\omega}%
	_{k}=-i\Gamma -2J_{\mathrm{eff}}\sin k$, where the decay rate is the
	constant $\Gamma $. Similarly, for $2J<\Gamma $%
	\begin{eqnarray}
		&&\Sigma_f(\omega )  \notag \\
		&=&\Omega ^{2}\int_{\left\vert z\right\vert =\sqrt{z_{\text{max}}}}\frac{dz}{%
			2\pi iz}\frac{1}{\omega +i\Gamma -i(J-\frac{\Gamma }{2})z+i(J+\frac{\Gamma }{%
				2})z^{-1}}  \notag \\
		&=&\Omega ^{2}\int_{\left\vert z\right\vert =1}\frac{dz}{2\pi iz}\frac{1}{%
			\omega +i\Gamma +iJ_{\mathrm{eff}}z+iJ_{\mathrm{eff}}z^{-1}}  \notag \\
		&=&\Omega ^{2}\int \frac{dk}{2\pi }\frac{1}{\omega +i\Gamma +2iJ_{\mathrm{eff%
			}}\cos k},\notag
	\end{eqnarray}%
	where $J_{\mathrm{eff}}=\sqrt{\Gamma ^{2}/4-J^{2}}$. For this case, the
	effective bath has the spectrum $\bar{\omega}_{k}=-i\Gamma -2iJ_{\mathrm{eff}%
	}\cos k$, where the dispersion relation becomes trivial.
	
	The advantage of the analytic continuation is to collapse the complex
	elliptical branch cut to the line connecting the foci of the ellipse in the
	second RS. The contour integrals in $\Sigma_f(\omega )$ can be
	obtained efficiently as%
	\begin{eqnarray}
		\Sigma_f(\omega ) &=&\frac{\Omega ^{2}\text{sgn}[1-\left\vert z_{-}(\omega
			)\right\vert ]}{\sqrt{(\omega +i\Gamma )^{2}-4J_{\mathrm{eff}}^{2}}},  \notag
		\\
		z_{-}(\omega ) &=&-i\bigg[\frac{\omega +i\Gamma }{2J_{\mathrm{eff}}}-\sqrt{\frac{%
				(\omega +i\Gamma )^{2}}{4J_{\mathrm{eff}}^{2}}-1}\bigg]\notag
	\end{eqnarray}%
	for $2J>\Gamma $ and%
	\begin{eqnarray}
		\Sigma_f(\omega ) &=&\frac{i\Omega ^{2}\text{sgn}[1-\left\vert
			z_{-}(\omega )\right\vert ]}{\sqrt{-(\omega +i\Gamma )^{2}-4J_{\mathrm{eff}%
				}^{2}}},  \notag \\
		z_{-}(\omega ) &=&i\frac{\omega +i\Gamma }{2J_{\mathrm{eff}}}-\sqrt{-\frac{%
				(\omega +i\Gamma )^{2}}{4J_{\mathrm{eff}}^{2}}-1}\notag
	\end{eqnarray}%
	for $2J<\Gamma $. The self-energy $\Sigma_f(\omega )$ results in the
	analytic continuation $G_f(\omega )=1/[\omega -\Delta -\Sigma_f(\omega )]$ of $G^{R}(\omega )$.
	
	\begin{figure}[tb]
		\includegraphics[width=0.8\columnwidth]{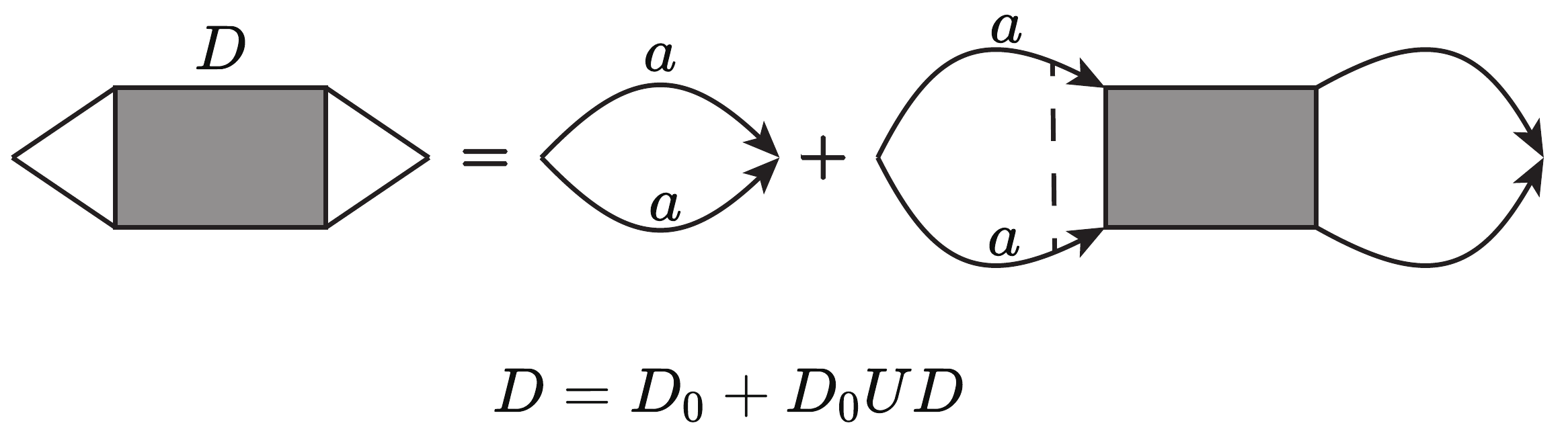}
		\caption{Feynman diagram for the two-excitation Green function. The first
			term is the convolution of two exact single-excitation Green functions. In
			the second term, the dashed line denotes the interaction of the emitter.}
		\label{SubFig3}
	\end{figure}
	
	The spontaneous decay of the excitation is described by the Fourier
	transform $G(t)=\int (d\omega /2\pi )e^{-i\omega t}G_f(\omega
	) $, which is determined by the behaviors in the vicinity of poles and
	branch cuts of $G_f$. For $2J>\Gamma $, the Green function%
	\begin{align}
		G_f(\omega )=&\sum_{s}\frac{Z_{s}}{\omega -\varepsilon _{s}}  +\int_{-2J_{\mathrm{eff}}}^{2J_{\mathrm{eff}}}\frac{dx}{\pi }\frac{1}{%
			\omega -x+i\Gamma }\notag\\ 
		&\times\frac{\Omega ^{2}\sqrt{4J_{\mathrm{eff}}^{2}-x^{2}}}{(4J_{%
				\mathrm{eff}}^{2}-x^{2})(x-\Delta -i\Gamma )^{2}+\Omega ^{4}}\notag
	\end{align}%
	is determined by the poles $\varepsilon _{s}$ [i.e., $G_f
	^{-1}(\varepsilon _{s})=0$], the corresponding residues%
	\begin{equation}
		Z_{s}=\frac{1}{1-\left. \partial _{\omega }\Sigma_f\right\vert
			_{\omega =\varepsilon _{s}}},\label{Zs}
	\end{equation}%
	and the contribution from the branch cut, whose Fourier transform can be
	performed straightforwardly as%
	\begin{align}
		G(t)=&-i\sum_{s}Z_{s}e^{-i\varepsilon _{s}t}  -i\int_{-2J_{\mathrm{eff}}}^{2J_{\mathrm{eff}}}\frac{dx}{\pi }\notag\\
		&\times\frac{\Omega
			^{2}\sqrt{4J_{\mathrm{eff}}^{2}-x^{2}}}{(4J_{\mathrm{eff}}^{2}-x^{2})(x-%
			\Delta -i\Gamma )^{2}+\Omega ^{4}}e^{-ixt-\Gamma t}.\notag
	\end{align}%
	For $2J<\Gamma $, the Fourier transform of the Green function%
	\begin{align}
		G_f(\omega )=&\sum_{s}\frac{Z_{s}}{\omega -\varepsilon _{s}}-\int_{-2J_{\mathrm{eff}}}^{2J_{\mathrm{eff}}}\frac{dx}{\pi }\frac{1}{%
			\omega -ix+i\Gamma }\notag\\
		&\times\frac{\Omega ^{2}\sqrt{4J_{\mathrm{eff}}^{2}-x^{2}}}{%
			(4J_{\mathrm{eff}}^{2}-x^{2})(x-\Gamma +i\Delta )^{2}+\Omega ^{4}}\notag
	\end{align}%
	gives%
	\begin{align}
		G(t)=&-i\sum_{s}Z_{s}e^{-i\varepsilon _{s}t}  +i\int_{-2J_{\mathrm{eff}}}^{2J_{\mathrm{eff}}}\frac{dx}{\pi }\notag\\
		&\times\frac{\Omega
			^{2}\sqrt{4J_{\mathrm{eff}}^{2}-x^{2}}}{(4J_{\mathrm{eff}}^{2}-x^{2})(x-%
			\Gamma +i\Delta )^{2}+\Omega ^{4}}e^{-(\Gamma -x)t}.  \label{eq:GRt}
	\end{align}
	
	\subsection{Two emitters}
	
	For two emitters in the open bath, the retarded Green function%
	\begin{equation}
		G^{R}(\omega )=\frac{1}{\omega -\Delta -\Sigma (\omega )}\notag
	\end{equation}%
	of emitters is determined by the self-energy matrix $\Sigma(\omega )$
	whose element is%
	\begin{equation}
		\Sigma _{ll^{\prime }}(\omega )=\frac{\Omega ^{2}}{N}\sum_{k}\frac{%
			e^{ik(l-l^{\prime })}}{\omega -\varepsilon _{k}+i\gamma _{k}}.\label{sigllp}
	\end{equation}%
	In the thermodynamic limit, the matrix element becomes%
	\begin{align}
		&\Sigma _{ll^{\prime }}(\omega )\notag\\
		=& \Omega ^{2}\int \frac{dk}{2\pi }\frac{%
			e^{ikd_{ll^{\prime }}}}{\omega -\varepsilon _{k}+i\gamma _{k}}  \notag \\
		=& \Omega ^{2}\int_{\left\vert z\right\vert =1}\frac{dz}{2\pi iz}\frac{%
			z^{d_{ll^{\prime }}}}{\omega -\lambda (z)}  \notag \\
		=& \Omega ^{2}\int_{\left\vert z\right\vert =1}\frac{dz}{2\pi iz}\frac{%
			z^{d_{ll^{\prime }}}}{\omega +i\Gamma -i(J-\frac{\Gamma }{2})z+i(J+\frac{%
				\Gamma }{2})z^{-1}}, \notag
	\end{align}%
	where $d_{ll^{\prime }}=l-l^{\prime }$.
	
	The analytic continuation can also be applied here as%
	\begin{eqnarray}
		&&\Sigma_{f,ll^{\prime }}(\omega )  \notag \\
		&=&\Omega ^{2}\int_{\left\vert z\right\vert =\sqrt{z_{\text{max}}}}\frac{dz}{%
			2\pi iz}\frac{z^{d_{ll^{\prime }}}}{\omega +i\Gamma -i(J-\frac{\Gamma }{2}%
			)z+i(J+\frac{\Gamma }{2})z^{-1}}  \notag \\
		&=&\Omega ^{2}{z_{\text{max}}^{d_{ll^{\prime }}/2}}\int_{\left\vert
			z\right\vert =1}\frac{dz}{2\pi iz}\frac{z^{d_{ll^{\prime }}}}{\omega
			+i\Gamma -iJ_{\mathrm{eff}}z+iJ_{\mathrm{eff}}z^{-1}}  \notag \\
		&=&\Omega ^{2}z_{\text{max}}^{d_{ll^{\prime }}/2}\int \frac{dk}{2\pi }\frac{%
			e^{ikd_{ll^{\prime }}}}{\omega +i\Gamma +2J_{\mathrm{eff}}\sin k}  \notag \\
		&=&i^{d_{ll^{\prime }}}\Omega ^{2}z_{\text{max}}^{d_{ll^{\prime }}/2}\int
		\frac{dk}{2\pi }\frac{e^{ik\left\vert d_{ll^{\prime }}\right\vert }}{\omega
			+i\Gamma +2J_{\mathrm{eff}}\cos k} \notag
	\end{eqnarray}%
	for $2J>\Gamma $ and%
	\begin{eqnarray}
		&&\Sigma _{f,ll^{\prime }}(\omega )  \notag \\
		&=&\Omega ^{2}\int_{\left\vert z\right\vert =\sqrt{z_{\text{max}}}}\frac{dz}{%
			2\pi iz}\frac{z^{d_{ll^{\prime }}}}{\omega +i\Gamma +i(\frac{\Gamma }{2}%
			-J)z+i(\frac{\Gamma }{2}+J)z^{-1}}  \notag \\
		&=&\Omega ^{2}{z_{\text{max}}^{d_{ll^{\prime }}/2}}\int_{\left\vert
			z\right\vert =1}\frac{dz}{2\pi iz}\frac{z^{d_{ll^{\prime }}}}{\omega
			+i\Gamma +iJ_{\mathrm{eff}}z+iJ_{\mathrm{eff}}z^{-1}}  \notag \\
		&=&\Omega ^{2}z_{\text{max}}^{d_{ll^{\prime }}/2}\int \frac{dk}{2\pi }\frac{%
			e^{ik\left\vert d_{ll^{\prime }}\right\vert }}{\omega +i\Gamma +2iJ_{\mathrm{%
					eff}}\cos k}\notag
	\end{eqnarray}%
	for $2J<\Gamma $. In the matrix form, the self-energy reads%
	\begin{align}
		\Sigma_f(\omega )=&\Omega ^{2}\int \frac{dk}{2\pi }\frac{1}{\omega
			+i\Gamma +2J_{\mathrm{eff}}\cos k} +\Omega ^{2}\int \frac{dk}{2\pi }\notag\\
		&\times\frac{e^{ikd}}{\omega +i\Gamma +2J_{%
				\mathrm{eff}}\cos k}\left(
		\begin{array}{cc}
			0 & i^{-d}z_{\text{max}}^{-d/2} \\
			i^{d}z_{\text{max}}^{d/2} & 0%
		\end{array}%
		\right)\notag
	\end{align}%
	for $2J>\Gamma $ and%
	\begin{align}
		\Sigma_f(\omega )=&\Omega ^{2}\int \frac{dk}{2\pi }\frac{1}{\omega
			+i\Gamma +2iJ_{\mathrm{eff}}\cos k}   +\Omega ^{2}\int \frac{dk}{2\pi }\notag\\
		&\times\frac{e^{ikd}}{\omega +i\Gamma +2iJ_{%
				\mathrm{eff}}\cos k}\left(
		\begin{array}{cc}
			0 & z_{\text{max}}^{-d/2} \\
			z_{\text{max}}^{d/2} & 0%
		\end{array}%
		\right)\notag
	\end{align}%
	for $2J<\Gamma $, where $d$ is the distance between two emitters.
	
	By comparing the self-energies $\Sigma_f$ and $\Sigma$, we can
	write the effective Hamiltonians%
	\begin{align}
		\bar{H}_{\mathrm{eff}} =&\Delta \sum_{l}a_{l}^{\dagger }a_{l}+\sum_{k}(-2J_{%
			\mathrm{eff}}\cos k-i\Gamma )b_{k}^{\dagger }b_{k}  \notag \\
		&+\frac{\Omega }{\sqrt{N}}\sum_{k}[(a_{1}^{\dagger }+i^{d}z_{\text{max}%
		}^{d/2}e^{ikd}a_{2}^{\dagger })b_{k}  \notag \\
		&+b_{k}^{\dagger }(a_{1}+i^{-d}z_{\text{max}}^{-d/2}e^{-ikd}a_{2})] \notag
	\end{align}%
	for $2J>\Gamma $ and%
	\begin{align}
		\bar{H}_{\mathrm{eff}} =&\Delta \sum_{l}a_{l}^{\dagger }a_{l}+\sum_{k}(-2iJ_{%
			\mathrm{eff}}\cos k-i\Gamma )b_{k}^{\dagger }b_{k}  \notag \\
		&+\frac{\Omega }{\sqrt{N}}\sum_{k}[(a_{1}^{\dagger }+z_{\text{max}%
		}^{d/2}e^{ikd}a_{2}^{\dagger })b_{k}  \notag \\
		&+b_{k}^{\dagger }(a_{1}+z_{\text{max}}^{-d/2}e^{-ikd}a_{2})] \notag
	\end{align}%
	for $2J<\Gamma $.
	
	The analytic continuation and effective models allow us to diagonalize the
	Green function analytically as follows. For $2J>\Gamma $, the Green function%
	\begin{eqnarray}
		G_f(\omega ) &=&\frac{1}{\omega -\Delta -\Sigma_f(\omega )}
		\notag \\
		&=&S\left(
		\begin{array}{cc}
			G_{+}(\omega ) & 0 \\
			0 & G_{-}(\omega )%
		\end{array}%
		\right) S^{-1} \label{GR2}
	\end{eqnarray}%
	is diagonalized in the \textquotedblleft $\pm $\textquotedblright\ channels
	with eigenvalues%
	\begin{equation}
		G_{\pm }(\omega )=\frac{1}{\omega -\Delta -\Sigma_{\pm }(\omega )},
		\label{GRpm}
	\end{equation}%
	where the transformation%
	\begin{equation}
		S=\frac{1}{\sqrt{2}}\left(
		\begin{array}{cc}
			i^{-d/2}z_{\text{max}}^{-d/4} & i^{-d/2}z_{\text{max}}^{-d/4} \\
			i^{d/2}z_{\text{max}}^{d/4} & -i^{d/2}z_{\text{max}}^{d/4}%
		\end{array}\notag
		\right) 
	\end{equation}%
	and the self-energy in the \textquotedblleft $\pm $\textquotedblright\
	channels can be obtained as%
	\begin{eqnarray}
		\Sigma_{\pm }(\omega ) &=&\Omega ^{2}\int \frac{dk}{2\pi }\frac{1\pm e^{ikd}%
		}{\omega +i\Gamma +2J_{\mathrm{eff}}\cos k}  \notag \\
		&=&\left\{
		\begin{array}{c}
			-\frac{\Omega ^{2}[1\pm z_{-}^{d}(\omega )]}{\sqrt{(\omega +i\Gamma
					)^{2}-4J_{\mathrm{eff}}^{2}}},\left\vert z_{-}(\omega )\right\vert <1, \\
			\frac{\Omega ^{2}[1\pm z_{+}^{d}(\omega )]}{\sqrt{(\omega +i\Gamma )^{2}-4J_{%
						\mathrm{eff}}^{2}}},\left\vert z_{+}(\omega )\right\vert <1,%
		\end{array}%
		\right.   \notag \\
		z_{\pm }(\omega ) &=&-\frac{\omega +i\Gamma }{2J_{\mathrm{eff}}}\pm \sqrt{%
			\frac{(\omega +i\Gamma )^{2}}{4J_{\mathrm{eff}}^{2}}-1}.\notag
	\end{eqnarray}
	For $2J<\Gamma $, the Green functions have the same forms as Eqs. (\ref{GR2}%
	) and (\ref{GRpm}), where%
	\begin{equation}
		S=\frac{1}{\sqrt{2}}\left(
		\begin{array}{cc}
			z_{\text{max}}^{-d/4} & z_{\text{max}}^{-d/4} \\
			z_{\text{max}}^{d/4} & -z_{\text{max}}^{d/4}%
		\end{array}
		\right) \notag
	\end{equation}%
	and the self-energy is%
	\begin{eqnarray}
		\Sigma_{\pm }(\omega ) &=&\Omega ^{2}\int \frac{dk}{2\pi }\frac{1\pm e^{ikd}%
		}{\omega +i\Gamma +2iJ_{\mathrm{eff}}\cos k}  \notag \\
		&=&\left\{
		\begin{array}{c}
			i\frac{\Omega ^{2}[1\pm z_{-}^{d}(\omega )]}{\sqrt{-(\omega +i\Gamma
					)^{2}-4J_{\mathrm{eff}}^{2}}},\left\vert z_{-}(\omega )\right\vert <1, \\
			-i\frac{\Omega ^{2}[1\pm z_{+}^{d}(\omega )]}{\sqrt{-(\omega +i\Gamma
					)^{2}-4J_{\mathrm{eff}}^{2}}},\left\vert z_{+}(\omega )\right\vert <1,%
		\end{array}%
		\right.   \notag \\
		z_{\pm }(\omega ) &=&i\frac{\omega +i\Gamma }{2J_{\mathrm{eff}}}\pm \sqrt{-%
			\frac{(\omega +i\Gamma )^{2}}{4J_{\mathrm{eff}}^{2}}-1}.\notag
	\end{eqnarray}%
	Eventually, the dynamics of emitters is completely determined by the
	analytic structure of $G_{\pm }(\omega )$ that is obtained analytically.
	
	For $2J>\Gamma $, the Fourier transform of the Green functions%
	\begin{align}
		&G_{\sigma }(\omega )  \notag \\
		=&\sum_{s}\frac{Z_{s}^{\sigma }}{\omega -\varepsilon _{s}^{\sigma }}%
		+\int_{-2J_{\mathrm{eff}}}^{2J_{\mathrm{eff}}}\frac{dx}{2\pi i}\frac{\sqrt{%
				4J_{\mathrm{eff}}^{2}-x^{2}}}{\omega -x+i\Gamma }  \notag \\
		&\times\{\frac{1}{\sqrt{4J_{\mathrm{eff}}^{2}-x^{2}}(x-i\Gamma -\Delta )-i\Omega
			^{2}[1+\sigma z_{-}^{d}(x-i\Gamma )]}  \notag \\
		&-\frac{1}{\sqrt{4J_{\mathrm{eff}}^{2}-x^{2}}(x-i\Gamma -\Delta )+i\Omega
			^{2}[1+\sigma z_{+}^{d}(x-i\Gamma )]}\}\notag
	\end{align}%
	in the $\sigma =\pm $ channels results in%
	\begin{align}
		&G_{\sigma }(t)  \notag \\
		=&-i\sum_{s}Z_{s}^{\sigma }e^{-i\varepsilon _{s}^{\sigma }t}-\int_{-2J_{%
				\mathrm{eff}}}^{2J_{\mathrm{eff}}}\frac{dx}{2\pi }e^{-i(x-i\Gamma )t}\sqrt{%
			4J_{\mathrm{eff}}^{2}-x^{2}}  \notag \\
		&\times\{\frac{1}{\sqrt{4J_{\mathrm{eff}}^{2}-x^{2}}(x-i\Gamma -\Delta )-i\Omega
			^{2}[1+\sigma z_{-}^{d}(x-i\Gamma )]}  \notag \\
		&-\frac{1}{\sqrt{4J_{\mathrm{eff}}^{2}-x^{2}}(x-i\Gamma -\Delta )+i\Omega
			^{2}[1+\sigma z_{+}^{d}(x-i\Gamma )]}\},\notag
	\end{align}%
	where $\varepsilon _{s}^{\sigma }$ and $Z_{s}^{\sigma }$ are the poles and
	the corresponding residues%
	\begin{equation}
		Z_{s}^{\sigma }=\frac{1}{1-\left. \partial _{\omega }\Sigma _{\sigma
			}(\omega )\right\vert _{\omega =\varepsilon _{s}^{\sigma }}}\notag
	\end{equation}%
	of $G_{\sigma }$.
	
	For $2J<\Gamma $, the Fourier transform of the Green functions%
	\begin{align}
		&G_{\sigma }(\omega ) \notag\\
		=&\sum_{s}\frac{Z_{s}^{\sigma }}{\omega -\varepsilon
			_{s}^{\sigma }}-\int_{-2J_{\mathrm{eff}}}^{2J_{\mathrm{eff}}}\frac{dx}{2\pi }%
		\frac{\sqrt{4J_{\mathrm{eff}}^{2}-x^{2}}}{\omega -ix+i\Gamma }  \notag \\
		&\times\{\frac{1}{\sqrt{4J_{\mathrm{eff}}^{2}-x^{2}}(x-\Gamma +i\Delta )+i\Omega
			^{2}[1+\sigma z_{-}^{d}(ix-i\Gamma )]}  \notag \\
		&-\frac{1}{\sqrt{4J_{\mathrm{eff}}^{2}-x^{2}}(x-\Gamma +i\Delta )-i\Omega
			^{2}[1+\sigma z_{+}^{d}(ix-i\Gamma )]}\}\notag
	\end{align}%
	leads to%
	\begin{align}
		&G_{\sigma }(t)  \notag \\
		=&-i\sum_{s}Z_{s}^{\sigma }e^{-i\varepsilon _{s}^{\sigma }t}+\int_{-2J_{%
				\mathrm{eff}}}^{2J_{\mathrm{eff}}}\frac{dx}{2\pi }e^{-(\Gamma -x)t}\sqrt{4J_{%
				\mathrm{eff}}^{2}-x^{2}}  \notag \\
		&\times\{\frac{1}{\sqrt{4J_{\mathrm{eff}}^{2}-x^{2}}(x-\Gamma +i\Delta )-i\Omega
			^{2}[1+\sigma z_{+}^{d}(ix-i\Gamma )]}  \notag \\
		&-\frac{1}{\sqrt{4J_{\mathrm{eff}}^{2}-x^{2}}(x-\Gamma +i\Delta )+i\Omega
			^{2}[1+\sigma z_{-}^{d}(ix-i\Gamma )]}\}.\notag
	\end{align}
	
	\subsection{Two excitations}
	
	In the two-excitation subspace of a single emitter, the dynamics is
	determined by the retarded Green function%
	\begin{eqnarray}
		D(t) &=&-i\frac{1}{2}\left\langle [a^{2}(t),a^{\dagger
			2}(0)]\right\rangle \theta (t)  \notag \\
		&=&-i\frac{1}{2}\left\langle a^{2}(t)a^{\dagger 2}(0)\right\rangle \theta
		(t),\notag
	\end{eqnarray}%
	where the second equation is valid for the initial vacuum state. As shown in
	Fig. \ref{SubFig3}, the Fourier transform $D^{R}(\omega )=\int dte^{i\omega
		t}D(t)$ of $D(t)$\ can be obtained by\ the Dyson expansion of the
	on-site interaction $U$ as%
	\begin{equation}
		D^{R}(\omega )=\frac{1}{\Pi^{-1}(\omega )-U},\notag
	\end{equation}%
	where%
	\begin{equation}
		\Pi(\omega )=i\int \frac{d\omega ^{\prime }}{2\pi }G^{K}(\omega ^{\prime
		})G^{R}(\omega -\omega ^{\prime })\notag
	\end{equation}%
	is the convolution of $G^{K}$ and $G^{R}$. For the general nonequilibrium
	problem, $G^{K}$, $G^{R}$, and $G^{A}$ are three independent Green
	functions, and the fluctuation-dissipation theorem $G^{K}(\omega )=[1\pm
	2n(\omega)][G^{R}(\omega )-G^{A}(\omega )]$ is satisfied only in the
	equilibrium state. In the present case, the nature of the bath in zero
	temperature results in the relation $G^{K}(\omega )=G^{R}(\omega
	)-G^{A}(\omega )$ that can also be checked directly from Eq. (\ref{Gs}). As a
	result, the convolution becomes%
	\begin{eqnarray}
		\Pi(\omega ) &=&i\int \frac{d\omega ^{\prime }}{2\pi }[G^{R}(\omega
		^{\prime })-G^{A}(\omega ^{\prime })]G^{R}(\omega -\omega ^{\prime })  \notag
		\\
		&=&i\int \frac{d\omega ^{\prime }}{2\pi }G^{R}(\omega ^{\prime
		})G^{R}(\omega -\omega ^{\prime }),  \label{Do}
	\end{eqnarray}%
	where in the second equation the causality condition is used.
	
	To perform the Fourier transform $D(t)=\int (d\omega/2\pi)%
	D^{R}(\omega )e^{-i\omega t}$ efficiently, we introduce the analytic
	continuation%
	\begin{equation}
		D_f(\omega )=\frac{1}{\Pi_f^{-1}(\omega )-U}  \label{eq:ficD}
	\end{equation}%
	of $D^{R}(\omega )$, where%
	\begin{equation}
		\Pi_f(\omega )=i\int \frac{d\omega ^{\prime }}{2\pi }G_f(\omega ^{\prime })G_f(\omega -\omega ^{\prime }).\notag
	\end{equation}%
	Since in the integral contour of the Fourier transform $D^{R}(\omega )=D_f(\omega )$, the Fourier transform $D(t)=\int (d\omega/2\pi)%
	D_f(\omega )e^{-i\omega t}$. Compared with the original convolution (%
	\ref{Do}), $\Pi_f(\omega )$ has the simple structure of branch cuts.
	With knowing the analytic structure of $G_f$, i.e., as shown in the
	first subsection the simple behaviors in the vicinity of poles and branch
	cuts, we can calculate $\Pi_f(\omega )$ and $D^{R}(\omega )$ as well as
	the Fourier transform efficiently. 
	
	\bigskip
	
\section{Steady-state correlation functions of the driven emitter}\label{sec:steady}
	
In this section, we present the approach to systematically calculate the quantum correlation functions of the weakly driven emitter in the steady state of the master equation (\ref{eq:rho}). First, in Appendix~\ref{undriven}, we consider an undriven emitter and show that the spontaneous emission of $n$ excitations is fully determined by the retarded $n$-particle Green function (of the emitter) in the vacuum state. Then, in Appendix~\ref{driven}, we take into account the weak driving field in the master equation. Following Refs.~\cite{Shitao2015,Yue2016}, we develop a perturbative solution and connect the physical observables of the driven emitter with Green functions of the undriven case. 

\subsection{Master equation without driving}\label{undriven}

Our starting point is the master equation (\ref{eq:rho}) without the drive ($\epsilon=0$), i.e., 
\begin{eqnarray}
\dot{\rho} &=&\mathcal{L}_{0}\rho \nonumber\\
&=&-i[H_0,\rho]-\frac{\Gamma}{2}\sum_j\left(\{O_j^{\dagger }O_j,\rho \}+2 O_j\rho O_j^{\dagger }\right). \label{M}
\end{eqnarray}
Here, $H_0=H_\textrm{emit}+H_\textrm{sb}+H_{b}$, where the Hamiltonian for an undriven emitter is $H_\textrm{emit}=\Delta a^\dag a+(U/2)a^\dag a^\dag a a$, the emitter-bath coupling $H_\textrm{sb}=\Omega a^\dag b_0+\textrm{H.c.}$, and $H_{b}=\sum_j Je^{i\theta} b_j^\dag b_{j+1}+\textrm{H.c}$. The jump operator in Eq.~(\ref{M}) is $O_j=b_j+b_{j+1}$. We stress that the effective Hamiltonian 
\begin{equation}
H_\textrm{eff}=H_0-i\frac{\Gamma}{2} \sum_jO_j^\dag O_j \label{Heff}
\end{equation} 
commutes with the total particle number $N=a^\dag a+\sum_j b_j^\dag b_j$:
\begin{equation}
[H_\textrm{eff},N]=0. \label{HN}
\end{equation}
Thus, the number of excitations is preserved in the nonunitary time evolution driven by $H_\textrm{eff}$. Consequently, the steady state of Eq.~(\ref{M}) at zero temperature is the vacuum state $|0\rangle$.

We now show that the dynamics for the initial pure state $\rho
(0)=\left\vert \psi (0)\right\rangle \left\langle \psi (0)\right\vert $ with
finite $n$ excitations can be efficiently studied using $H_{\mathrm{eff}}$. Exploiting the property (\ref{HN}), the Dyson expansion gives rise to the formal solution%
\begin{widetext}
\begin{eqnarray}
\rho (t) 
&=&e^{-iH_{\mathrm{eff}}t}\rho (0)e^{iH_{\mathrm{eff}%
}^\dag t}+\Gamma\sum_{j_1}\int_{0}^{t}ds_{1}e^{-iH_{\mathrm{eff}}(t-s_{1})}O_{j_1}e^{-iH_{\mathrm{eff}%
}s_{1}}\rho (0)e^{iH_{\mathrm{eff}}^\dag s_{1}}O_{j_1}^{\dagger }e^{iH_{\mathrm{eff}%
}^\dag (t-s_{1})}  \notag \\
&&+\Gamma^2\sum_{j_1,j_2}\int_{0}^{t}ds_{1}\int_{0}^{s_{1}}ds_{2}e^{-iH_{\mathrm{eff}%
}(t-s_{1})}O_{j_1}e^{-iH_{\mathrm{eff}}(s_{1}-s_{2})}O_{j_2}e^{-iH_{\mathrm{eff}%
}s_{2}}\rho (0)e^{iH_{\mathrm{eff}}^\dag s_{2}}O_{j_2}^{\dagger }e^{iH_{\mathrm{eff}%
}^\dag (s_{1}-s_{2})}O_{j_1}^{\dagger }e^{iH_{\mathrm{eff}}^\dag (t-s_{1})}+\cdots \nonumber \\
&=&\left\vert \psi ^{(0)}(t)\right\rangle \left\langle \psi
^{(0)}(t)\right\vert +\Gamma\sum_{j_1}\int_{0}^{t}ds_{1}\left\vert \psi_{j_1}
^{(1)}(t,s_{1})\right\rangle \left\langle \psi_{j_1} ^{(1)}(t,s_{1})\right\vert \nonumber\\
&+&\Gamma^2\sum_{j_1,j_2}\int_{0}^{t}ds_{1}\int_{0}^{s_{1}}ds_{2}\left\vert \psi_{j_1,j_2}
^{(2)}(t,s_{1},s_{2})\right\rangle \left\langle \psi_{j_1,j_2}
^{(2)}(t,s_{1},s_{2})\right\vert +\cdots . \label{fs} 
\end{eqnarray}%
\end{widetext}
Here the density matrix $\rho (t)$ is
an incoherent superposition of states $\left\vert \psi
^{(0)}(t)\right\rangle $, $\left\vert \psi_{j_1} ^{(1)}(t,s_{1})\right\rangle $,
etc., where each of them is governed by the effective Hamiltonian $H_{%
\mathrm{eff}}$ in the corresponding subspace, i.e., 
\begin{widetext}
\begin{eqnarray}
\left\vert \psi ^{(0)}(t)\right\rangle &=&e^{-iH_{\mathrm{eff}}t}\left\vert
\psi (0)\right\rangle ,  \notag \\
\left\vert \psi_{j_1} ^{(1)}(t,s_{1})\right\rangle &=&e^{-iH_{\mathrm{eff}%
}(t-s_{1})}O_{j_1}e^{-iH_{\mathrm{eff}}s_{1}}\left\vert \psi (0)\right\rangle ,
\notag \\
\left\vert \psi_{j_1,j_2} ^{(2)}(t,s_{1},s_{2})\right\rangle &=&e^{-iH_{\mathrm{eff}%
}(t-s_{1})}O_{j_1}e^{-iH_{\mathrm{eff}}(s_{1}-s_{2})}O_{j_2}e^{-iH_{\mathrm{eff}%
}s_{2}}\left\vert \psi (0)\right\rangle. \label{psi}
\end{eqnarray}
\end{widetext}
Since the jump operator always depletes excitations from the system, the
series in Eq. (\ref{fs}) is automatically truncated after acting the jump
operator $n+1$ times. As a result, we have to study only the dynamics of the
initial pure state $\left\vert \psi (0)\right\rangle $ governed by the
effective Hamiltonian $H_{\mathrm{eff}}$, where the transition between the
subspaces with different excitation numbers is described by the jump operator.

It follows from Eqs.~(\ref{fs}) and (\ref{psi}) that the spontaneous
emission probability of $n$ excitations in the undriven emitter is given by 
\begin{equation}
P_{n}=tr[\left\vert n\right\rangle \left\langle n\right\vert \rho
(t)]=\left\vert \left\langle n\right\vert e^{-iH_{\mathrm{eff}}t}\left\vert
n\right\rangle \right\vert ^{2}.
\end{equation}%
This is just the norm square of the $n$-particle Green
function $G_{n}^{R}$ \textit{in the vacuum state}, which is completely determined by $H_{\mathrm{eff}}$. 

In summary, we can solve the full dynamics of the emitter first in subspaces with
different excitation numbers governed by the effective Hamiltonian $H_{%
\mathrm{eff}}$ and then connect the results using the jump operator.
Interestingly, for the trivial steady state $\left\vert 0\right\rangle $, since $
H_{\mathrm{eff}}|0\rangle=0$, the dynamics in the subspace with $n$ excitations is described by 
\begin{eqnarray}
G_{n}^{R}(\omega ) &=&-i\frac{1}{n!}\int dte^{i\omega t}\left\langle
0\right\vert [a^{n}(t),a^{\dagger n}(0)]\left\vert 0\right\rangle \theta (t)
\notag \\
&=&-i\frac{1}{n!}\int_{0}^{\infty }dte^{i\omega t}\left\langle 0\right\vert
a^{n}(t)a^{\dagger n}(0)\left\vert 0\right\rangle \theta (t). \label{Gnt}
\end{eqnarray}%
Thus, the fluctuation-dissipation theorem holds in the subspace for Green functions of the emitter in the vacuum state, even though the system 
state is out of equilibrium. 

\bigskip

\subsection{Weakly driven system}\label{driven}

With the above results, we now consider the case when the emitter is driven by an external field with the driving frequency $\omega_d$ and the driving strength $\epsilon\neq 0$. In the rotating framework with respect to $\omega_d$, the master equation reads
\begin{eqnarray}
\dot{\rho}  &=&\mathcal{L}_{0}\rho +\mathcal{L}_{\varepsilon }(t)\rho \label{Md1}
\end{eqnarray}
where $\mathcal{L}_{0}$ is the Liouvillian for the undriven system and $\mathcal{L}_{\varepsilon }$ describes the driving field with the strength $\epsilon$:
\begin{equation}
\mathcal{L}_{\varepsilon }(t)\rho =-i[\varepsilon (A^{\dagger }+A),\rho ], \label{Le}
\end{equation}
with the operator $A=a e^{i\omega_d t}$ in our case. 

In general, one has to solve the master equation (\ref{Md1}) numerically to
study the full dynamics. However, there are also special cases which allow
us to obtain the steady state and time evolution analytically. It turns out that, for the weak driving strength $\varepsilon $ much smaller
than the spectral gap of the Liouvillian $\mathcal{L}$ without the driving
term and the steady state of $\mathcal{L}$ is not degenerate, the dynamics
can also be studied using the Green functions $G_{n}^{R}$ in subspaces with
different excitations. This statement has been proven in Ref.~\cite{Shitao2015} and applied to study photon pair generation in Ref.~\cite{Shitao2016}. In the following, we use it for our case. 

Specifically, we can expand $\rho
(t)$ to some order of $\varepsilon $ for corresponding problems. For our purpose of calculating the second-order correlation function, we expand $\rho(t)$ up to the fourth order of $\varepsilon$ (the $n$th-order term is denoted by $\rho_n$, $n=0,1,...,4$). For the vacuum steady state $\rho _{0}=\left\vert 0\right\rangle \left\langle
0\right\vert $ of the undriven system, we obtain
 \begin{widetext}
\begin{eqnarray}
\rho (t) &=&\mathcal{T}e^{\int_0^t [\mathcal{L}_0+\mathcal{L}_\varepsilon(t_1)]dt_1}\rho_0\nonumber\\
&=&\rho _{0}+\rho _{1}(t)+\rho _{2}(t)+\rho _{3}(t)+\rho _{4}(t)\nonumber\\
&=&e^{\mathcal{L}_{0}t}\rho _{0}+\int_{0}^{t}ds_1 e^{\mathcal{L}%
_{0}(t-s_1)}\mathcal{L}_{\varepsilon }e^{\mathcal{L}_{0}s_1}\rho _{0} +\int_{0}^{t}ds_{1}\int_{0}^{s_{1}}ds_{2}e^{\mathcal{L}_{0}(t-s_{1})}%
\mathcal{L}_{\varepsilon }e^{\mathcal{L}_{0}(s_{1}-s_{2})}\mathcal{L}%
_{\varepsilon }e^{\mathcal{L}_{0}s_{2}}\rho _{0}  \notag \\
&&+\int_{0}^{t}ds_{1}\int_{0}^{s_{1}}ds_{2}\int_{0}^{s_{2}}ds_{3}e^{\mathcal{%
L}_{0}(t-s_{1})}\mathcal{L}_{\varepsilon }e^{\mathcal{L}_{0}(s_{1}-s_{2})}%
\mathcal{L}_{\varepsilon }e^{\mathcal{L}_{0}(s_{2}-s_{3})}\mathcal{L}%
_{\varepsilon }e^{\mathcal{L}_{0}s_{3}}\rho _{0}  \notag \\
&&+\int_{0}^{t}ds_{1}\int_{0}^{s_{1}}ds_{2}\int_{0}^{s_{2}}ds_{3}%
\int_{0}^{s_{3}}ds_{4}e^{\mathcal{L}_{0}(t-s_{1})}\mathcal{L}_{\varepsilon
}e^{\mathcal{L}_{0}(s_{1}-s_{2})}\mathcal{L}_{\varepsilon }e^{\mathcal{L}%
_{0}(s_{2}-s_{3})}\mathcal{L}_{\varepsilon }e^{\mathcal{L}_{0}(s_{3}-s_{4})}%
\mathcal{L}_{\varepsilon }e^{\mathcal{L}_{0}s_{4}}\rho _{0}.\label{rhoe}
\end{eqnarray}
\end{widetext}

By using Eq.~(\ref{fs}), and keeping in mind that the undriven effective Hamiltonian $H_\textrm{eff}$ [see Eq.~(\ref{Heff})] is number conserving, we calculate Eq.~(\ref{rhoe}) as%
\\
 \begin{widetext}
\begin{eqnarray}
\rho (t) 
&=&\rho _{0}-i\varepsilon \int_{0}^{t}ds_1\Big[e^{-iH_{\mathrm{eff}%
}(t-s_1)}A^{\dagger }\rho _{0}-\rho _{0}Ae^{iH_{\mathrm{eff}}^{\dagger }(t-s_1)}\Big]\nonumber
\\
&&-\varepsilon ^{2}\int_{0}^{t}ds_{1}\int_{0}^{s_{1}}ds_{2}\Big[e^{-iH_{\mathrm{%
eff}}(t-s_{1})}A^{\dagger }e^{-iH_{\mathrm{eff}}(s_{1}-s_{2})}A^{\dagger
}\rho _{0}+\rho _{0}Ae^{iH_{\mathrm{eff}}^{\dagger }(s_{1}-s_{2})}Ae^{iH_{%
\mathrm{eff}}^{\dagger }(t-s_{1})}\Big] \nonumber\\
&&+\varepsilon ^{2}\Gamma \sum_{j_1}
\int_{0}^{t}ds_{1}\int_{0}^{s_{1}}ds_{2}\int_{0}^{t-s_{1}}ds_{3}e^{-iH_{%
\mathrm{eff}}(t-s_{1}-s_{3})}O_{j_1}e^{-iH_{\mathrm{eff}}s_{3}}\Big[e^{-iH_{\mathrm{eff%
}}(s_{1}-s_{2})}A^{\dagger }\rho _{0}A \nonumber\\
&&+A^{\dagger }\rho _{0}Ae^{iH_{\mathrm{eff}}^{\dagger
}(s_{1}-s_{2})}\Big]e^{iH_{\mathrm{eff}}^{\dagger }s_{3}}O_{j_1}^{\dagger }e^{iH_{%
\mathrm{eff}}^{\dagger }(t-s_{1}-s_{3})} \nonumber\\
&&+\varepsilon ^{2}\int_{0}^{t}ds_{1}\int_{0}^{s_{1}}ds_{2}\Big[e^{-iH_{\mathrm{%
eff}}(t-s_{2})}A^{\dagger }\rho _{0}Ae^{iH_{\mathrm{eff}}^{\dagger
}(t-s_{1})}+e^{-iH_{\mathrm{eff}}(t-s_{1})}A^{\dagger }\rho _{0}Ae^{iH_{%
\mathrm{eff}}^{\dagger }(t-s_{2})}\Big] \nonumber\\
&&-\varepsilon ^{2}\int_{0}^{t}ds_{1}\int_{0}^{s_{1}}ds_{2}\Big[\rho _{0}Ae^{iH_{%
\mathrm{eff}}^{\dagger }(s_{1}-s_{2})}A^{\dagger }+Ae^{-iH_{\mathrm{eff}%
}(s_{1}-s_{2})}A^{\dagger }\rho _{0}\Big] +\rho _{3}(t)+\rho _{4}(t), \label{rho4}
\end{eqnarray}
\end{widetext}
where the third-order and fourth-order terms are useful for the excitation
statistics. Finally, the steady state $\rho_\textrm{ss}$ is obtained as taking $t\rightarrow \infty $.

Note that one can benchmark the above derivations using a two-level system with $A=\sigma^{-}$, $H_{0}=0$, and the jump operator $\sigma^{-}$. In this case, following similar steps, 
the steady-state density matrix up to $\varepsilon^2 $ is derived as%
\begin{equation}
\rho _{\mathrm{ss}}\sim (1-\frac{4\varepsilon ^{2}}{\gamma ^{2}})\left\vert
g\right\rangle \left\langle g\right\vert -i\frac{2\varepsilon }{\gamma }%
(\left\vert e\right\rangle \left\langle g\right\vert -\left\vert
g\right\rangle \left\langle e\right\vert )+\frac{4\varepsilon ^{2}}{\gamma
^{2}}\left\vert e\right\rangle \left\langle e\right\vert, \nonumber
\end{equation}%
which is exactly the steady-state solution%
\begin{equation}
\!\!\!\!\!\rho _{\mathrm{ss}}=\frac{4\varepsilon ^{2}+\gamma ^{2}}{8\varepsilon
^{2}+\gamma ^{2}}\left\vert g\right\rangle \left\langle g\right\vert -\frac{%
2i\varepsilon \gamma }{8\varepsilon ^{2}+\gamma ^{2}}(\left\vert
e\right\rangle \left\langle g\right\vert -\left\vert g\right\rangle
\left\langle e\right\vert )+\frac{4\varepsilon ^{2}}{8\varepsilon
^{2}+\gamma ^{2}}\left\vert e\right\rangle \left\langle e\right\vert \nonumber
\end{equation}%
of the Liouvillian $\mathcal{L}_{0}+\mathcal{L}_{\varepsilon }$ up to the order $\varepsilon^2$.

Now we are ready to calculate for our case the first- and second-order correlation functions of the driven emitter in the steady state. Keeping the second-order terms in Eq.~(\ref{rho4}), we obtain%
\begin{equation}
\left\langle a^{\dagger }a\right\rangle _{\mathrm{ss}}=\varepsilon ^{2}\left\vert \left\langle 0\right\vert a\frac{1}{\omega_d-H_{\mathrm{%
eff}}}a^{\dagger }\left\vert 0\right\rangle \right\vert ^{2}. \label{G1}
\end{equation}%
Keeping the fourth-order terms in Eq.~(\ref{rho4}), we obtain
\begin{widetext}
\begin{eqnarray}
&&\left\langle a^{\dagger }a^{\dagger }(\tau )a(\tau
)a\right\rangle _{\mathrm{ss}}  \notag \\
&=&\textrm{Tr}\Big[ae^{\mathcal{L}\tau }(a\rho _{\mathrm{ss}}a^{\dagger })a^{\dagger }\Big]
\notag \\
&=&\textrm{Tr}\Big[ae^{\mathcal{L}_{0}\tau }\int_{0}^{\tau }ds_{1}\int_{0}^{s_{1}}ds_{2}%
\mathcal{L}_{\varepsilon }(s_{1})\mathcal{L}_{\varepsilon }(s_{2})(a\rho _{2}a^{\dagger })a^{\dagger }]+\textrm{Tr}\Big[ae^{\mathcal{L}_{0}\tau }\int_{0}^{\tau }ds_{1}
\mathcal{L}_{\varepsilon }(s_{1})(a\rho _{3}a^{\dagger })a^{\dagger }\Big]+\textrm{Tr}\Big[ae^{\mathcal{L}_{0}\tau }(a\rho _{\mathrm{4}}a^{\dagger })a^{\dagger }\Big]
\notag \\
&=&\varepsilon ^{4}\left\vert \left\langle 0\right\vert a\frac{1-e^{i(\omega_{d}-H_{%
\mathrm{eff}})\tau }}{\omega_{d}-H_{\mathrm{eff}}}a^{\dagger }\left\vert 0\right\rangle
\left\langle 0\right\vert a\frac{1}{\omega_{d}-H_{\mathrm{eff}}}a^{\dagger }\left\vert
0\right\rangle +\left\langle 0\right\vert ae^{i(\omega_d-H_{\mathrm{eff}})\tau }a\frac{%
1}{2\omega_{d}-H_{\mathrm{eff}}}a^{\dagger }\frac{1}{\omega_{d}-H_{\mathrm{eff}}}a^{\dagger
}\left\vert 0\right\rangle \right\vert ^{2},\label{G2}
\end{eqnarray}
\end{widetext}
where the quantum regression theorem is used in the second row.  

The correlation functions (\ref{G1}) and (\ref{G2}) can be written in terms of Green functions on the vacuum state. In the compact form, we have
\begin{eqnarray}
\!\!\!\!\!\!\!\!\!\!\!\!\!\!\!\frac{\left\langle a^{\dagger }a\right\rangle _{\mathrm{ss}}}{\varepsilon ^{2}} &=&\left\vert \int_{-\infty}^\infty dt e^{-i\omega _{d%
		}t}\left\langle 0\right\vert \mathcal{T}
a(0)a^{\dagger }(t)\left\vert 0\right\rangle \right\vert ^{2}\nonumber\\
\!\!\!\!\!\!\!\!\!\!\!\!\!\!\!\frac{\left\langle a^{\dagger }a^{\dagger }(\tau )a(\tau)a\right\rangle _{\mathrm{ss}}}{\varepsilon ^{4}}&=&\left|\int_{-\infty }^{+\infty }dt_{1}dt_{2}e^{-i\omega _{d%
		}(t_{1}+t_{2})}G(\tau ;t_{1},t_{2})\right|^2,\label{G12}\notag \\
\end{eqnarray}
with $G(\tau; t_{1},t_{2})=-i\left\langle 0\right\vert \mathcal{T%
	}a(\tau )a(0)a^{\dagger }(t_{1})a^{\dagger
	}(t_{2})\left\vert 0\right\rangle /2$,
where $a(t)=e^{iH_{\mathrm{eff}}^\dag t}ae^{-iH_{\mathrm{eff}}t}$ is governed by
the undriven effective Hamiltonian and the Green functions are all defined on the
vacuum state. 

From the above analysis, we show explicitly that, even though
our system is weakly driven, the relevant physical observables of the system can be calculated from
Green functions of the undriven system in the vacuum state, which is governed by $H_{\mathrm{eff}}$. In the explicit form, the normalized second-order correlation function is found as
\begin{widetext}
\begin{equation}
g^{(2)}(\tau )=\frac{\left\vert \left\langle 0\right\vert a\frac{1-e^{i(\omega_{d}-H_{%
\mathrm{eff}})\tau }}{\omega_{d}-H_{\mathrm{eff}}}a^{\dagger }\left\vert
0\right\rangle \left\langle 0\right\vert a\frac{1}{\omega_{d}-H_{\mathrm{eff}}}%
a^{\dagger }\left\vert 0\right\rangle +\left\langle 0\right\vert ae^{i(\omega_{d}-H_{%
\mathrm{eff}})\tau }a\frac{1}{2\omega_{d}-H_{\mathrm{eff}}}a^{\dagger }\frac{1%
}{\omega_d-H_{\mathrm{eff}}}a^{\dagger }\left\vert 0\right\rangle \right\vert
^{2}}{\left\vert \left\langle 0\right\vert a\frac{1}{\omega_{d}-H_{\mathrm{eff}}}%
a^{\dagger }\left\vert 0\right\rangle \right\vert ^{4}}.
\end{equation}
\end{widetext}
It is completely determined by the spectral properties of the effective Hamiltonian $H_{%
\mathrm{eff}}$ in the single- or two-
excitation subspaces, respectively. Therefore, once the
eigenproblem of $H_{\mathrm{eff}}$ in the undriven case is solved, we can obtain $g^{(2)}(\tau )$.

Based on Eq.~(\ref{G12}), we can now apply the Dyson expansion as illustrated in Fig.~\ref{SubFig3} to calculate the right side and obtain
\begin{equation}
g^{(2)}(\tau )=\left\vert 1+\bigg[i\int \frac{d\omega ^{\prime }}{2\pi }%
G^{R}(\omega ^{\prime })G^{R}(-\omega ^{\prime })e^{-i\omega ^{\prime }\tau
}\bigg]T(0)\right\vert ^{2}.
\end{equation}%
Here, the poles and branch cuts of%
\begin{equation}
T(\omega )=\frac{1}{U^{-1}-i\int \frac{d\omega ^{\prime }}{2\pi }%
G^{R}(\omega ^{\prime })G^{R}(\omega -\omega ^{\prime })}\nonumber\\
\end{equation}%
correspond to the quasibound state and continuum of $H_{\mathrm{eff}}$ in the two-excitation subspace. This eventually leads to Eq.~(\ref{g2}) in our text. 

\bigskip
	
	\section{Scaling behaviors in quantum Zeno regimes}
	\label{sec:scaling}
	
	In this section, we study the scaling behaviors in the quantum Zeno regime.
	In parallel with Appendix~\ref{sec:G}, we investigate the analytic structure of $G_f(\omega )$ and $G_{\sigma }(\omega )$ for the large $\Gamma $ in Appendixes~\ref{sec:detailone} and \ref{sec:detailtwo}. The approximate analytical expressions of Green functions $%
	G^{R}(\omega )$ in the time domain are achieved, which agree with the exact
	results quantitatively. In Appendix~\ref{sec:detail2ex}, the effective Hamiltonian for
	two excitations is derived via the perturbation theory, which gives rise to
	the scaling behavior in a good agreement with the exact solution.
	
		\begin{figure}[tb]
		\centering
		\includegraphics[width=0.85\columnwidth]{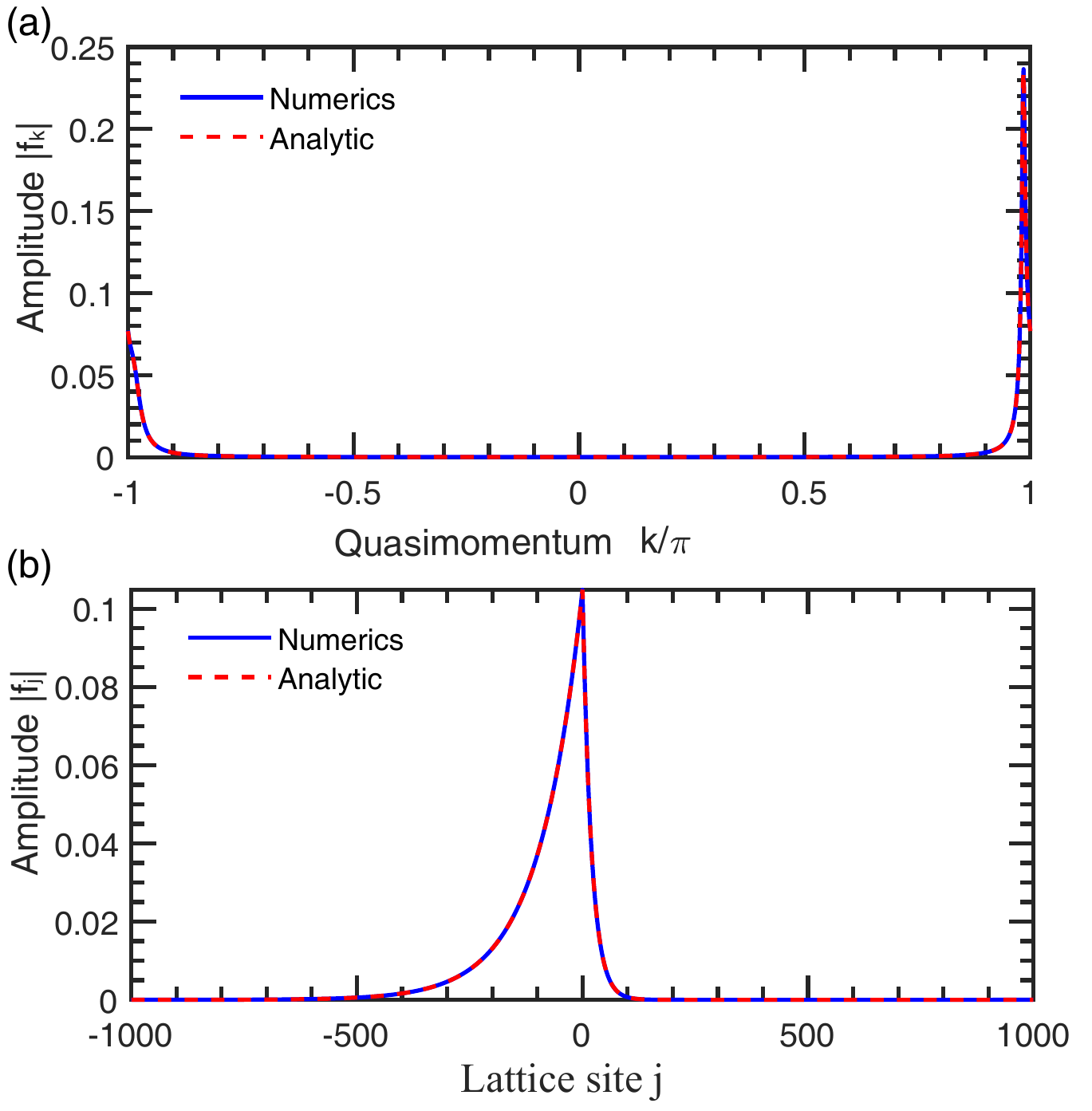}
		\caption{Comparison between the analytic and numerical results for the bath component of the single-emitter quasibound state in (a) momentum space and (b) real space representation. In both (a) and (b), the numerical results (blue solid curves) are obtained from diagonalizing the effective Hamiltonian $H_\textrm{eff}$ with the open-bath size $N_b=2000$, the emitter detuning $\Delta=0$, bath dissipation strength $\Gamma/J=100$, and the coupling strength $\Omega/J=1$. The analytical results (red dashed curves) in (a) are obtained from Eq.~(\ref{fk}) and in (b) are obtained from Eq.~(\ref{fj}). }
		\label{Fig:wavefunction}
	\end{figure}

	\subsection{Single emitter}\label{sec:detailone}
	
	In the large $\Gamma $ limit, the leading order in the Taylor expansion of $%
	G_f^{-1}(\omega )$ leads to two poles $\epsilon _{s=\pm }=i(\Omega
	^{4}/2)^{1/3}e^{i(2\pi s/3)}\Gamma ^{-1/3}$ for the resonant
	case $\Delta =0$ and two poles%
	\begin{eqnarray}
		\Sigma_f(\omega ) &=&\frac{i\Omega ^{2}\textrm{sgn}[1-\left\vert
			z_{-}(\omega )\right\vert ]}{\sqrt{-(\omega +i\Gamma )^{2}-4J_{\mathrm{eff}%
				}^{2}}}, \notag \\
		z_{-}(\omega ) &=&i\frac{\omega +i\Gamma }{2J_{\mathrm{eff}}}-\sqrt{-\frac{%
				(\omega +i\Gamma )^{2}}{4J_{\mathrm{eff}}^{2}}-1} \notag \\
		\omega &=&i(\frac{\Omega ^{4}}{2\Gamma })^{1/3}e^{i(2\pi/3)s}\notag \\
		\textrm{sgn}[1-\left\vert z_{-}(\omega )\right\vert ] &=&-1 \notag
	\end{eqnarray}%
	\begin{eqnarray}
		\varepsilon _{1} &=&\Delta -i\frac{\Omega ^{2}}{\sqrt{2\Delta }}e^{i(\pi/4)}\Gamma ^{-1/2},  \notag \\
		\varepsilon _{2} &=&-i(\frac{\Omega ^{4}}{2\Delta ^{2}}+2J^{2})\frac{1}{%
			\Gamma }\notag
	\end{eqnarray}%
	for the finite $\Delta$.
	
	 We can also analytically obtain the wave function of the quasibound states with the complex energy $\epsilon_s$. Without loss of generality, here we focus on the case $\Delta=0$. In the momentum space representation, the quasibounds state is $|B\rangle=c_1 a_1^\dag|0\rangle+\sum_k f_k b_k^\dag|0\rangle$, where $c_1$ and $f_k$ are coefficients. The $c_1$ can be obtained via the residue $Z_s$ of the Green function at $\omega=\epsilon_s$ [see Eq.~(\ref{Zs})], giving
\begin{equation}
c_1=Z_s=\frac{\epsilon_s^2+2i\Gamma\epsilon_s-4J^2}{2\epsilon_s^2+3i\Gamma\epsilon_s-4J^2}. \label{c1}
\end{equation}
In the limit $\Gamma/(2J)\rightarrow \infty$, $c_1\rightarrow 2/3$. The $f_k$ can be calculated using the self-energy matrix element in Eq.~(\ref{sigllp}) with $\omega=\epsilon_s$, giving 
\begin{equation}
f_k=\frac{\Omega c_1}{\sqrt{N_b}}\frac{1}{\epsilon_s-(\epsilon_k-i\gamma_k)},\label{fk}
\end{equation}
where in our case $\epsilon_k=2J\sin(k)$ and $\gamma_k=\Gamma[1+\cos(k)]$. In Fig.~\ref{Fig:wavefunction}(a), we show the numerical result (solid curve) of $|f_k|$ by diagonalizing the effective Hamiltonian $H_\textrm{eff}$ and compare it with Eq.~(\ref{fk}). A good agreement is found.

In real space, the quasibound state is $|B\rangle=c_1 a_1^\dag|0\rangle+\sum_j f_j b_j^\dag|0\rangle$, where $f_j$ is the amplitude of the bath component at lattice site $j$. Noting that the emitter is locally coupled to $b_0$, the $f_j$ can be obtained from the self-energy matrix element (\ref{sigllp}) with $\omega=\epsilon_s$, which yields 
 \begin{equation}
 f_j=\frac{\Sigma_{j0}(\epsilon_s)}{\Omega}c_1, \label{fj}
 \end{equation}
where for $\Gamma/(2J)>1$ we have 
 \begin{eqnarray}
		\Sigma_{j0}(\omega ) &=&\left\{
		\begin{array}{c}
			-i\frac{\Omega^2 e^{j\ln\beta_{+}}}{\sqrt{-(\omega +i\Gamma
					)^{2}-4J_{\mathrm{eff}}^{2}}},\hspace{2mm} (j>0) \\
			-i\frac{\Omega^2 e^{-j\ln\beta_{-}}}{\sqrt{-(\omega +i\Gamma
					)^{2}-4J_{\mathrm{eff}}^{2}}},\hspace{2mm} (j<0)
		\end{array}\right. \nonumber\\
		\beta_\pm(\omega)&=&\frac{i(\omega+i\Gamma)+\sqrt{-(\omega+i\Gamma)^2-4J_\textrm{eff}^2}}{\Gamma\pm 2J} \label{sigj0}
\end{eqnarray}
with $J_\textrm{eff}=\sqrt{\Gamma^2/4-J^2}$. In Fig.~\ref{Fig:wavefunction}(b), we numerically calculate $|f_j|$ by diagonalizing $H_\textrm{eff}$ in real space. We see that the numerical results (solid curve) agree very well with that obtained from Eq.~(\ref{fj}). Note that, in general, $\beta_+\neq \beta_-$, and, therefore, the quasibound state has a different localization length on the left and right sides of the emitter.

In the limit $\Gamma/(2J)\rightarrow \infty$, where $\beta_+\approx \beta_-$, one can obtain a simplified expression for the average momentum $k_b$ and the average localization length $l_b$ of the bath component. In this limit, we find $\ln\beta_+\approx  i\pi-\sqrt{-2i\epsilon_s/\Gamma}$. Using, e.g., $\epsilon_s=\frac{1}{2}( \sqrt{3}-i)(\Omega^4/2)^{1/3}\Gamma^{-1/3}$, we obtain $
e^{j\ln\beta_+}=e^{-ijk_b-j/l_b}$, where 
\begin{eqnarray}
k_b&=&\pi-\frac{\sqrt{3}}{2^{2/3}}\left(\frac{\Omega}{\Gamma}\right)^{2/3}, \label{kb}\\
l_{b}&=&{2^{2/3}}\left(\frac{\Gamma}{\Omega}\right)^{2/3}. \label{lb}
\end{eqnarray}
Thus, when $\Gamma/(2J)\rightarrow \infty$, $k_b\rightarrow \pi$, and $l_b$ increases as $(\Gamma/\Omega)^{2/3}$. In Fig.~\ref{Fig:wavefunction}(a), the momentum at which the peak occurs is well described by Eq.~(\ref{kb}).

	\subsection{Two emitters}\label{sec:detailtwo}
	
	For two emitters, we solve the equations $1/G_{\pm }(\varepsilon )=0$ to
	obtain the poles for the resonant case $\Delta =0$. Here, we focus on only
	the even distance $d$, since for the odd distance it turns out that the
	physics is the same by interchanging the $+$ and $-$ channels. In the even
	channel, the leading term in the Taylor expansion of $1/G_{+}(\varepsilon )$
	results in the same scaling behavior $\varepsilon _{s=\pm }=i(2\Omega
	^{4})^{1/3}e^{i(2\pi s/3)}\Gamma ^{-1/3}$ with the prefactor
	enhanced by a factor of $4^{1/3}$. In the odd channel, two poles are%
	\begin{equation}
		\varepsilon _{s}=sR-i\frac{\Omega ^{2}d}{\Gamma },\notag
	\end{equation}%
	where the positive $R\sim \Gamma ^{-2}$ for $\Omega >\sqrt{2/d}J$. For the
	small $\Omega <\sqrt{2/d}J$, two poles merge into a single pole $\varepsilon
	=-i\Omega ^{2}d/\Gamma $.
	
	In the $+$ channel, the residues $Z_{s}\sim 2/3$ and the contribution from
	the branch cut is negligible, which leads to%
	\begin{equation}
		G_{+}(t)\sim -i\frac{4}{3}e^{-(\Omega ^{4}/4\Gamma )^{1/3}t}\cos \bigg[%
		\sqrt{3}\bigg(\frac{\Omega ^{4}}{4\Gamma }\bigg)^{1/3}t\bigg].\notag
	\end{equation}%
	In the $-$ channel, the residues $Z_{s}\sim 1$ and the contribution from the
	branch cut cancels that from one residue for $\Omega >\sqrt{2/d}J$. For the
	small $\Omega <\sqrt{2/d}J$, the contribution from the branch cut is
	negligible, and only a single pole with residue $Z_{s}\sim 1$ survives. As a
	result, the Green function reads%
	\begin{equation}
		G_{-}(t)\sim -ie^{-(\Omega ^{2}d/\Gamma)t}.\notag
	\end{equation}%
	The transformation $S$ leads to%
	\begin{align}
		iG_{11}^{R}(t) =&\frac{2}{3}e^{-(\Omega ^{4}/4\Gamma )^{1/3}t}\cos \bigg[%
		\sqrt{3}\bigg(\frac{\Omega ^{4}}{4\Gamma }\bigg)^{1/3}t\bigg]+\frac{1}{2}e^{-(\Omega^{2}d/\Gamma )t},  \notag \\
		iG_{12}^{R}(t) =&\frac{2}{3}e^{-(\Omega ^{4}/4\Gamma )^{1/3}t}\cos \bigg[%
		\sqrt{3}\bigg(\frac{\Omega ^{4}}{4\Gamma }\bigg)^{1/3}t\bigg]-\frac{1}{2}e^{-(\Omega^{2}d/\Gamma )t}.\notag
	\end{align}
	
	\subsection{Effective models of two excitations}\label{sec:detail2ex}
	
	\begin{figure}[tb]
		\centering
		\includegraphics[width=0.95\columnwidth]{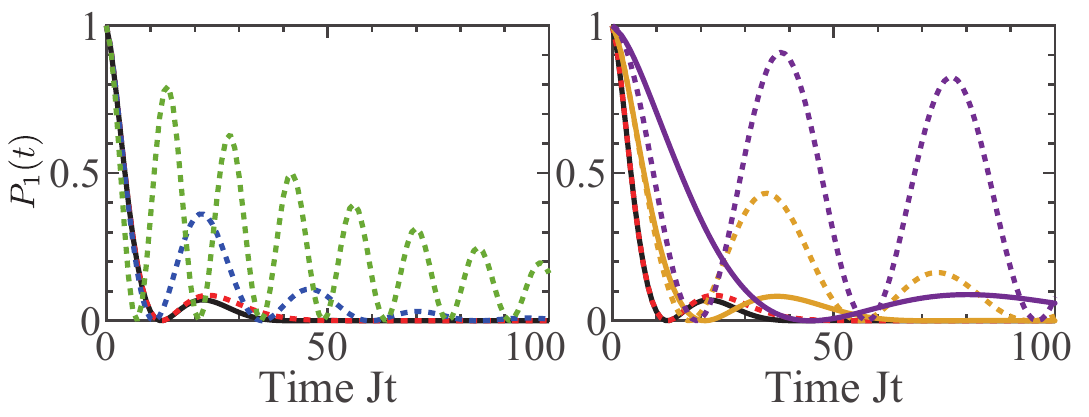}
		\caption{Spontaneous emission of single excitation in an emitter coupled to
			a strongly open bath with the finite size $N_b$. The solid curves
			denote the result obtained via the Green function approach, and the dashed
			line denotes the results obtained using the corresponding non-Hermitian
			Hamiltonian with a finite $N_b$, detuning $\Delta/J=0$ and Rabi frequency $%
			\Omega/J=1$. (a) Results for $N_b=20,60,150$ (green, blue, and red line, respectively) and the bath
			dissipation rate $\Gamma/2J=100$. The corresponding $\protect\alpha$ in Eq.~(%
			\protect\ref{alpha}) is $\protect\alpha=0.1,0.9,5.7$. (b) Results for $%
			\Gamma/2J=100,500,5000$ (red, yellow, and purple line, respectively) and $N_b=150$, corresponding
			to $\protect\alpha=5.7,1.1,0.1$. }
		\label{Fig:compare}
	\end{figure}

\begin{figure*}[tb]
	\centering
	\includegraphics[width=0.8\textwidth]{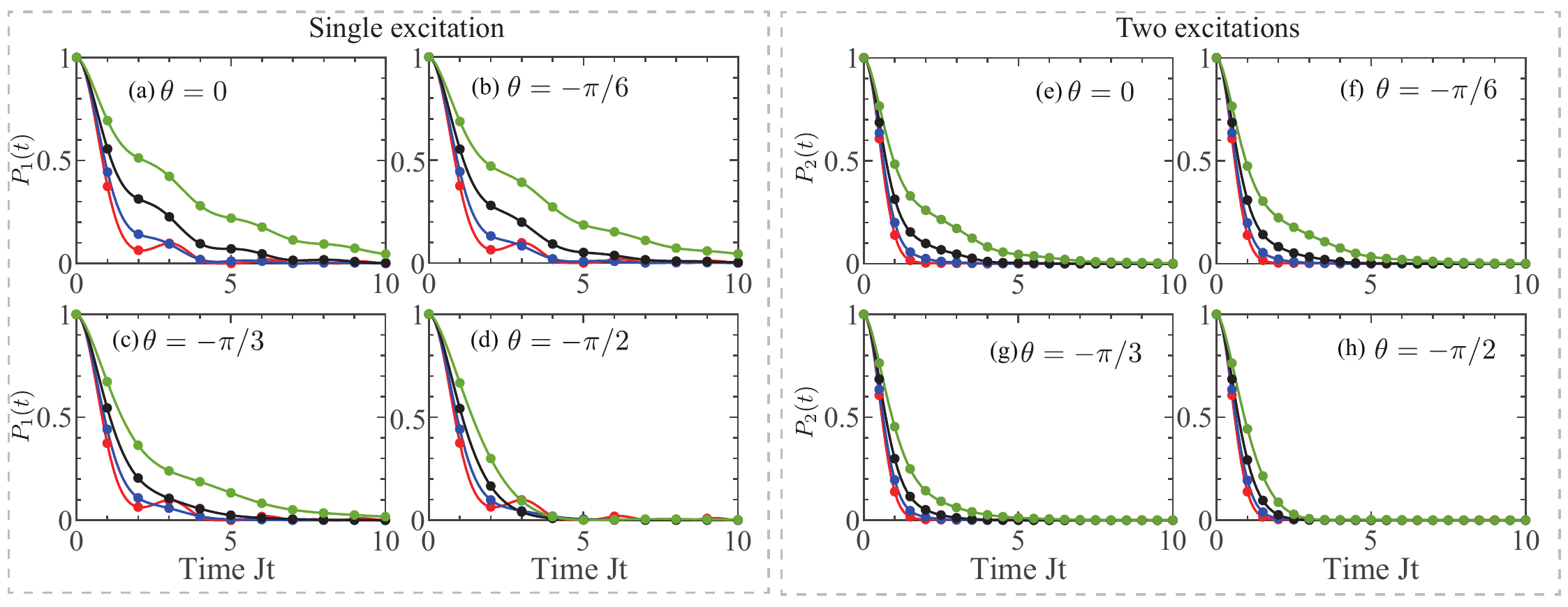}
	\caption{Spontaneous emission of (a) single excitation and (b) two
		excitations in an emitter coupled to a finite-size bath with various
		dissipation and coupling phase. In all panels, the solid line denotes the
		result obtained via the Green function approach with $\Omega/J=1$, and $%
		\Gamma/2J=0.005, 0.3, 1.005, 3$ (red, blue, black, and green line, respectively); the dots denote
		the results obtained using the corresponding non-Hermitian Hamiltonian with
		the finite size $N_b=60$. In (a), the single-excitation population $P_1(t)$
		is obtained as a function of time $t$ for $\Delta/J=0$. In (b), the two-excitation population $P_2(t)$ is obtained for $\Delta/J=-U/J=-1$. }
	\label{Fig:other}
\end{figure*}
	
	In this subsection, we derive the effective Hamiltonian in the two-excitation subspace using the perturbation theory, where $\Delta $ is finite
	and the resonance condition $U=-\Delta $ is assumed. The two-excitation
	sector is spanned in the basis $\{a^{\dagger 2}\left\vert 0\right\rangle /%
	\sqrt{2}\equiv \left\vert d\right\rangle ,a^{\dagger }b_{k}^{\dagger
	}\left\vert 0\right\rangle \equiv \left\vert k\right\rangle _{e%
	},b_{k}^{\dagger }b_{k^{\prime }}^{\dagger }\left\vert 0\right\rangle \equiv
	\left\vert kk^{\prime }\right\rangle \}$. Under the condition $U=-\Delta $,
	the doublon state $\left\vert d\right\rangle $ and states $\left\vert
	k\right\rangle _{e}$ are nearly degenerate. We can adiabatically
	eliminate the states $\left\vert kk^{\prime }\right\rangle $ as%
	\begin{eqnarray}
		{H}_{2}^{\prime\prime} &=&H_{0}+\sum_{k}\frac{\Omega ^{2}}{N}\sum_{p}\frac{1%
		}{\Delta -{\bar{\omega}_{p}}}\left\vert k\right\rangle _{e%
		}\left\langle k\right\vert  \notag \\
		&&+\frac{\Omega ^{2}}{2N}\sum_{kk^{\prime }}(\frac{1}{\Delta -\bar{\omega}%
			_{k}}+\frac{1}{\Delta -\bar{\omega}_{k^{\prime }}})\left\vert k\right\rangle
		_{e}\left\langle k^{\prime }\right\vert ,\notag
	\end{eqnarray}%
	where the unperturbed Hamiltonian%
	\begin{eqnarray}
		H_{0} &=&(2\Delta +U)\left\vert d\right\rangle \left\langle d\right\vert
		+\sum_{k}(\Delta +\bar{\omega}_{k})\left\vert k\right\rangle _{e%
		}\left\langle k\right\vert  \notag \\
		&&+\frac{\sqrt{2}\Omega }{\sqrt{N}}\sum_{k}(\left\vert k\right\rangle _{%
			e}\left\langle d\right\vert +\mathrm{H.c.})\notag
	\end{eqnarray}%
	and $\bar{\omega}_{k}=-i\Gamma -2iJ_{\mathrm{eff}}\cos k$.
	
	The adiabatic elimination of two-excitation states $\left\vert kk^{\prime
	}\right\rangle $ leads to the constant energy shift%
	\begin{equation}
		\delta \omega =\frac{\Omega ^{2}}{N}\sum_{p}\frac{1}{\Delta -\bar{\omega}_{p}%
		}\sim -i\frac{\Omega ^{2}e^{i(\pi/4)\text{sgn}(\Delta )}}{\sqrt{2\left\vert
				\Delta \right\vert \Gamma }}\notag
	\end{equation}%
	and the effective potential in the momentum space. The bound state%
	\begin{equation}
		\left\vert \Psi _{B}\right\rangle =u\left\vert d\right\rangle
		+\sum_{k}f_{k}\left\vert k\right\rangle _{e}\notag
	\end{equation}%
	of the effective Hamiltonian $H_{\mathrm{eff}}^{(2)}$ satisfies%
	\begin{eqnarray}
		E_{b}u&=&(2\Delta +U)u+\frac{\sqrt{2}\Omega }{\sqrt{N}}\sum_{k}f_{k},
		\notag \\
		E_{b}f_{k}&=&(\Delta +\bar{\varepsilon}_{k})f_{k}+\frac{ \sqrt{2}\Omega }{\sqrt{N}}u\notag\\
		&&+\frac{\Omega ^{2}}{2N}\sum_{p}(\frac{1}{ \Delta -\bar{\omega}_{k}}+\frac{1}{%
			\Delta -\bar{\omega}_{p}})f_{p} ,  \label{B2}
	\end{eqnarray}%
	where the bound state energy $E_{b}$ corresponds to the poles of $D_f(\omega )$ and $\bar{\varepsilon}_{k}=\bar{\omega}_{k}+\delta \omega $.
	
	The leading term of poles is determined by the Hamiltonian $H_{0}$. For the
	resonant case $U=-\Delta $, $H_{0}$ is exactly the Hamiltonian in the single-excitation sector with $\Delta =0$ and $\Omega \rightarrow \sqrt{2}\Omega $.
	Thus, the poles of $D_f(\omega )$ scale as $\Gamma ^{-1/3}$. The
	subleading correction can be obtained by the solution of Eq. (\ref{B2}). The
	solution%
	\begin{equation}
		u=\frac{\sqrt{2}\Omega }{\varepsilon _{b}\sqrt{N}}\sum_{k}f_{k}\notag
	\end{equation}%
	of $u$ leads to the equation%
	\begin{equation}
		\bar{\varepsilon}_{k}f_{k}+\frac{\Omega ^{2}}{2N}\sum_{p}(\frac{1}{\Delta -%
			\bar{\omega}_{k}}+\frac{1}{\Delta -\bar{\omega}_{p}})f_{p}+\frac{2\Omega ^{2}%
		}{\varepsilon _{b}N}\sum_{p}f_{p}=\varepsilon _{b}f_{k},\notag
	\end{equation}%
	where $\varepsilon _{b}=E_{b}-\Delta $.

	The formal solution%
	\begin{equation}
		f_{k}=\Omega ^{2}\bigg[\frac{\frac{2}{\varepsilon _{b}}C_{0}+\frac{1}{2}C_{1}}{%
			\varepsilon _{b}-\bar{\varepsilon}_{k}}+\frac{C_{0}}{2(\Delta -\bar{\omega}%
			_{k})(\varepsilon _{b}-\bar{\varepsilon}_{k})}\bigg]\notag
	\end{equation}%
	is determined by two constants%
	\begin{equation}
		C_{0}=\frac{1}{N}\sum_{p}f_{p},C_{1}=\frac{1}{N}\sum_{p}\frac{1}{\Delta -%
			\bar{\omega}_{p}}f_{p}.\notag
	\end{equation}%
	The constants obey the self-consistent equations%
	\begin{eqnarray}
		C_{0} &=&\bigg[\frac{2}{\varepsilon _{b}}I_{1}(\varepsilon _{b})+\frac{1}{2}%
		I_{2}(\varepsilon _{b})\bigg]C_{0}+\frac{1}{2}I_{1}(\varepsilon _{b})C_{1},
		\notag \\
		C_{1} &=&\bigg[\frac{2}{\varepsilon _{b}}I_{2}(\varepsilon _{b})+\frac{1}{2}%
		I_{3}(\varepsilon _{b})\bigg]C_{0}+\frac{1}{2}I_{2}(\varepsilon _{b})C_{1},\notag
	\end{eqnarray}%
	where the integrals%
	\begin{eqnarray}
		I_{1}(\varepsilon _{b}) &=&\Omega ^{2}\int \frac{dk}{2\pi }\frac{1}{%
			\varepsilon _{b}-\bar{\varepsilon}_{k}}=\Sigma_f(\varepsilon
		_{b}-\delta \omega ),  \notag \\
		I_{2}(\varepsilon _{b}) &=&\Omega ^{2}\int \frac{dk}{2\pi }\frac{1}{(\Delta -%
			\bar{\omega}_{k})(\varepsilon _{b}-\bar{\varepsilon}_{k})}  \notag \\
		&=&\frac{\Sigma_f(\Delta )-\Sigma_f(\varepsilon _{b}-\delta
			\omega )}{\varepsilon _{b}-\delta \omega -\Delta },  \notag \\
		I_{3}(\varepsilon _{b}) &=&\Omega ^{2}\int \frac{dk}{2\pi }\frac{1}{(\Delta -%
			\bar{\omega}_{k})^{2}(\varepsilon _{b}-\bar{\varepsilon}_{k})}  \notag \\
		&=&-\partial _{\Delta }I_{2}(\varepsilon _{b})\notag
	\end{eqnarray}%
	are evaluated analytically. Eventually, the poles of $D^{R}$ can be
	determined by%
	\begin{equation}
		\det \left(
		\begin{array}{cc}
			\frac{2}{\varepsilon _{b}}I_{1}(\varepsilon _{b})+\frac{1}{2}%
			I_{2}(\varepsilon _{b})-1 & \frac{1}{2}I_{1}(\varepsilon _{b}) \\
			\frac{2}{\varepsilon _{b}}I_{2}(\varepsilon _{b})+\frac{1}{2}%
			I_{3}(\varepsilon _{b}) & \frac{1}{2}I_{2}(\varepsilon _{b})-1%
		\end{array}%
		\right) =0.\notag
	\end{equation}

\section{Scalings for arbitrary open baths}
\label{sec:all scaling}
	
	In this section, we present the detailed derivations leading to the results in Table~{\ref{table1}}, for the open bath with arbitrary dissipative dispersions and dimensions.  
	
	We consider a tight-binding open bath with the dissipative band structure $\gamma({\bf k})$ at dimensions $d=1,2,3$. We concentrate on the purely dissipative bath and set $\epsilon({\bf k})=0$. We denote the minimum dissipation rate as $\gamma_\textrm{min}\equiv\textrm{min}[\gamma({\bf k})]$, which of course sits at the dissipation band edge, at some lattice momentum ${\bf k}_0$. We assume the bath is spatially homogeneous. Near the band gap (edge) $\gamma_\textrm{min}$, the dissipative dispersion can be approximated as
	\begin{equation}
\gamma({\bf k})\approx \gamma_\textrm{min}+c\Gamma |{\bf k}-{\bf k}_0|^\mu, \label{eq:gammak0}
\end{equation}
where the power $\mu$ depends on the specific form of $\gamma({\bf k})$, the coefficient $c\propto a^\mu$ with $a$ the lattice constant, and $\Gamma$ is some dissipation energy scale. We define the number of the bath modes with the decay rate $\gamma$ (i.e., dDOS) as $D_s(\gamma):=\int[d^d k/(2\pi)^d]\delta [\gamma-\gamma({\bf k})]$. Near $\gamma_\textrm{min}$, it follows from Eq.~(\ref{eq:gammak0}) that 
\begin{equation}
D_s(\gamma)= \frac{A}{(2\pi)^d}\frac{(\gamma-\gamma_\textrm{min})^{-1+(d/\mu)}}{\mu(c\Gamma)^{d/\mu}},\label{eq:Dgamma0}
\end{equation}
with $A=2,2\pi,4\pi$ for dimension $d=1,2,3$, respectively. Therefore, when $d/\mu<1$, $\lim_{\gamma\rightarrow\gamma_\textrm{min}}D_s(\gamma)\rightarrow \infty$, whereas for $d/\mu>1$, $\lim_{\gamma\rightarrow\gamma_\textrm{min}}D_s(\gamma)=0$. 

Now suppose an emitter with the on-site energy $\Delta$ is coupled to the open bath with a coupling rate $\Omega$. We begin with $\Delta=0$. Since the bath is purely dissipative, for convenience, we introduce $s=-i\omega$, so the poles of the single-particle Green function, corresponding to the quasibound states, are determined by 
\begin{equation}
s+\Sigma(s)=0, \label{eq:b}
\end{equation}
where the self-energy function $\Sigma(s)$ associated with the purely dissipative open bath is given by
\begin{equation}
\Sigma(s)=\Omega^2\int \frac{d^d k}{(2\pi)^d} \frac{1}{s+\gamma({\bf k})}. \label{eq:Sigma00}
\end{equation}

In the following, we analytically calculate Eq.~(\ref{eq:Sigma00}). Since the main contribution to the long-time emitter dynamics comes from the vicinity of $\gamma_\textrm{min}$, we expand $\gamma({\bf k})$ near $\gamma_\textrm{min}$ via Eq.~(\ref{eq:gammak0}) and change the integral variable from the quasimomentum to $\gamma$ via Eq.~(\ref{eq:Dgamma0}). We obtain
\begin{equation}
\Sigma=\frac{A\Omega^2}{(2\pi)^d}\frac{1}{\mu(c\Gamma)^{d/\mu}}\int_{\gamma_\textrm{min}}^{\Lambda}\frac{1}{s+\gamma}\frac{1}{(\gamma-\gamma_\textrm{min})^{1-(d/\mu)}}d{\gamma}, \label{eq:sig0}
\end{equation}
where we introduce a cutoff $\Lambda$. It follows from Eq.~(\ref{eq:gammak0}) that $\Lambda\propto \Gamma$. Denoting $s'=s+\gamma_\textrm{min}$ and $\Lambda'=\Lambda-\gamma_\textrm{min}$, Eq. (\ref{eq:sig0}) is calculated as
\begin{eqnarray}
\Sigma&=&\frac{A\Omega^2}{(2\pi)^d}\frac{(\Lambda')^{d/\mu}}{\mu(c\Gamma)^{d/\mu}}\frac{1}{s'}\int_0^{1}\left(1+\frac{\Lambda'}{s'}t\right)^{-1}t^{-1+(d/\mu)}dt\nonumber\\
&=&C\left(\frac{\Lambda'}{\Gamma}\right)^{d/\mu} \frac{1}{s'}F\left(1,\frac{d}{\mu},\frac{d}{\mu}+1;-\frac{\Lambda'}{s'}\right). \label{eq:Sigma}
\end{eqnarray}
Here, $F(\alpha,\beta,\zeta; z)$ is the hypergeometric function, and the precoefficient $C=B(\frac{d}{\mu},1)A\Omega^2/[(2\pi)^d\mu c^{{d}/{\mu}}]$ with $B$ the beta function. 

By substituting Eq.~(\ref{eq:Sigma}) into Eq.~(\ref{eq:b}), we can solve for the quasibound state solutions. We are interested in the regime $|{\Lambda'}/{s'}|\gg 1$ of Eq.~(\ref{eq:Sigma}), corresponding to when $\Gamma$ is the largest energy scale compared to all the other relevant scales as discussed in our manuscript. As such, we can expand the hypergeometric function in terms of the small parameter $|s'/{\Lambda'}|\ll 1$ and derive the analytical solutions at the leading order. In the following, we analyze the cases where $\gamma_\textrm{min}\neq 0$ and $\gamma_\textrm{min}=0$, respectively. 

\begin{itemize}
\item The bath has a dissipative gap ($\gamma_\textrm{min}\neq 0$)
\end{itemize}

(i) When $d/\mu<1$, we substitute $F[1,(d/\mu),(d/\mu)+1;-(\Lambda'/s')]=[1+(\Lambda'/s')]^{-d/\mu}F[(d/\mu),(d/\mu),(d/\mu)+1;1/(1+\frac{s'}{\Lambda'})]$ in Eq.~(\ref{eq:Sigma}). At the leading order of $|{s'}/{\Lambda'}|\ll 1$, we obtain $
\Sigma(s)=C'(s+\gamma_\textrm{min})^{-1+(d/\mu)}\Gamma^{-d/\mu}$, where  $C'=CF[(d/\mu),(d/\mu),(d/\mu)+1;1]$ and we use $s=s'-\gamma_\textrm{min}$. Inserting it into Eq.~(\ref{eq:b}), we find two quasibound states residing in the dissipative gap
\begin{eqnarray}
s_1&=&-C'\gamma^{-1+(d/\mu)}_\textrm{min}\Gamma^{-d/\mu},\label{eq:s1}\\
s_2&=&-\gamma_\textrm{min}+\alpha\Gamma^{-1/(\mu/d-1)},
\end{eqnarray}
with some coefficient $\alpha$. The long-term emitter dynamics is determined by $s_1$ in Eq.~(\ref{eq:s1}). 

(ii) When $d/\mu>1$, we insert $F[1,(d/\mu),(d/\mu)+1;-(\Lambda'/s')]=[1+(\Lambda'/s')]^{-1}F(1,1,(d/\mu)+1;1/(1+\frac{s'}{\Lambda'})]\approx (s'/\Lambda')F[1,1,(d/\mu)+1;1]$ into Eq.~(\ref{eq:Sigma}). Thus, at the leading order of $|{s'}/{\Lambda'}|\ll 1$, we obtain $\Sigma(s)=C^{''}({\Lambda'}^{-1+\frac{d}{\mu}}/\Gamma^{\frac{d}{\mu}})= \tilde{C}^{''}\Gamma^{-1}$, with $C^{''}=CF[1,1,(d/\mu)+1;1]$. Here, we exploit the fact that $\Lambda'=\Lambda-\gamma_\textrm{min}\propto \Gamma$ according to Eq.~(\ref{eq:Dgamma0}), with some constant $\tilde{C}^{''}$. In this case, we find one solution to Eq.~(\ref{eq:b}), describing a quasibound state in the dissipative gap:
\begin{equation}
s=-\tilde{C}^{''}\Gamma^{-1}.\label{eq:s2}
\end{equation}

(iii) When $d/\mu=1$, we substitute $F[1,1,2;-(\Lambda'/s')]=[\ln(1+\Lambda'/s')/(\Lambda'/s')]$ into Eq.~(\ref{eq:Sigma}). At the leading order of $|s'/\Lambda'|$, we have 
$\Sigma(s)= C({\Lambda'}^{\frac{d}{\mu}-1}/\Gamma^{\frac{d}{\mu}})\ln(\Lambda'/s')$. Since $\Lambda'\propto \Gamma$ as mentioned before, for $|s|\ll \gamma_\textrm{min}\ll \Gamma$, we have $\Sigma\approx\tilde{C}(1/\Gamma)\ln(\Lambda/\gamma_\textrm{min})$. Inserting it into Eq.~(\ref{eq:b}), we find one quasibound state in the dissipative gap:
\begin{equation}
s= -\tilde{C}\frac{\ln (\frac{\Gamma}{\gamma_\textrm{min}})}{\Gamma}. \label{eq:s3}
\end{equation}

\begin{itemize}
\item The bath has gapless dissipation ($\gamma_\textrm{min}=0$)
\end{itemize}

(i) When $d/\mu<1$, at the leading order of $|s/{\Lambda}|$, we obtain $
\Sigma(s)= C's^{-1+(d/\mu)}\Gamma^{-d/\mu}$, with $C'=CF[(d/\mu),(d/\mu),(d/\mu)+1;1]$. Therefore, Equation~(\ref{eq:b}) has solutions
\begin{equation}
s=(-C')^{1/(2-d/\mu)}\Gamma^{-1/(2\mu/d-1)}. \label{eq:s10}
\end{equation}

(ii) When $d/\mu>1$, we obtain the leading expression $\Sigma(s)=\tilde{C}^{''}\Gamma^{-1}$, with $C^{''}=CF(1,1,(d/\mu)+1;1)$. In this case, we find 
\begin{equation}
s=-\tilde{C}^{''}\Gamma^{-1},\label{eq:s20}
\end{equation}
which is the same as the gapped case. 

(iii) When $d/\mu=1$, we have the leading expression
$\Sigma(s)= C({\Lambda}^{\frac{d}{\mu}-1}/\Gamma^{\frac{d}{\mu}})\ln(\Lambda/s)$. Considering $\Lambda\propto \Gamma$, we find 
\begin{equation}
s= -\tilde{C}\frac{\ln (\frac{\Gamma}{\eta})}{\Gamma} \label{eq:s30}
\end{equation}
with some energy scale $\eta$.

When $\Delta\neq 0$, the quasibound state solutions are determined by 
\begin{equation}
s-i\Delta+\Sigma(s)=0. \label{eq:b}
\end{equation}
We can redefine $\tilde{s}=s-i\Delta$ and then follow similar steps as before. 

In summary, the above derivations lead to the results in Table \ref{table1} in the text. 

\bigskip

\section{Generic open bath with a finite size}
\label{sec:condition}
	
	In this section, we show that the emitter dynamics obtained from the Green
	function approach in Appendix~\ref{sec:G}, though developed in the thermodynamic
	limit, remains valid for a finite-size bath under the condition
	\begin{equation}
		\alpha\equiv\frac{\Omega}{\delta\gamma}=\frac{\Omega}{2\pi^2}\frac{N_b^2}{%
			\Gamma}\gg 1.  \label{alpha}
	\end{equation}
	Here, $\Omega$ is the Rabi frequency, $N_b$ is the finite size of the bath, $%
	\delta\gamma$ is the dissipative gap, and $\Gamma$ is the bath dissipation
	rate. Moreover, apart from the strong dissipation regime, we illustrate the
	emitter dynamics for a generic bath dissipation from the coherent limit $%
	\Gamma/2J\rightarrow 0$ to the intermediate regime $\Gamma/2J\sim 1$ and
	for various coupling phases $\theta\in[0, -\pi/2]$.
	
	We first validate the Green function approach for a finite-size, strongly
	dissipative, bath in three scenarios: $\alpha<1$, $\alpha\sim 1$, and $%
	\alpha>1$, respectively. For the spontaneous emission of a single excitation
	in an emitter, we compare the results obtained from two approaches: (i) We
	compute the population dynamics of single excitation, $P_1(t)$, by
	numerically evolving the state via the non-Hermitian emitter-bath Hamiltonian in Eq.~(\ref{eq:H}) with $U=0$; (ii) we use $P_1(t)=|G(t)|^2
	$ with $G(t)$ in Eq.~(\ref{eq:GRt}). We remark that Eq.~(\ref{alpha})
	indicates $\alpha$ depends on both $\Gamma$ and $N_b$. In Fig.~\ref%
	{Fig:compare}(a), we change $N_b$ and fix $\Gamma$. The black curve denotes
	the emitter dynamics from the Green function function, i.e., as in the
	thermodynamic limit, while the dashed curves denote the results from the
	approach (i) for various $N_b$. We see that when $\alpha<1$, the finite-size
	dynamics is significantly different from that in the thermodynamic limit,
	but the two becomes compatible with each other when $N_b$ is increased to $%
	\alpha\gg 1$. In Fig.~\ref{Fig:compare}(b), we change $\Gamma$ and fix $N_b$%
	. As shown, the two approaches agree well with each other by choosing a
	smaller $\Gamma$ so that $\alpha\gg 1$.
	
	Although our key results in the main text are for the strongly open
	bath, our developed approach applies to arbitrary parameter regimes of the
	bath. In Fig.~\ref{Fig:other}(a), we present the dynamics of single
	excitation with the detuning $\Delta=0$, when a bath of the size $N_b=60$ is
	nearly coherent $\Gamma/2J\rightarrow 0$, weakly dissipative $\Gamma/2J<1$,
	and in the intermediate regimes $\Gamma/2J\gtrsim 1$. We consider $%
	\theta=0,-\pi/6,-\pi/3,-\pi/2$, respectively. As $\alpha\gg 1$, we see that
	the results from the Green function approach (solid curves) agree well with
	the time evolution via the non-Hermitian (dots). In Fig.~\ref{Fig:other}(b),
	we show the spontaneous emissions of two excitations in an emitter coupled
	to the bath with the finite size $N_b=60$ and various $\Gamma$. The solid
	curves denote the results obtained from $P_2(t)=|D(t)|^2$, where $D(t)$
	is the Fourier transform of the fictitious two-particle Green function (\ref%
	{eq:ficD}). The dotted lines denote the calculation of $P_2(t)$ via the
	non-Hermitian Hamiltonian in Eq.~(\ref{H2}) of the main text. Again, we find
	perfect good agreement between the two.


\begin{thebibliography}{99}
		
		
		
		\bibitem{Bender1998}{C. M. Bender and S. Boettcher, Real Spectra in Non-Hermitian Hamiltonians Having PT Symmetry, Phys.
		Rev. Lett. \textbf{80}, 5243 (1998).}
		
		\bibitem{Wang2018} { S. Yao and Z. Wang, Edge States and Topological Invariants of Non-Hermitian Systems, Phys. Rev. Lett. \textbf{121},
			086803 (2018).}
		
		\bibitem{Kunst2018} {F. K. Kunst, E. Edvardsson, J. C. Budich, and E. J. Bergholtz,
			Biorthogonal Bulk-Boundary Correspondence in Non-Hermitian Systems, Phys. Rev. Lett. \textbf{121}, 026808 (2018).}
		
		\bibitem{Gong2018} {Z. Gong, Y. Ashida, K. Kawabata, K. Takasan, S.
			Higashikawa, and M. Ueda, Topological Phases of Non-Hermitian Systems, Phys. Rev. X \textbf{8}, 031079 (2018).}
		
		\bibitem{Zhou2018} {H. Zhou, C. Peng, Y. Yoon, C. W. Hsu, K. A. Nelson, L. Fu,
			J. D. Joannopoulos, M. Soljačić, and B. Zhen, Observation
			of Bulk Fermi Arc and Polarization Half Charge from
			Paired Exceptional Points, Science \textbf{359}, 1009 (2018).}
		
		\bibitem{Kawabata2019} {K. Kawabata, K. Shiozaki, M. Ueda, and M. Sato, Symmetry
			and Topology in Non-Hermitian Physics, Phys. Rev. X \textbf{9},
			041015 (2019).}
		
		\bibitem{Xue2020} {L. Xiao, T. Deng, K. Wang, G. Zhu, Z. Wang, W. Yi, and P.
			Xue, Non-Hermitian Bulk-Boundary Correspondence in
			Quantum Dynamics, Nat. Phys. \textbf{16}, 761 (2020).}
		
		\bibitem{Nunnenkamp2020} {C. C. Wanjura, M. Brunelli, and A. Nunnenkamp, Topological Framework for Directional Amplification in Driven-Dissipative Cavity Arrays, Nat. Commun. \textbf{11}, 3149 (2020).}
		
		
		\bibitem{Wang2021} {K. Wang, A. Dutt, C. C. Wojcik, and S. Fan, Topological
			Complex-Energy Braiding of Non-Hermitian Bands, Nature
			(London) \textbf{598}, 59 (2021).}
		
		\bibitem{JanReview2021} {E. J. Bergholtz, J. C. Budich, and F. K. Kunst, Exceptional
			Topology of Non-Hermitian Systems, Rev. Mod. Phys. \textbf{93},
			015005 (2021).}
		
		
	
		
		
		\bibitem{Fang2017}{ K. Fang, J. Luo, A. Metelmann, M. H. Matheny, F. Marquardt,
			A. A Clerk, and O. Painter, Generalized Non-reciprocity
			in an Optomechanical Circuit via Synthetic Magnetism and
			Reservoir Engineering, Nat. Phys. \textbf{13}, 465 (2017).}
		
		\bibitem{Xiao2017}{L. Xiao, X. Zhan, Z. H. Bian, K. K. Wang, X. Zhang, X. P.
			Wang, J. Li, K. Mochizuki, D. Kim, N. Kawakami, W. Yi,
			H. Obuse, B. C. Sanders, and P. Xue, Observation of
			Topological Edge States in Parity-Time-Symmetric Quantum Walks, Nat. Phys. \textbf{13}, 1117 (2017).}
		
		\bibitem{Ozturk2021}{F. E. Öztürk, T. Lappe, G. Hellmann, J. Schmitt, J. Klaers, F.
			Vewinger, J. Kroha, and M. Weitz, Observation of a Non-Hermitian Phase Transition in an Optical Quantum Gas,
			Science \textbf{372}, 88 (2021).}
		
		
		\bibitem{antiPT2016} { P. Peng, W. Cao, C. Shen, W. Qu, J. Wen, L. Jiang, and Y.
			Xiao, Anti-Parity–Time Symmetry with Flying Atoms, Nat.
			Phys. \textbf{12}, 1139 (2016).}
		
		\bibitem{Li2019} { J. Li, A. K. Harter, J. Liu, L. d. Melo, Y. N. Joglekar, and L.
			Luo, Observation of Parity-Time Symmetry Breaking Transitions in a Dissipative Floquet System of Ultracold Atoms,
			Nat. Commun. \textbf{10}, 855 (2019).}
		
		\bibitem{Takasu2020}{Y. Takasu, T. Yagami, Y. Ashida, R. Hamazaki, Y. Kuno,
			and Y. Takahashi, PT-Symmetric Non-Hermitian Quantum
			Many-Body System Using Ultracold Atoms in an Optical
			Lattice with Controlled Dissipation, Prog. Theor. Exp.
			Phys. \textbf{2020}, ptaa094 (2020).}
			
		
		\bibitem{Dongdong2022}{D. Hao, L. Wang, X. Lu, X. Cao, S. Jia, Y. Hu, and Y. Xiao,
			Topological Atomic Spin Wave Lattices by Dissipative
			Couplings, Phys. Rev. Lett. \textbf{130}, 153602 (2023).}
		
		
		\bibitem{Yanbo2022} {Q. Liang, D. Xie, Z. Dong, H. Li, H. Li, B. Gadway, W. Yi,
			and B. Yan, Dynamic Signatures of Non-Hermitian Skin
			Effect and Topology in Ultracold Atoms, Phys. Rev. Lett.
			\textbf{129}, 070401 (2022). }
		
		
		\bibitem{Wu2019}{Y. Wu, W. Liu, J. Geng, X. Song, X. Ye, C.-K. Duan, X.
			Rong, and J. Du, Observation of Parity-Time Symmetry
			Breaking in a Single-Spin System, Science \textbf{364}, 878
			(2019).}
		
		
		\bibitem{Naghiloo2019}{M. Naghiloo, M. Abbasi, Y. N. Joglekar, and K. W. Murch,
			Quantum State Tomography across the Exceptional Point in
			a Single Dissipative Qubit, Nat. Phys. \textbf{15}, 1232 (2019).}
		
		
		\bibitem{FanSH2021} {C. Leefmans, A. Dutt, J. Williams, L. Yuan, M. Parto, F.
			Nori, S. Fan, and A. Marandi, Topological Dissipation in a
			Time-Multiplexed Photonic Resonator Network, Nat. Phys.
			\textbf{18}, 442 (2022).}
		
		
		\bibitem{Pickup2020}{L. Pickup, H. Sigurdsson, J. Ruostekoski, and P. G.
			Lagoudakis, Synthetic Band-Structure Engineering in
			Polariton Crystals with Non-Hermitian Topological
			Phases, Nat. Commun. \textbf{11}, 4431 (2020).}
		
		
		\bibitem{Pernet2022}{N. Pernet, P. St-Jean, D. D. Solnyshkov, G. Malpuech, N. C.
			Zambon, Q. Fontaine, B. Real, O. Jamadi, A. Lemaître, M.
			Morassi, L. L. Gratiet, T. Baptiste, A. Harouri, I. Sagnes, A.
			Amo, S. Ravets, and J. Bloch, Gap Solitons in a One-Dimensional Driven-Dissipative Topological Lattice, Nat.
			Phys. \textbf{18}, 678 (2022).}
		
		
		
		
		\bibitem{Longhi2016}{S. Longhi, Quantum Decay and Amplification in a Non-Hermitian Unstable Continuum, Phys. Rev. A \textbf{93}, 062129
			(2016).}
				
		\bibitem{Nakagawa2018}{ M. Nakagawa, N. Kawakami, and M. Ueda, Non-Hermitian
			Kondo Effect in Ultracold Alkaline-Earth Atoms, Phys. Rev.
			Lett. \textbf{121}, 203001 (2018).}
		
		
		\bibitem{Yamamoto2019}{ K. Yamamoto, M. Nakagawa, K. Adachi, K. Takasan, M.
			Ueda, and N. Kawakami, Theory of Non-Hermitian Fermionic Superfluidity with a Complex-Valued Interaction,
			Phys. Rev. Lett. \textbf{123}, 123601 (2019).}
		
		
		
		\bibitem{Gopalakrishnan2021}{S. Gopalakrishnan and M. J. Gullans, Entanglement and
			Purification Transitions in Non-Hermitian Quantum
			Mechanics, Phys. Rev. Lett. \textbf{126}, 170503 (2021).}
		
		
		\bibitem{Roccati2022}{ F. Roccati, S. Lorenzo, G. Calajò, G. M. Palma, A. Carollo,
			and F. Ciccarello, Exotic Interactions Mediated by a Non-Hermitian Photonic Bath, Optica \textbf{9}, 565 (2022).}
		
		\bibitem{Gong202201}{Z. Gong, M. Bello, D. Malz, and F. K. Kunst, Bound States
			and Photon Emission in Non-Hermitian Nanophotonics,
			Phys. Rev. A \textbf{106}, 053517 (2022).}
		
		\bibitem{Gong202202}{Z. Gong, M. Bello, D. Malz, and F. K. Kunst, Anomalous
			Behaviors of Quantum Emitters in Non-Hermitian Baths,
			Phys. Rev. Lett. \textbf{129}, 223601 (2022).}
		
		
		
		
		\bibitem{Poyatos1996}{J. F. Poyatos, J. I. Cirac, and P. Zoller, Quantum Reservoir
			Engineering with Laser Cooled Trapped Ions, Phys. Rev.
			Lett. \textbf{77}, 4728 (1996).}
		
		\bibitem{Diehl2008}{ S. Diehl, A. Micheli, A. Kantian, B. Kraus, H. P. Büchler, and
			P. Zoller, Quantum States and Phases in Driven Open
			Quantum Systems with Cold Atoms, Nat. Phys. \textbf{4}, 878 (2008).}
		
		\bibitem{Verstraete2009}{F. Verstraete, M. M. Wolf, and J. I. Cirac, Quantum Computation and Quantum-State Engineering Driven by Dissipation, Nat. Phys. \textbf{5}, 633 (2009).}
		
		\bibitem{Diehl2011}{S. Diehl, E. Rico, M. A. Baranov, and P. Zoller, Topology by
			Dissipation in Atomic Quantum Wires, Nat. Phys. \textbf{7}, 971
			(2011).}
		
		
		\bibitem{Muller2012}{M. Müller, S. Diehl, G. Pupillo, and P. Zoller, Engineered
			Open Systems and Quantum Simulations with Atoms and
			Ions, Adv. At. Mol. Opt. Phys. \textbf{61}, 1 (2012).}
		
		
		\bibitem{Tomadin2012}{A. Tomadin, S. Diehl, M. D. Lukin, P. Rabl, and P. Zoller,
			Reservoir Engineering and Dynamical Phase Transitions in
			Optomechanical Arrays, Phys. Rev. A \textbf{86}, 033821 (2012).}
		
		\bibitem{Douglas2015}{J. S. Douglas, H. Habibian, C.-L. Hung, A. V. Gorshkov,
			H. J. Kimble, and D. E. Chang, Quantum Many-Body
			Models with Cold Atoms Coupled to Photonic Crystals,
			Nat. Photonics \textbf{9}, 326 (2015).}
		
		\bibitem{Ramos2014}{T. Ramos, H. Pichler, A. J. Daley, and P. Zoller, Quantum
			Spin Dimers from Chiral Dissipation in Cold-Atom Chains,
			Phys. Rev. Lett. \textbf{113}, 237203 (2014).}
		
		\bibitem{Chang2018} {D. E. Chang, J. S. Douglas, A. González-Tudela, C.-L.
			Hung, and H. J. Kimble, Colloquium: Quantum Matter
			Built from Nanoscopic Lattices of Atoms and Photons,
			Rev. Mod. Phys. \textbf{90}, 031002 (2018).}
		
		\bibitem{Cian2019}{Z.-P. Cian, G. Zhu, S.-K. Chu, A. Seif, W. DeGottardi, L.
			Jiang, and Mohammad Hafezi, Photon Pair Condensation
			by Engineered Dissipation, Phys. Rev. Lett. \textbf{123}, 063602
			(2019).}
		
		\bibitem{Harrington2022}{ P. M. Harrington, E. J. Mueller, and K. W. Murch,
			Engineered Dissipation for Quantum Information Science,
			Nat. Rev. Phys. \textbf{4}, 660 (2022). }
		
		
		
		\bibitem{Paul1982}{H. Paul, Photon Antibunching, Rev. Mod. Phys. \textbf{54}, 1061
			(1982).}
		
		
		
		\bibitem{Bamba2011}{M. Bamba, A. Imamoğlu, I. Carusotto, and C. Ciuti, Origin
			of Strong Photon Antibunching in Weakly Nonlinear Photonic Molecules, Phys. Rev. A \textbf{83}, 021802(R) (2011).}
		
		\bibitem{Kong2022}{X. Kong, C. Navarrete-Benlloch, and Y. Chang, Accessing
			Strongly-Coupled Systems without Compromising Them,
			arXiv:2204.04212.} 
		
		
		
		
		
		
		
		\bibitem{Balatsky2006}{A. V. Balatsky, I. Vekhter, and J.-X. Zhu, Impurity-Induced
			States in Conventional and Unconventional Superconductors, Rev. Mod. Phys. \textbf{78}, 373 (2006).}
		
		\bibitem{Hur2012}{K. L. Hur, Kondo Resonance of a Microwave Photon, Phys.
			Rev. B \textbf{85}, 140506(R) (2012). }
		
		\bibitem{Goldstein2013}{ M. Goldstein, M. H. Devoret, M. Houzet, and L. I. Glazman,
			Inelastic Microwave Photon Scattering off a Quantum
			Impurity in a Josephson-Junction Array, Phys. Rev. Lett.
			\textbf{110}, 017002 (2013).}
		
		\bibitem{Shitao2016} {T. Shi, Y-H. Wu, A. González-Tudela, and J. I. Cirac, Bound
			States in Boson Impurity Models, Phys. Rev. X \textbf{6}, 021027
			(2016).}
		
		
		
		
		
		
		
		\bibitem{John1990}{S. John and J. Wang, Quantum Electrodynamics near a
			Photonic Band Gap: Photon Bound States and Dressed
			Atoms, Phys. Rev. Lett. \textbf{64}, 2418 (1990).}
		
		\bibitem{Nakazato1996}{ H. Nakazato, M. Namiki, and S. Pascazio, Temporal
			Behavior of Quantum Mechanical Systems, Int. J. Mod.
			Phys. B \textbf{10}, 247 (1996).}
		
		\bibitem{fQHE2012} {R. O. Umucalılar and I. Carusotto, Fractional Quantum
			Hall States of Photons in an Array of Dissipative Coupled
			Cavities, Phys. Rev. Lett. \textbf{108}, 206809 (2012).}
		
		\bibitem{Tudela2015} {A. González-Tudela, V. Paulisch, D. E. Chang, H. J.
			Kimble, and J. I. Cirac, Deterministic Generation of Arbitrary Photonic States Assisted by Dissipation, Phys. Rev.
			Lett. \textbf{115}, 163603 (2015).}
			
		\bibitem{Sollner2015} {I. Söllner, S. Mahmoodian, S. L. Hansen, L. Midolo,
			A. Javadi, G. Kiršanskė, T. Pregnolato, H. El-Ella, E. H.
			Lee, J. D. Song, S. Stobbe, and P. Lodahl, Deterministic
			Photon-Emitter Coupling in Chiral Photonic Circuits,
			Nat. Nanotechnol. \textbf{10}, 775 (2015).}
		
		\bibitem{Ramos2016}{T. Ramos, B. Vermersch, P. Hauke, H. Pichler, and P.
			Zoller, Non-Markovian Dynamics in Chiral Quantum
			Networks with Spins and Photons, Phys. Rev. A \textbf{93},
			062104 (2016).}
		
		\bibitem{Liu2016} {Y. Liu and A. A. Houck, Quantum Electrodynamics near a
			Photonic Bandgap, Nat. Phys. \textbf{13}, 48 (2017).}
		
		\bibitem{Shi2017} { T. Shi, H. J. Kimble, and J. I. Cirac, Topological Phenomena in Classical Optical Networks, Proc. Natl. Acad. Sci.
			U.S.A. \textbf{114}, E8967 (2017).}
		
		
		
		\bibitem{Tudela2018} {A. González-Tudela and J. I. Cirac, Exotic Quantum
			Dynamics and Purely Long-Range Coherent Interactions
			in Dirac Conelike Baths, Phys. Rev. A \textbf{97}, 043831 (2018).} 
		
		
		
		\bibitem{Shi2018} {T. Shi, Y-H. Wu, A. González-Tudela, and J. I. Cirac,
			Effective Many-Body Hamiltonians of Qubit-Photon Bound
			States, New J. Phys. \textbf{20}, 105005 (2018).}
		
		\bibitem{Bello2019} {M. Bello, G. Platero, J. I. Cirac, and A. González-Tudela,
			Unconventional Quantum Optics in Topological Waveguide
			QED, Sci. Adv. \textbf{5}, eaaw0297 (2019).}
		
		\bibitem{Luengo2019} { J. Argüello-Luengo, A. González-Tudela, T. Shi, P. Zoller,
			and J. I. Cirac, Analogue Quantum Chemistry Simulation,
			Nature (London) \textbf{574}, 215 (2019).}
		
		\bibitem{Luengo2020} {J. Argüello-Luengo, A. González-Tudela, T. Shi, P. Zoller,
			and J. I. Cirac, Quantum Simulation of Two-Dimensional
			Quantum Chemistry in Optical Lattices, Phys. Rev. Res. \textbf{2},
			042013(R) (2020).}
		
		\bibitem{Luengo2021} {J. Argüello-Luengo, T. Shi, and A. González-Tudela,
			Engineering Analog Quantum Chemistry Hamiltonians
			Using Cold Atoms in Optical Lattices, Phys. Rev. A \textbf{103},
			043318 (2021).} 
		
		
		\bibitem{Shitao2015}{T. Shi, D. E. Chang, and J. I. Cirac, Multiphoton-Scattering
			Theory and Generalized Master Equations, Phys. Rev. A
			\textbf{92}, 053834 (2015).}
		
		\bibitem{Yue2016}{Y. Chang, A. González-Tudela, C. S. Muñoz, C. Navarrete-Benlloch, and T. Shi, Deterministic Down-Converter and
			Continuous Photon-Pair Source within the Bad-Cavity
			Limit, Phys. Rev. Lett. \textbf{117}, 203602 (2016).}
		
		
		
		
		\bibitem{Dalibard1992} {J. Dalibard, Y. Castin, and K. Mølmer, Wave-Function
			Approach to Dissipative Processes in Quantum Optics,
			Phys. Rev. Lett. \textbf{68}, 580 (1992).}
		
		\bibitem{Daley2014}{A. J. Daley, Quantum Trajectories and Open Many-Body
			Quantum Systems, Adv. Phys. \textbf{63}, 77 (2014).}
		
		
		
		
		
		
		\bibitem{Rammer2007}{J. Rammer, Quantum Field Theory of Non-equilibrium
			States (Cambridge University Press, Cambridge, England,
			2007).}
		
		\bibitem{Sieberer2016}{L. M. Sieberer, M. Buchhold, and S. Diehl, Keldysh Field
			Theory for Driven Open Quantum Systems, Rep. Prog.
			Phys. \textbf{79}, 096001 (2016).}
		
		
		
		
		
		\bibitem{Misra1977} {B. Misra and E. C. G. Sudarshan, The Zeno’s Paradox in
			Quantum Theory, J. Math. Phys. (N.Y.) \textbf{18}, 756 (1977).}
		
		\bibitem{Itano1990}{W. M. Itano, D. J. Heinzen, J. J. Bollinger, and D. J.
			Wineland, Quantum Zeno Effect, Phys. Rev. A \textbf{41}, 2295
			(1990).}
		
		\bibitem{Scully1997} {M. O. Scully and M. S. Zubairy, Quantum Optics
			(Cambridge University Press, Cambridge, England, 1997).}
		
		\bibitem{Wang2008} {Xiang-Bin Wang, J. Q. You, and F. Nori, Quantum Entanglement via Two-Qubit Quantum Zeno Dynamics, Phys.
			Rev. A \textbf{77}, 062339 (2008).}
		
		\bibitem{Maniscalco2008} {S. Maniscalco, F. Francica, R. L. Zaffino, N. L. Gullo, and F.
			Plastina, Protecting Entanglement via the Quantum Zeno
			Effect, Phys. Rev. Lett. \textbf{100}, 090503 (2008).}
		
		\bibitem{Syassen2008}{ N. Syassen, D. M. Bauer, M. Lettner, T. Volz, D. Dietze, J. J.
			García-Ripoll, J. I. Cirac, G. Rempe, and S. Dürr, Strong
			Dissipation Inhibits Losses and Induces Correlations in
			Cold Molecular Gases, Science \textbf{320}, 1329 (2008).}
		
		\bibitem{Han2009}{Y.-J. Han, Y.-H. Chan, W. Yi, A. J. Daley, S. Diehl, P. Zoller,
			and L.-M. Duan, Stabilization of the p-Wave Superfluid State
			in an Optical Lattice, Phys. Rev. Lett. \textbf{103}, 070404 (2009).}
			
		\bibitem{Signoles2014} {A. Signoles, A. Facon, D. Grosso, I. Dotsenko, S. Haroche,
			Jean-Michel Raimond, M. Brune, and S. Gleyzes, Confined
			Quantum Zeno Dynamics of a Watched Atomic Arrow, Nat.
			Phys. \textbf{10}, 715 (2014).}
		
		\bibitem{Hu2015}{ Y. Hu, Z. Cai, M. A. Baranov, and P. Zoller, Majorana
			Fermions in Noisy Kitaev Wires, Phys. Rev. B \textbf{92}, 165118
			(2015).}
		
		
		\bibitem{Froml2020}{H. Fröml, C. Muckel, C. Kollath, A. Chiocchetta, and S.
			Diehl, Ultracold Quantum Wires with Localized Losses:
			Many-Body Quantum Zeno Effect, Phys. Rev. B \textbf{101},
			144301 (2020).}
		
		

		
		
		
		
		\bibitem{Kessler2012}{E. M. Kessler, Generalized Schrieffer-Wolff Formalism for
			Dissipative Systems, Phys. Rev. A \textbf{86}, 012126 (2012).}
		
		\bibitem{Reiter2012}{ F. Reiter and A. S. Sørensen, Effective Operator Formalism
			for Open Quantum Systems, Phys. Rev. A \textbf{85}, 032111
			(2012).}
		
		
		

		\bibitem{SM} {See details in the Appendixes~\ref{sec:G} and~\ref{sec:scaling}.}
		
		
		\bibitem{Shi2011}{T. Shi, S. Fan, and C. P. Sun, Two-Photon Transport in a
			Waveguide Coupled to a Cavity in a Two-Level System,
			Phys. Rev. A \textbf{84}, 063803 (2011).}
		
		\bibitem{Shi2013}{T. Shi and S. Fan, Two-Photon Transport through a
			Waveguide Coupling to a Whispering-Gallery Resonator
			Containing an Atom and Photon-Blockade Effect, Phys.
			Rev. A \textbf{87}, 063818 (2013).}
		

		\bibitem{Zhang2019}{Y.-X. Zhang and K. Mølmer, Theory of Subradiant States of
			a One-Dimensional Two-Level Atom Chain, Phys. Rev.
			Lett. \textbf{122}, 203605 (2019).}
		
		
		\bibitem{footnote}{This approximation is generally good if the major contribution to $\int (d\omega/2\pi)G_f(\omega)e^{-i\omega t}$ comes from the residue associated with $\epsilon_s^{(1)}$. }
				
		\bibitem{Goldman2016}{N. Goldman, J. C. Budich, and P. Zoller, Topological
			Quantum Matter with Ultracold Gases in Optical Lattices,
			Nat. Phys. \textbf{12}, 639 (2016).}
		
        

		
		
		



		
		
		
		
		
		
		
		
		
		
		
		
		

		

		
		
		
		
\end{thebibliography}
	\end{document}